\documentclass[%
 reprint,
 amsmath,amssymb,
 aps,
rmp,
floatfix,
]{revtex4-2}
\usepackage{graphicx}
\usepackage{dcolumn}
\usepackage{bm}

\begin{document}

\preprint{APS/123-QED}

\title{Optical diagnostics of laser-produced plasmas}

\author{S. S. Harilal}
\email{hari@pnnl.gov}
\affiliation{Pacific Northwest National Laboratory, Richland, Washington 99352, USA.}

\author{M. C. Phillips}
\affiliation{James C. Wyant College of Optical Sciences, University of Arizona, Tucson, Arizona 85721, USA.}

\author{D. H. Froula}
\affiliation{Laboratory for Laser Energetics, University of Rochester, Rochester, New York 14623, USA}

\author{K. K. Anoop}
\author{R. C. Issac}
\affiliation{Department of Physics, Cochin University of Science and Technology, Cochin, Kerala 682022, India}

\author{F. N. Beg}
\affiliation{Center for Energy Research, University of California San Diego, La Jolla, California 92093, USA}

\date{\today}

\begin{abstract}
Laser-produced plasmas (LPPs) engulf exotic and complex conditions ranging in temperature, density, pressure, magnetic and electric fields, charge states, charged particle kinetics, and gas-phase reactions, based on the irradiation conditions, target geometries, and the background cover gas. The application potential of the LPP is so diverse that it generates considerable interest for both basic and applied research areas. The fundamental research on LPPs can be traced back to the early 1960s, immediately after the invention of the laser. In the 1970s, the laser was identified as a tool to pursue inertial confinement fusion, and since then, several other technologies emerged out of LPPs. These applications prompted the development and adaptation of innovative diagnostic tools for understanding the fundamental nature and spatiotemporal properties of these complex systems. Although most of the  traditional characterization techniques developed for other plasma sources can be used to characterize the LPPs, care must be taken to interpret the results because of their small size, transient nature, and inhomogeneities. The existence of the large spatiotemporal density and temperature gradients often necessitates non-uniform weighted averaging over distance and time. Among the various plasma characterization tools, optical-based diagnostic tools play a key role in the accurate measurements of LPP parameters. The optical toolbox contains optical spectroscopy (emission, absorption, and fluorescence), passive and active imaging and optical probing methods (shadowgraphy, Schlieren, interferometry, Thomson scattering, deflectometry, and velocimetry). Each technique is useful for measuring a specific property, and its use is limited to a certain time span during the LPP evolution because of the sensitivity issues related to the selected measuring tool. Therefore, multiple diagnostic tools are essential for a comprehensive insight into the entire plasma behavior. In recent times, the improvements in performance in the lasers and detector systems expanded the capability of the aforementioned passive and active diagnostics tools. This review provides an overview of optical diagnostic tools frequently employed for the characterization of the LPPs and emphasizes techniques, associated assumptions, and challenges. Considering most of the industrial and other applications of the LPP belong to low to moderate laser intensities ($10^8-10^{15}$ Wcm$^{-2}$), this review focuses on diagnostic tools pertaining to this regime.  

\end{abstract}

\maketitle

\tableofcontents

\section{\label{sec:Introduction}Introduction}

Plasma is the fourth state of matter, and there are several ways to generate plasmas in the laboratory. One of the methods commonly used is to focus an intense pulsed laser on a matter of interest. Among the various laboratory plasmas, the laser-produced plasma (LPP) may be one of the most complex systems because of its transient nature combined with spatial inhomogeneities \cite{Radziemski1983}. LPPs are also characterized by high temperatures and high densities of electrons and ions. Although the history of fundamental research on LPPs can be traced back to the early 1960s with the earliest article about LPPs appearing immediately after the invention of the laser \cite{Brech1962, Linlor1962}, most advances in using LPPs for various applications emerged after lasers were proposed as drivers for inertial confinement fusion. Currently, LPPs find applications in a wide variety of fields such as material science \cite{Chrisey1994}, analytical instrumentation  \cite{russo2013laser}, spectroscopy \cite{Musazzi2014Book}, planetary science \cite{Singh2020-LIBSbook}, geology \cite{fabre2020advances}, agriculture \cite{nicolodelli2019recent}, high energy density physics (HEDP) \cite{Drake2006Book}, laser ion source \cite{yeates2010dcu}, laser ablation propulsion \cite{phipps2010laser}, laser processing (micromachining, cutting, etc.) \cite{gattass2008femtosecond}, and medicine \cite{gitomer1991laser}, among others. 

 All LPP applications require the availability of reliable lasers with different characteristics (laser energy, pulse width, wavelength, beam profile) and a deeper understanding of the LPP properties by developing and using state-of-the-art diagnostic tools in conjunction with modeling efforts. In materials science and nanotechnology, pulsed laser deposition (PLD) is a well-established method for fabricating thin films of complex oxides \cite{singh1990pulsed, willmott2000pulsed} where a strong correlation between the dynamics of the LPP and the quality of thin films exists \cite{Chrisey1994, kwok1997correlation}. In spectroscopic applications, such as in laser-induced breakdown spectroscopy (LIBS), the LPP is generated via stoichiometric ablation, and the subsequent light emission from the plasma is used for the qualitative and quantitative elemental and isotopic analysis of multi-element samples \cite{Miziolek2006Book, Cremers2013Book, Musazzi2014Book, 2018-APR-Hari}. The LPP is also used as the front end for various analytical tools such as laser ablation inductively coupled plasma mass spectrometry (LA-ICP-MS) \cite{2015-SR-Nicole}, LA inductively coupled plasma optical emission spectroscopy (LA-ICP-OES) \cite{trejos2013forensic}, LA time-of-flight mass spectrometry (LA-TOF-MS) \cite{ahmad2018qualitative}, LA laser-induced fluorescence (LA-LIF) \cite{miyabe2015ablation}, and LA laser-absorption spectroscopy (LA-LAS) \cite{tarallo2016bah}.

The LPP can produce high-brightness extreme ultraviolet (EUV) radiation, which is currently being used as a photon  source in nanolithography \cite{gwyn1998extreme, stamm2004extreme, banine2011physical}. LPPs are also considered a potential radiation source for water-window microscopy \cite{kondo1994optimization}. Higher harmonic generation from an LPP is a promising tool for generating coherent EUV sources \cite{singhal2010high}. LPPs are also recognized as a promising medium for generating intense pulsed X-ray and gamma-ray radiation sources \cite{norreys1999observation, rajeev2003metal, issac2004ultra, cipiccia2011gamma}, collimated ion beams \cite{fews1994plasma, li2019ionization}, and plasma-based particle accelerators \cite{joshi2003plasma, corde2013femtosecond}. The laser-plasma accelerators (LPAs) are capable of producing fields $10^4$ times those of conventional accelerators, and advanced LPAs are going to play a significant role in accelerator physics, radiotherapy, and high-energy physics applications in the upcoming years \cite{malka2005laser, brunetti2010low, bartal2012focusing, subiel2014dosimetry, weichman2020laser}. The LPP is also a powerful and compact source of multi-MeV ions \cite{krushelnick2000ultrahigh}. Some of the recent scientific achievements obtained with LPPs are neutron source development \cite{mirfayzi2020proof}, extreme ionization of heavy atoms \cite{hollinger2020extreme}, fusion reactions \cite{labaune2013fusion}, and terahertz generation \cite{Herzer2018}. An LPP-based emission spectroscopy system has been deployed in the Mars Curiosity and Perseverance rovers for elemental analysis of rocks and soil \cite{clegg2017recalibration, manrique2020supercam}. In addition to these, the LPP is useful for recreating astrophysical plasmas in the laboratory and used as a surrogate for studying plasma chemistry occurring in high-explosion and nuclear events \cite{ledingham2003applications, Kimblin2017, Kautz2021LaserinducedFO}.

\begin{figure}[t]
\includegraphics[width=\linewidth]{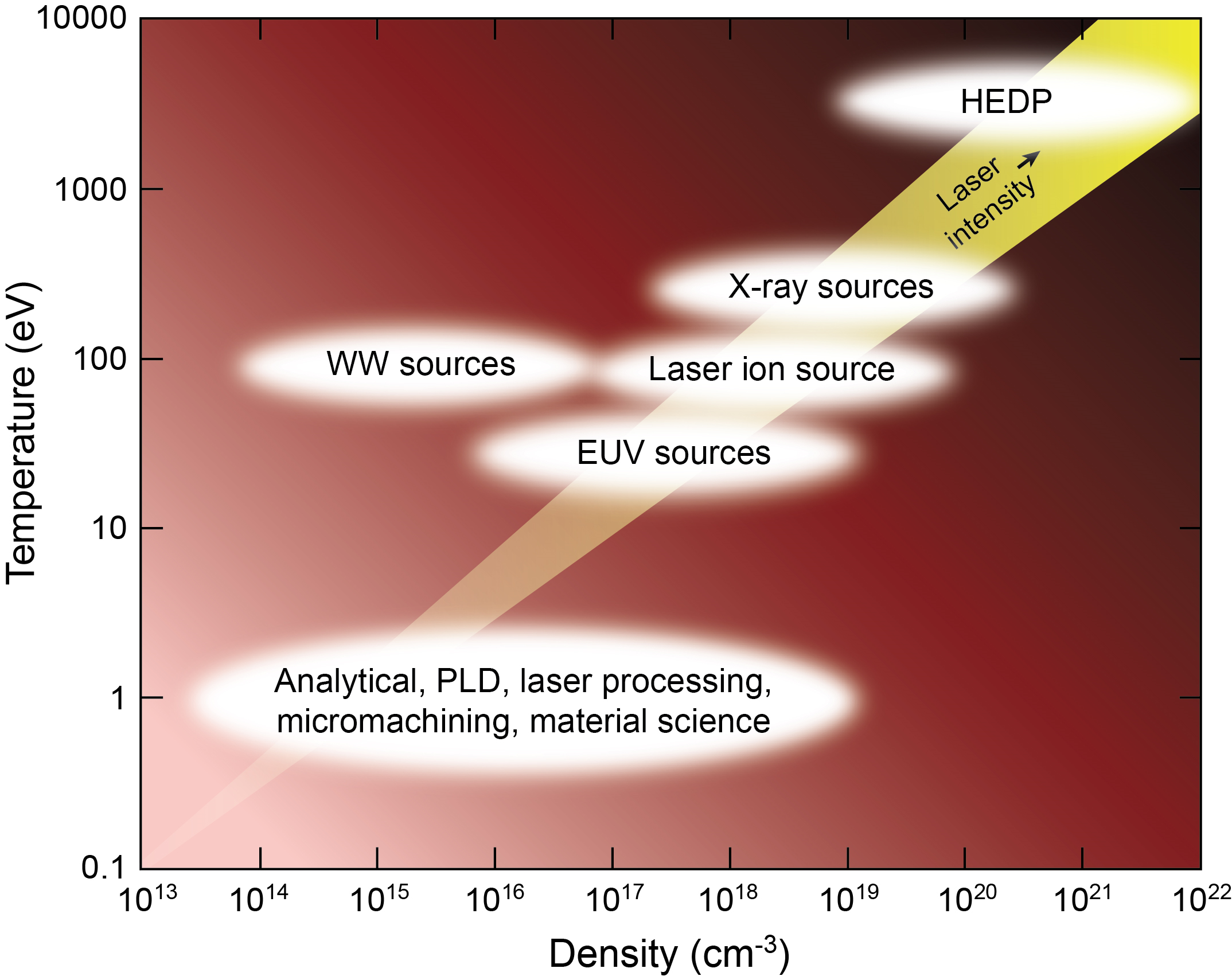}
\caption{\label{fig:1} Typical properties of LPPs used for various applications.}
\end{figure}

Plasmas are traditionally defined as partially ionized gaseous mediums with total charge neutrality and are characterized and classified according to their temperature and density. Most of the other plasma parameters, \emph{viz.} particle kinetics, opacity, pressure, the energy of the shocks, etc., are directly and indirectly connected to the temperature and density of the plasma system. The laser parameters such as energy, wavelength, and pulse length turn out to be crucial control knobs for changing the fundamental parameters of the LPP. For example, by changing the laser power density, the initial peak temperature of the LPP can be tuned from a few thousand Kelvin to millions of Kelvin. The peak density of LPP systems also changes by several orders of magnitude based on laser power density and ambient conditions. Fig.~\ref{fig:1} shows an approximate map of the temperature and density range of LPPs used in various applications. Regardless of the initial conditions of the LPP, the temperature and density decrease as the plasma expands and eventually reaches equilibrium with the surrounding environment. Thus, the physical conditions within the LPP can span a wide range of temperatures, pressures, and atomic/electron densities, with a corresponding change in chemical composition  (ionized atoms, neutral atoms, molecules, clusters, etc.).

The physics of the LPP generation and subsequent evolution changes significantly with laser intensity. The physical properties of plasmas produced by similar laser intensities with varying pulse widths will also be vastly different due to differences in the physics of plasma generation.    For analytical applications such as LIBS, the plasma source should provide a copious amount of light from the excited-state population of atomic species where the required temperature and density are in the range of 1-3 eV and $10^{15}-10^{17}cm^{-3}$ \cite{Singh2020-LIBSbook}, respectively, and hundreds of µJ to tens of mJ of laser pulse energies are therefore used for plasma production \cite{russo2013laser, chen2015comparison}. For LPP sources for EUV lithography, a bright plasma source emitting at a narrow band wavelength region around 13.5 nm is needed, which requires an optically thin plasma with temperatures $\approx$ 30 eV \cite{stamm2004extreme, 2007-JAP-Tao, Versolato2019}. For water-window microscopy, a bright source emitting in the spectral region of 2.2-4.4 nm is required, which necessitates a plasma with a temperature of $\approx$100 eV. In HEDP, the temperature and density of the plasmas reaches above 1 keV and near-solid density, respectively. Therefore, tens of kJ to MJ pulse energies are used where plasma pressure exceeds 100 GPa \cite{glenzer2009x, bartal2012focusing}. HEDPs  play an important role in national security applications, high-brightness sources of gamma and X-rays, neutrons, and protons \cite{norreys1999observation, clark2000measurements}. 

\begin{figure}[t]
\includegraphics[width=0.45\textwidth]{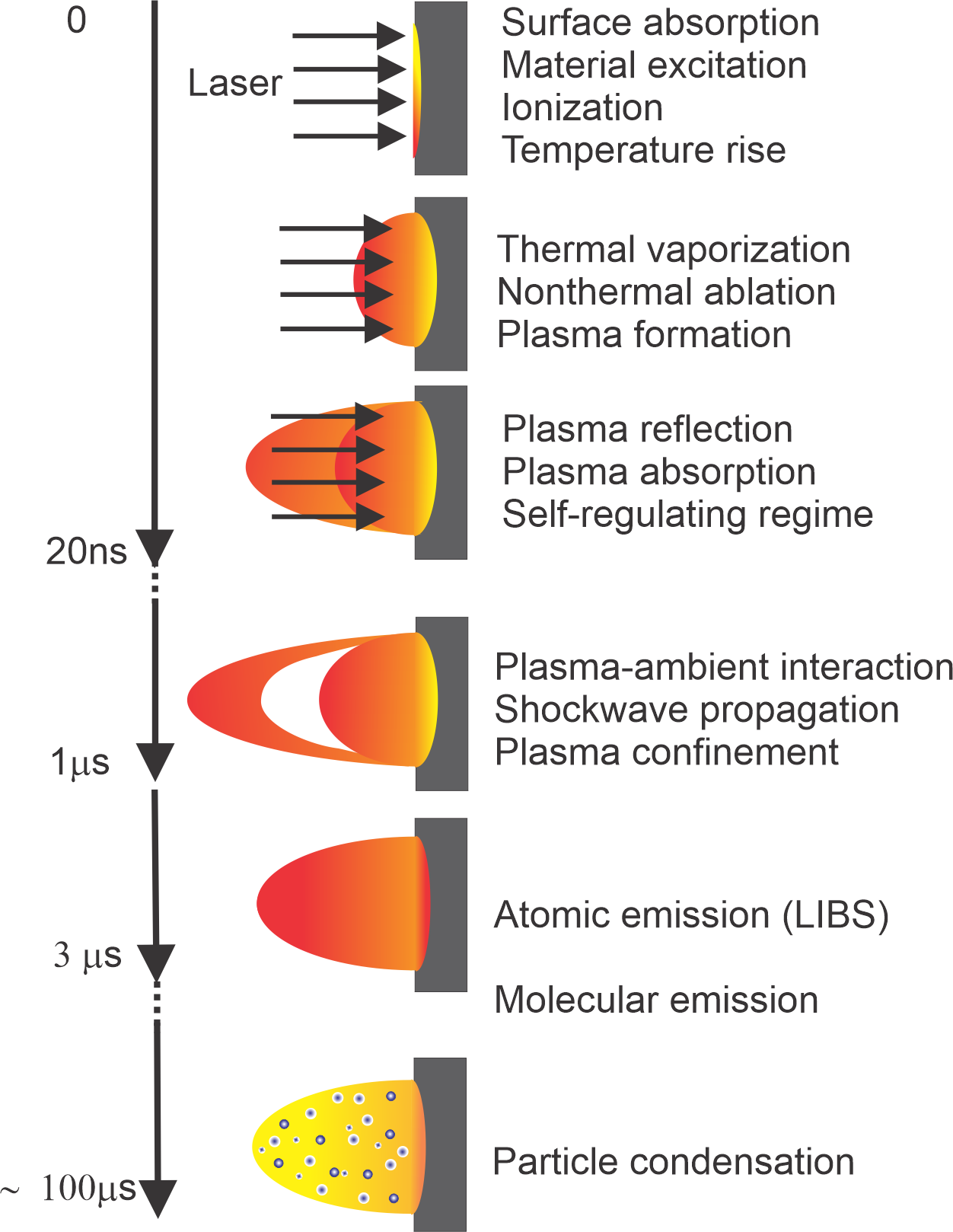}
\caption{\label{fig:2} Various physical processes happening during a nanosecond laser-produced plasma generation and expansion into an ambient gas medium are given.}
\end{figure}

Although the procedure of generating an LPP is relatively simple - focus a pulsed laser with intensities above the laser ablation threshold ($\gtrsim 10^8$ Wcm$^{-2}$) on a material of interest  - the physics involved in such a process is notoriously complex.  Numerous interrelated processes happen during the generation and subsequent expansion of LPP. Some of them are heating, ionization, melting, vaporization, phase explosion, ejection of particles, plasma creation, laser-plasma interaction, instabilities, plasma hydrodynamic expansion, shock wave generation, and confinement, to list a few \cite{Radziemski1989Book}. A schematic example of various processes during a nanosecond LPP generation and expansion into an ambient is given in Fig.~\ref{fig:2}. The lifecycle of an LPP spans over several orders of magnitude in time, and its fundamental properties change many orders during its lifecycle. The energy, pulse width, and wavelength of the laser will affect both laser-target and laser-plasma interactions \cite{le2004influence, sunku2013femtosecond, 2016-JAP-Anoop}.  For longer-pulsed lasers ($\sim$ ns-µs), the LPP generation happens during the leading edge of the laser pulse, and a large fraction of energy is used for heating the plasma. The laser-plasma heating is absent for shorter pulsed lasers (e.g., fs laser pulses). However, prompt ionization can occur during the ultrafast laser pulse interaction with the target in the high-intensity regime, which is the basis of laser wakefield acceleration \cite{esarey2009physics, leemans2006gev}.  

Nanosecond pulsed lasers are commonly used for LPP generation, but ultrashort femtosecond LPPs have numerous advantages in material science and analytical applications \cite{labutin2016femtosecond, rao2016femtosecond, 2018-APR-Hari, vanraes2021laser}. A detailed account of laser-produced plasma generation physics is well-documented in many review articles and textbooks \cite{Radziemski1989Book, Gamaly2011Book, hahn2012}.The physics of laser-matter interactions changes with the type of matter (solid, liquid, gas), and the laser intensity thresholds for generating plasmas depend strongly on the medium properties \cite{Cremers2013Book}. Even for solid targets, the ablation physics vary with target physical and chemical properties \cite{gamaly2002ablation, nica2017investigation}. In metals, the laser couples with free electrons. For semiconductors and insulators, the laser is coupled with bound electrons. The presence of an ambient medium or external magnetic field also significantly alters plasma properties  \cite{2020-AC-Liz, 2004-PRE-Hari, bashir2012influence}. So, the parameter space for controlling the LPP properties is vast.  However, many tuning variables are also helpful for defining the LPP properties for each application. Therefore, the broad appeal of LPP for many scientific and industrial applications can also be related to better control of LPP properties using a wide variety of parameters. 

 Plasma diagnostics play a vital role in understanding the physical properties of all plasmas, including  LPPs, as well as their optimization for various applications \cite{Huddelstone1965, Lochte-Holtgreven1968, Muraoka2001Book, Hutchinson2005, Kunze2009}.  Because of the large variation in fundamental parameters with space and time, different diagnostic tools are essential to capture details of the LPP during its entire lifecycle. Considering the transient nature combined with spatial inhomogeneity of LPPs, the diagnostics need to have high spatial and temporal resolutions. Therefore, in addition to general issues associated with the diagnostics of traditional steady-state plasmas, there exist certain challenges for a comprehensive characterization of LPPs because of their small size and transient and inhomogeneous nature. Hence, great care must be taken when collecting and interpreting results obtained using various diagnostic tools because of the large gradients in the physical and chemical properties of LPPs with time and space, which may result in  weighted averaging of measured properties over certain length and time scales. 

Some of the common optical diagnostics used for plasma characterization include optical spectroscopy (emission spectroscopy, absorption spectroscopy, and laser-induced fluorescence [LIF]), imaging and diagnostics employing an optical probe such as shadowgraphy, Schlieren, interferometry, Thomson scattering, deflectometry, and velocimetry. In addition to optical diagnostic methods, charged particle analysis tools such as ion probes \cite{doggett2009langmuir}, Faraday cup \cite{2015-JAP-Anoop}, electrostatic energy analyzer \cite{burdt2010laser}, and time-of-flight mass spectrometry \cite{wu2017diagnostic, ahmad2018qualitative} are also used regularly by the LPP community. Each diagnostic tool has its pros and cons and should be considered complementary. It is not possible to obtain complete information about the plasma from a single diagnostic tool because each is fundamentally constrained to operate only on certain temperature and density regimes of the plasma. Therefore, multiple tools are essential for global insight into plasma behavior. For example, imaging tools are very useful for understanding the hydrodynamics and morphology of the plume. Interferometry provides density measurements during plasma evolution. Emission spectroscopy is useful for tracking various species in the plume as well as measuring fundamental properties of the plume at intermediate time scales. Absorption and fluorescence methods provide valuable information at late times during LPP evolution.  Several assumptions are also made for characterizing the LPPs using different diagnostic tools (e.g., homogeneous plasma, thermodynamic equilibrium, geometry and symmetry considerations). However, these assumptions may not necessarily be valid for complex LPP systems.

Plasma characterization employing optical methods use either photons released by the plasma via spontaneous emission or the effects generated on/by an external photon source (e.g., laser, arc lamp) when it interacts with a plasma (e.g., scattering, deflection, phase shift, rotation of polarization, absorption, excitation, etc.). Considering the experimental methodologies and measurement assumptions, various plasma diagnostic tools can be classified into active and passive methods and as direct or indirect methods \cite{de1996comparison, Hahn2010}. Passive methods use radiation (photons) emitted from the plasma for measuring the parameters of interest. For example, emission spectroscopy and self-emission imaging belong to this category. In  active methods, an external source is used for LPP characterization, and examples of this category include all optical probing methods. Direct methods refer to obtaining plasma parameters without making assumptions about the plasma conditions such as thermodynamic equilibrium, electron, and ion velocity distributions, the geometry of the plasma, etc. On the other hand, indirect methods require certain assumptions of the conditions of the plasma for interpretation of plasma properties. The passive and active methods can simultaneously be direct or indirect.  

For a plasma scientist, it must be mentioned that accurate diagnosis of transient plasma systems such as LPP or pulsed power-driven plasmas (e.g., Z-pinch) is perhaps one of the most challenging tasks compared to other natural and man-made plasmas. The present review provides an overview of various optical diagnostic tools that can be used to accurately characterize transient LPPs. It uses an integrated approach combining various optical diagnostic tools for a comprehensive characterization by highlighting each method’s capabilities, limitations, and unique challenges. Since optical spectroscopy and Thomson scattering methods are capable of providing detailed measurements of various LPP parameters (temperature, density, plasma kinetics, etc.), further theoretical descriptions are furnished.  However, this review does not attempt to cover the details about the physics of laser-plasma generation. Since most of the existing applications of the LPP belong to laser intensities in the low to moderate range intensity regime ($10^8-10^{15}$ Wcm$^{-2}$), this review focuses on diagnostic tools pertaining to this regime. However, it must be mentioned that the fundamentals and scientific principles of the discussed diagnostic tools are similar for all laser intensity regimes, including HEDP plasmas.

The organization of this review is as follows. The optical diagnostics toolbox for LPPs contains a wide variety of passive and active techniques that are broadly separated into three categories in this review, \emph{viz.} optical spectroscopy, passive and active imaging tools, and optical probing.  Section \ref{sec:Optical_spectroscopy} gives the details of optical spectroscopic methods such as emission spectroscopy, absorption spectroscopy, and laser-induced fluorescence. In section \ref{sec:imaging}, the details of various imaging tools for LPP characterization are discussed, which include fast photography, and passive imaging methods. Section \ref{sec:optical probing} discusses the details of various optical probing techniques which include shadowgraphy, Schlieren, interferometry, and Thomson scattering.   A brief summary of other optical methods (e.g., deflectometry, and velocimetry) is given in Section \ref{sec:Other-methods}. This article ends with a summary (Section \ref{sec:summary})  that includes a table showing the pros and cons of each technique, utility of various diagnostic tools, assumptions, and challenges. Each section is self-contained and intended to be accessible to beginners and early career researchers.

\section{\label{sec:Optical_spectroscopy}Optical spectroscopy}

The basics of optical spectroscopy include analyzing atomic and molecular line radiations, which can be used for diagnosing the LPP through the knowledge of plasma spectroscopy. Emission spectroscopy - a passive method - uses spontaneously emitted light from atoms and molecules, typically excited by electrons. Absorption and fluorescence spectroscopy belong to the active sensing category, where an external light source is utilized for probing. In absorption spectroscopy, the amount of light transmitted (absorbed) is measured when the probe beam passes through the plasma. Fluorescence spectroscopy combines absorption and emission and monitors the change in spontaneous emission caused by the absorption of probe photons. 

Compared to absorption and fluorescence spectroscopy, emission spectroscopy is a widely utilized diagnostic tool in the LPP community because of its experimental simplicity and non-intrusive nature.   This is also partly due to LIBS' widespread use as an analytical tool, which inherently is a combination of the LPP and emission spectroscopy. Therefore, significant work exists in the literature about the characterization of the LPP using emission spectroscopy, including some excellent reviews \cite{Aragon2008, Hahn2010, Konjevic2010, Singh2020-LIBSbook, Zhang2014}. However, even though absorption and fluorescence spectroscopy are capable of providing highly accurate results for the study of LPP fundamentals, they have seen only modest use for diagnosing LPPs. This could be due to the active nature of these techniques combined with more demanding experimental efforts \cite{Whitty1998, King1999, Miyabe2012, 2017-SR-Mark, 2018-Nature, merten2022laser}.  

All spectral lines originating from an LPP are broadened due to various mechanisms such as Stark, Doppler, van der Waals, natural, etc. \cite{Griem1974, Kunze2009, 2018-APR-Hari,  Gornushkin1999}.  The lineshapes provided by each broadening mechanism vary: collisional-broadening mechanisms including Stark, resonance, and van der Waals provide a Lorentzian profile while Doppler contributions yield a Gaussian profile. Therefore, the recorded emission line profile from an LPP will be a convolution of various line-broadening contributions. The theory of various line-broadening mechanisms is given in detail elsewhere \cite{Griem1974}. Among the various broadening mechanisms, the Stark effect, which is caused by charged particles, is the dominant mechanism at the early times of LPP evolution. The Doppler effect, contributed by the thermal motion of the species with respect to the observer, is dominant in low background pressure environments and at late times of plasma evolution. When an LPP is expanding into moderate to high background pressures, Van der Waals broadening may be nontrivial \cite{2021-PRE-Hari}.  Resonance broadening may be important for high atomic number densities. Compared to other line-broadening mechanisms, the contribution of natural line broadening is insignificant. For example, for an atomic transition with a 10 ns spontaneous lifetime, the natural broadening width is 16 MHz.

In this section, details about the measurement of various physical parameters of the LPP using emission, absorption, and fluorescence techniques are given. For the sake of brevity, the details of line broadening, self-absorption, self-reversal, and theoretical aspects of line emission are not discussed in detail here. Instead, the experimental aspects and the application of optical spectroscopy for the measurement of various LPP properties such as temperature, density, kinetics, etc. are discussed with concerned equations, but if the reader would like to get more information on the subject, appropriate citations are given for further reading.

\subsection{Emission Spectroscopy}

Emission spectroscopy refers to the measurement of spontaneous emission of radiation as a function of wavelength. There are different types of emission spectroscopy: X-ray, EUV, UV, VIS, IR, etc. The experimental instrumentation requirements change significantly with respect to the choice of the wavelength region.  Regardless of the selection of the spectral region, a spectrometer with an appropriate wavelength dispersive element (grating, prism, crystal, etc.) or interference-based element (etalon, interferometer) is used for the separation of the light into its wavelength components. From the experimental point of view, the UV-VIS-NIR spectral range is more straightforward and widely used due to its operation in ambient air atmosphere, simple practical alignment, and availability of cost-effective instrumentation. Because most of the applications of LPPs are in the moderate laser intensity regime with temperatures $\leq$ 10 eV, the focus of the diagnostic discussion here pertains to the UV-VIS regime; however, the analysis methodologies for inferring plasma parameters are similar regardless of the spectral region selected for measurement. 

OES is perhaps the most used diagnostic tool for LPP characterization. Accurate information about the plasma temperature and density can be gathered using emission spectroscopy. Apart from temperature and density, other information such as the composition of various species in the plume (atoms, ions, molecules), kinetic distribution, line broadening, and insight into the plasma process or chemistry can also be gathered. OES is also a useful method for gathering fundamental spectral parameters such as oscillator strengths and Stark impact parameters \cite{Nishijima2015, Aberkane2020, 2019-POP-Milos}. Although the technique is simple because spectra can be easily measured, interpretation or measurement of physical parameters can be fairly complex, and the plasma conditions should meet specific requirements such as local thermodynamic equilibrium (LTE) \cite{Griem1964, Konjevic2002, Aragon2008, Kunze2009}.

In a plasma system, an atom can move from one electron configuration to another through the absorption or emission of a photon. The wavelength ($\lambda$) or frequency ($\nu$) of a photon involved in such an emission process corresponds to the energy difference between the two electronic levels: $E_j-E_i = h\nu = hc/\lambda$, where $h$ is Planck’s constant, $c$ is the speed of light, and $j (i)$ represents the upper (lower) energy level. The emission intensity of an electronic transition depends on the number density of the species in the upper level $n_j$ and the atomic transition probability $A_{ji}$  (Einstein coefficient) or, equivalently, the oscillator strength $f_{ji}$. All line transitions undergo broadening because of various mechanisms, and if $\chi(\nu)$ is the area-normalized line profile function, the spectral emission coefficient $I(\nu)_{j\to i}$ of a line with a central frequency $\nu$ is given by
\begin{equation}
    \label{eq:10}
    I(\nu)_{j\to i} = \dfrac{h\nu}{4\pi}A_{ji}n_j\chi(\nu)
\end{equation}
Assuming thermal equilibrium, the Boltzmann equation provides the distribution of the atoms in the same ionization state among the various energy levels as a function of energy and temperature:
\begin{equation}
    \label{eq:11}
    n_j=\dfrac{g_j n_{tot}}{U(T)} e^{-E_{j}/{k_bT_{ex}}}
\end{equation}
where $n_{tot}$ is the total atomic number density, $g_j$ is the degeneracy of level $j$, $k_b$ is the Boltzmann constant, $T_{ex}$ is the excitation temperature, and $U(T) = \sum_jg_j e^{-E_{j}/{k_bT}}$ is the partition function.  Eq.~\ref{eq:11} also gives the relationship between the level population, excitation energy, and temperature. 

The Saha equation provides the relative distribution of atoms among successive ionization states as a function of temperature and electron density \cite{Kunze2009}.  
\begin{equation}
    \label{eq:12}
   \dfrac{n_{Z+1}n_e}{n_Z} = 2\dfrac{U_{Z+1}(T)}{U_Z(T)} \left(\dfrac{2\pi m_e k_BT}{h^2}\right)^{3/2} e^{-E_{IP}/k_{b}T}
\end{equation}
where $z$ and $z+1$ represent successive ionization stages of a given element, $m_e$ is the electron mass, and $E_{IP}$ is the ionization potential from state $z$ to $z+1$. 

The following subsections provide details of emission spectroscopy instrumentation and spectral analysis considerations, methodologies used for density and temperature measurement, and LPP kinetics.  

\subsubsection{Instrumentation and analysis considerations}
\begin{figure}[t]
\includegraphics[width=\linewidth]{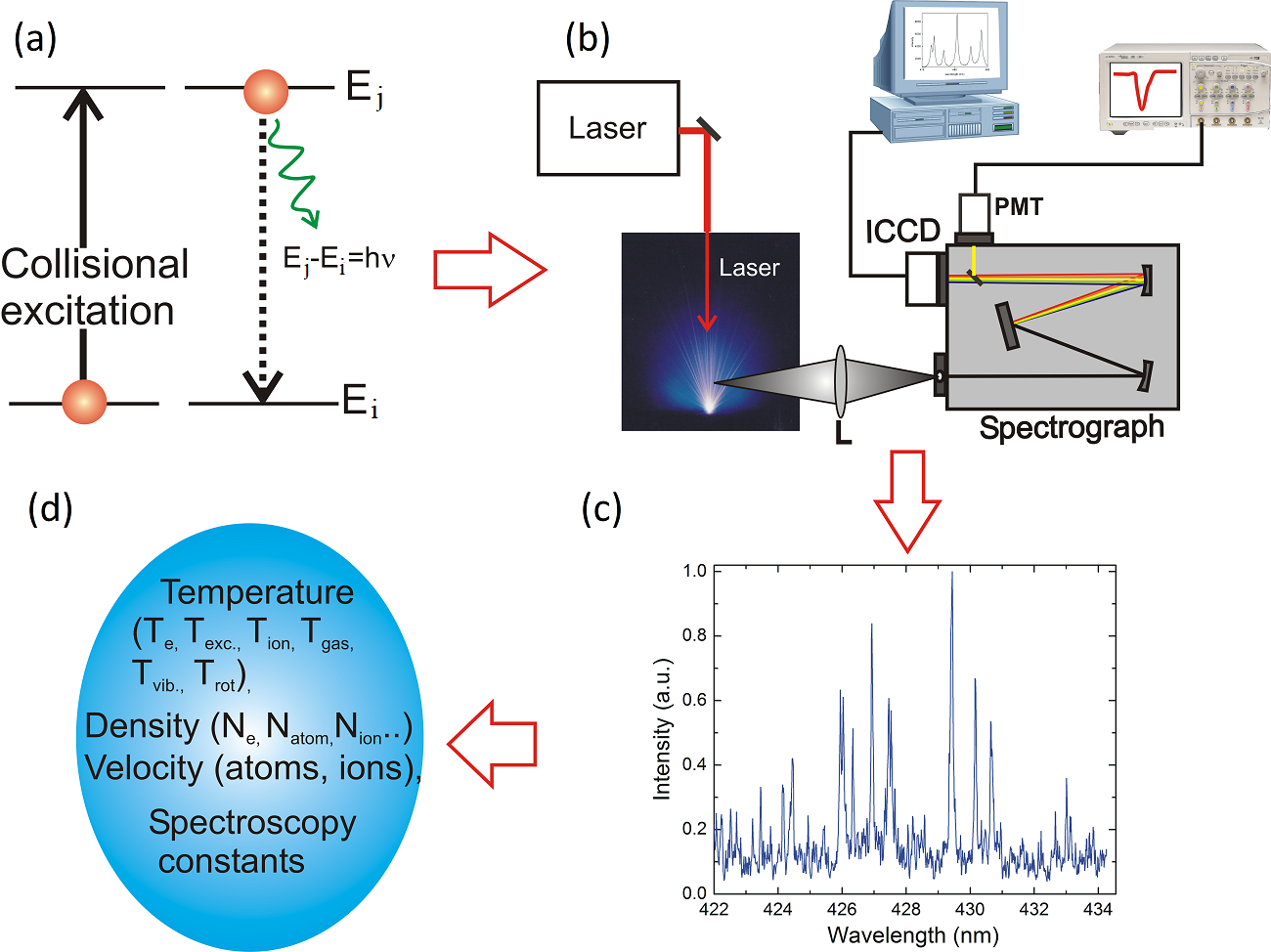}
\caption{\label{fig:13} (a) Emissions between two electronic levels in an atom, (b) schematic of an emission spectroscopic system employing a spectrograph-ICCD combination and/or monochromator-PMT combination, (c) an example of a spectrum collected from an LPP, and (d) the fundamental properties of the plasma that can be obtained from emission spectroscopy.}
\end{figure}

Emission spectroscopy instrumentation records the intensity of line radiation against the wavelength. A schematic of the emission process in a two-level system, OES experimental setup for the LPP characterization, an example of the emission spectrum, and potential physical parameters gathered using this technique are given in Fig.~\ref{fig:13}. A typical setup includes collection optics, a spectrometer, and a detector. Considering spatial and temporal gradients in an LPP, the light collection method and the specifications of emission spectroscopic instrumentation may influence the quality of the collected information. If one of the modalities (space or time) is integrated, only weighted average values of the measured properties are obtained.

Because an LPP expands orthogonal to the target surface irrespective of laser beam direction, the spatial information is gathered by imaging various locations of the plasma parallel to the target surface and onto the entrance slit of an imaging spectrograph. On the contrary, all plasma expansion features can be gathered when the plume expansion axis (target normal) is arranged parallel to the slit height. Depending on the orientation of the LPP plume relative to the spectrograph entrance slit, image rotation using prisms (e.g., Dove prism) \cite{2020-JAAS-Liz} or folding optical systems \cite{Siegel2004} may be necessary. Although positive lenses are commonly used for light collection and imaging, the use of mirrors can avoid chromatic aberration. The available spatial resolution is governed by the magnification of the optical system used to transfer the radiation from the plasma to the spectrograph and the slit width and height. For spatially integrated analysis, the emission from the entire plume is collected and analyzed. The latter scenario is preferred for LIBS analysis, especially with a standoff configuration \cite{Gottfried2008}.

The spectroscopic instrumentation used to analyze the LPP emission determines spectral resolution, bandwidth, and time resolution. Moderate to high spectral resolution spectrographs are essential for measuring lineshapes, linewidths, shifts, and the magnitude/area of a given emission peak. There exists a trade-off between the resolution of the spectrometer, the spectral bandwidth, and throughput. Czerny-Turner (CT) or Echelle-type spectrographs are the most common configurations. A CT spectrograph provides high throughput, but the bandwidth of the detection may be limited with the use of a high-resolution grating. As an example, a combination of a 0.5 m spectrograph and 2,400 grooves/mm grating provides a dispersion at the detector plane of $\approx$0.66 nm/mm at 500 nm, which corresponds to a bandwidth of $\approx$10 nm when using a detector that has an 18 mm width. Instead, Echelle spectrographs provide a large spectral range (200-900 nm) with reasonably good spectral resolution \cite{Munson2005, Cremers2012} though they come with reduced throughput compared to CT spectrographs. Spatial heterodyne spectrometers provide high spectral resolution and come with increased sensitivity and small size \cite{Gornushkin2014, Lenzner2016}.

The time resolution available from an emission spectroscopic system is determined by the detector coupled to the spectrograph. Depending on the selection of the detector, the wavelength dispersion instrument can be used as a spectrograph or a monochromator. Typically, multichannel detectors such as CCD and intensified CCD (ICCD) are used. Regular CCD or CMOS cameras can be operated in a continuous-acquisition mode or can be triggered (synchronized) with the LPP generation but cannot provide high-speed temporal gating for time-resolved emission measurements. ICCD detectors provide time delay/gating resolution down to a few ns and are preferred for time-resolved studies of LPPs. The addition of an electron multiplier (em) to CCD (emCCD) or ICCD (emICCD) is useful for low-light applications. The spectrograph is turned into a single channel analyzer or monochromator when a photomultiplier tube (PMT) is used. In this scenario, the spectral features can be obtained by wavelength scanning, and time-resolved data is easily obtained through high-speed digitization of the PMT signal. Compared to ICCDs, the PMTs provide a broader dynamic range. For quantitative analysis of the LPP, spectroscopic instrumentation (light collection system, spectrometer, and detector) should be radiometrically calibrated. The potential origins of noise associated with an emission spectroscopy measurement include source noise due to fluctuations in the laser-sample or laser-plasma interaction, shot noise due to the number of photons arriving on the detector, detector noise due to dark current, and instrumental drift due to thermal effects. 

 In addition to various plasma induced line-broadening mechanisms, the spectral lines can also be broadened because of artificial effects. For example, all emission spectroscopy detection systems have an inherent instrumental profile. So, knowledge of an instrumental profile is a prerequisite if its width is not negligible (e.g., $ \leq 10\times$ smaller than the contribution of other plasma broadening contributions). The shape of the instrumental broadening depends on several parameters associated with spectrometers (e.g., entrance and exit slits, pixel width in the case of a multi-channel detector, diffraction phenomena, quality of the system components, aberrations, alignment, etc.) and therefore will be a convolution of several individual terms that are typically neither completely Gaussian nor Lorentzian. The instrumental profile can be experimentally measured using a narrow linewidth laser (e.g., He-Ne) or a low-pressure discharge lamp. Because the instrumental profile depends on the spectrometer entrance slit width, it is important to measure the instrumental linewidth with the entrance slit used for the plasma lineshape measurement. 

 By knowing the instrumental profile of the spectrometer used, the contributions of various line-broadening mechanisms can be obtained through deconvolution. The broadening contributed by Lorentzian profiles adds linearly while the Gaussian profiles add quadratically, and the overall profile is a convolution of Gaussian and Lorentzian (Voigt function). Assuming the instrumental profile is Gaussian $(\Delta \lambda_{G(instru.)})$ and other broadening mechanisms are negligible except the Stark effect, the Stark width $(\Delta \lambda_{Stark})$ can be obtained from the measured full width half maximum (FWHM) $(\Delta \lambda_{mea.})$ using the following relation
\begin{equation}
    \label{eq:13}
   {\scriptstyle\Delta \lambda_{mea.} =  \dfrac{\Delta \lambda_{Stark}}{2}+\sqrt{{\left(\dfrac{\Delta {\lambda}_{Stark}}{2}\right)^2 + \left(\Delta {\lambda}_{G(instru.)}\right)^2}}}
\end{equation}
If the instrumental profile is Lorentzian $(\Delta \lambda_{L(instru.)})$, the measured profile will be a simple addition of Stark and instrumental broadening. 
\begin{equation}
    \label{eq:14}
    \Delta \lambda_{mea.} = \lambda_{Stark} + \Delta \lambda_{L(instru.)}
\end{equation}
At late times of plasma evolution, the Stark effect is negligible and the Doppler effect contributes to line broadening. If the instrumental broadening is Gaussian, then the Doppler contribution $\Delta\lambda_{Doppler}$ can be deduced from the following relation:
\begin{equation}
    \label{eq:15}
    \Delta \lambda_{mea.} = \sqrt{{\Delta\lambda_{Doppler}}^2 + {\Delta\lambda_{G(instru.)}}^2}
\end{equation}
It is recommended to use high-resolution spectrographs with spectral resolutions significantly better than the linewidth of the selected transition for plasma lineshape analysis. However, even with the use of a high-resolution spectrometer, it is challenging to separate crowded atomic transitions in a high-Z plasma system (e.g., W, U) with many overlapping emission lines and closely spaced molecular bands \cite{2020-JAAS-Liz, 2021-JAAS-HariPu}. 

The convolution or deconvolution of the spectral profiles discussed above is for spectral lineshapes measured under optically thin conditions. In addition to the instrumental profile, other factors such as self-absorption, self-reversal, etc. may distort the lineshape. For example, the self-absorption processes in the plasma may lead to the broadening of an emission line. This is because absorption is strongest at the line center and weakest at the wings of the spectral profile. However, it is difficult to evaluate the amount of self-absorption in an LPP by simply observing the lineshape. For inhomogeneous plasmas like LPPs, a central dip in the line profile can also be seen (self-reversal) for optically thick lines, which is caused by absorption in the cooler outer layers or coronal regions \cite{Cristoforetti2013, D'Angelo2015}. Such self-absorption and/or self-reversal in an LPP can cause artificial inflation in the measured linewidths as well as a decrease in peak height. In such a scenario, the correction of self-absorption is necessary for using lineshapes for plasma diagnostics \cite{Bulajic2002}. 

\begin{figure}[t]
\includegraphics[width=\linewidth]{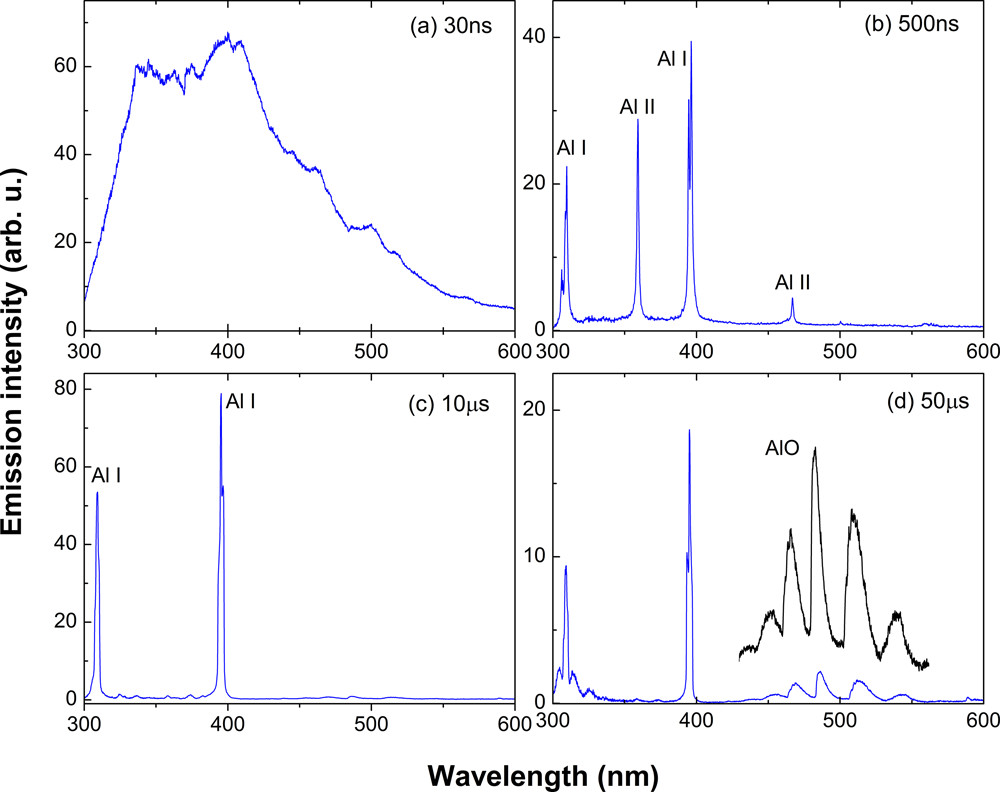}
\caption{\label{fig:14} Time-resolved plasma spectra recorded from a ns laser-generated Al plasma showing continuum, ionic, atomic, and molecular emissions at various times during its evolution. The gate delays used are marked in the spectral features. The insets in (d) correspond to the zoomed AlO emission spectral region. Adapted from \citealp{2016-AC}.
}
\end{figure}

 The spectral profiles recorded from an LPP are influenced by plasma physical conditions, and are therefore useful for inferring the plasma properties. An example of spectral features recorded at various times from an Al LPP is given in Fig.~\ref{fig:14} \cite{2016-AC} and it shows significant changes in spectral properties with time after the onset of plasma formation.  Typically the continuum radiation dominates at early times of LPP evolution because of free-free (Bremsstrahlung) and free-bound (recombination) transitions. As time evolves, the ionic and line radiation dominate the spectral features. Molecular species are generated at late times of plasma evolution \cite{Alessandro2017}.  Since the emission contribution from  the continuum, ions, neutrals, and molecules appears at various times after the plasma onset, their spectral features are useful for tracking temporal evolution of plasma physical conditions.  The various methods used for measuring density and temperature of an LPP system using emission spectroscopy are described below. 

\subsubsection{Electron density using Stark effect}
 Electron density measurements using Stark-broadened line profiles is an important spectroscopic diagnostic tool for the LPPs. Stark broadening arises from the proximity of an ion or electron to another particle that undergoes an optical transition. Stark broadening is mainly a density effect and does not depend sensitively on the temperature or on the electron velocity distribution, and it does not require the plasma to be in local thermodynamic equilibrium (LTE). So, Stark broadening can be used for measuring electron densities even in cases where the existence of LTE is doubtful, whereas some other methods would then become invalid.  In addition, measuring electron density from Stark broadening does not require the knowledge of absolute photon intensities, and just the relative lineshapes and widths are enough. The most important experimental considerations for choosing a spectral analysis for Stark-broadening measurements are line strength, separations from neighboring lines, and good spectral resolution of the measuring instrument. Because Stark broadening of a transition is significant when the densities of the plasmas $\geq 10^{16}$ cm$^{-3}$, a standard spectrograph with resolution $(\lambda/\Delta\lambda) \geq 10000$ is adequate for the measurement. The other factors to consider include (1) knowledge of the Stark broadening coefficient or impact parameter of the selected line with good accuracy; (2) optically thin lines having high intensity and good sensitivity to the Stark effect, and (3) deconvolution of the Stark-broadening linewidth from other broadening effects. 

The theory of the Stark effect describes that the presence of an external electrical field will cause a change in the energy of the emitting atom because of its interaction with the permanent dipole moment (Griem, 1964). For non-hydrogenic atoms with no permanent dipole moment, the change in energy of the emitting atoms will be proportional to the square of the electric field and known as the quadratic Stark effect. The broadening (FWHM) and shift of an atom transition due to the quadratic Stark effect are given by \cite{Kunze2009, Griem1964}.
\begin{equation}
    \label{eq:16}
    {\scriptstyle \Delta\lambda_{1/2}   \approx  2W\left(\dfrac{n_e}{10^{16}}\right) + 3.5A\left(\dfrac{n_e}{10^{16}}\right)^{1/4} \left[1 - \dfrac{3}{4}N_D^{-1/3}\right]W\left(\dfrac{n_e}{10^{16}}\right)}\\
\end{equation}
\begin{equation}
    \label{eq:17}
    {\scriptstyle \Delta\lambda_{shift}   \approx   D\left(\dfrac{n_e}{10^{16}}\right) \mp 2A\left(\dfrac{n_e}{10^{16}}\right)^{1/4}\left[1 - \dfrac{3}{4}N_D^{-1/3}\right]W\left(\dfrac{n_e}{10^{16}}\right)}\\
\end{equation}
where $W$ is the electron-impact parameter that can be incorporated at different temperatures, $A$ is the ion broadening parameter, and $N_D$ is the number of particles in the Debye sphere, which is given by 
\begin{equation}
    \label{eq:18}
    N_D = 1.72 \times 10^{12}\dfrac{(T(eV))^{3/2}}{(n_e(m^{-3}))^{1/2}}
\end{equation}

The first and second terms in Eq. (\ref{eq:16} and \ref{eq:17}) correspond to electron impact and ion correction factors, respectively. For non-hydrogenic atoms, Stark broadening is predominantly due to electron impact, and the perturbations caused by the ions can therefore be neglected. There exists an extensive library for Stark-broadening parameters for selected emission lines of low-Z and mid-Z elements \cite{Griem1974, Konjevic2002}; however, limited work is available for high-Z elements \cite{Nishijima2015, 2019-POP-Milos}.  An example of a Stark-broadened line profile as well as Stark shift for Ca I 585.74 nm transition at various times after the plasma onset is given in Fig.~\ref{fig:15}a \cite{Burger2016}.  Previous reports showed asymmetry in the Stark-broadened line profiles from LPPs \cite{Konjevic1999}, and the factors that introduce asymmetry in the lineshapes are the ion broadening and fine structure splitting. 

Because hydrogen and hydrogen-like atoms possess finite dipole moments, Stark broadening and shifts become very predominant, and the change in energy is proportional to the electric field and provides a linear Stark effect. Typical time evolution of Stark-broadened profiles of the $H_\alpha$ transition from an LPP are given in Fig.~\ref{fig:15}b \cite{Burger2016}. A set of empirical formulae are derived for measuring the electron density of LPPs in gases and solids using Stark-broadened $H_\alpha$  and $H_\beta$ transitions \cite{Parigger2019}:

\begin{figure}[t]
\includegraphics[width=0.75\linewidth]{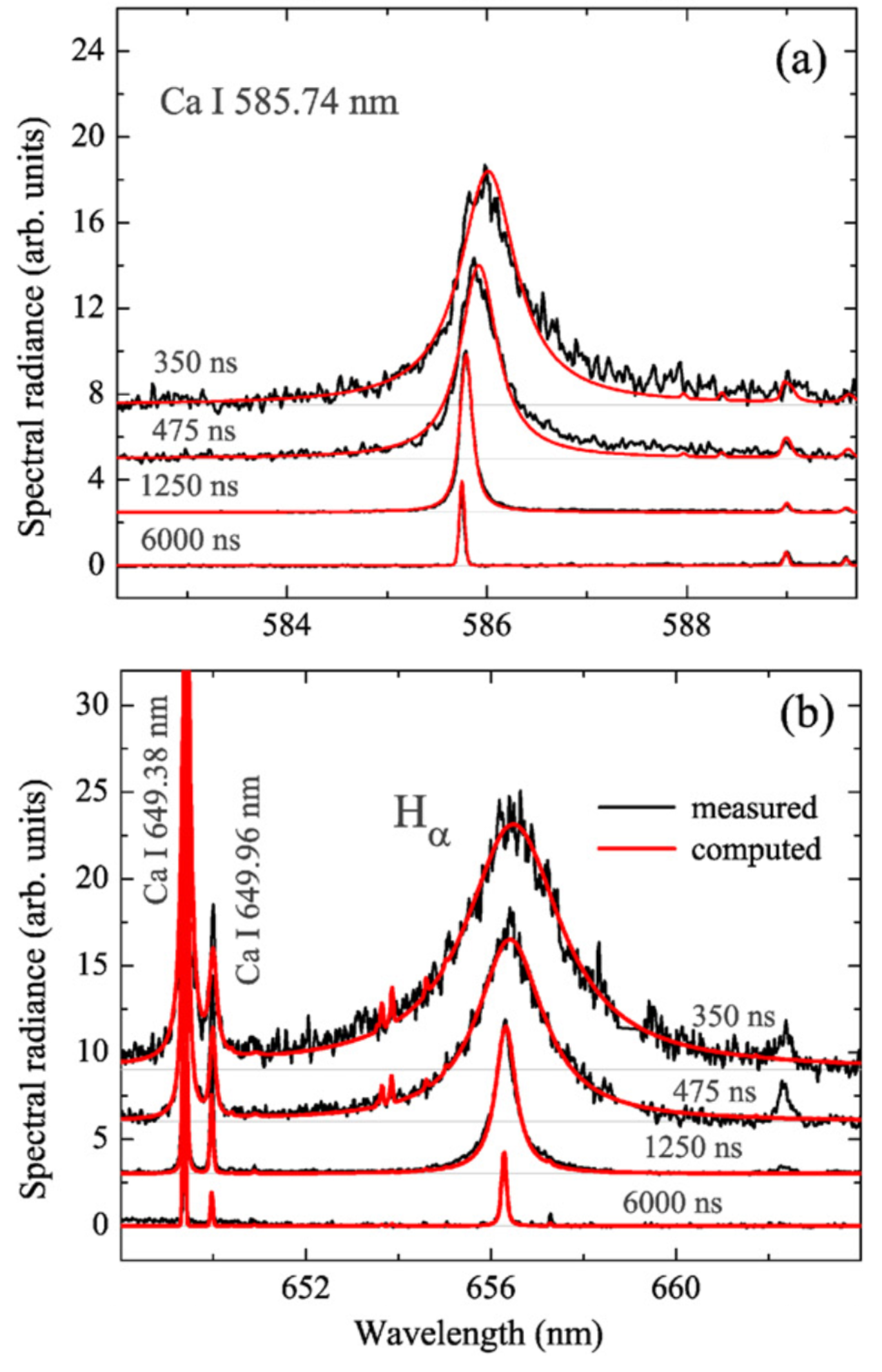}
\caption{\label{fig:15} Time evolution of the Stark-broadened profile of (a) Ca I  585.74 nm  and (b) H$_\alpha$ ~656.28 nm.  The smooth curves represent the spectral fits. Adapted from \citealp{Burger2016}}
\end{figure}

\begin{equation}
    \label{eq:19}
    \Delta\lambda_{H_\alpha}(nm) \approx  1.3\left(\dfrac{n_e(cm^{-3})}{10^{17}}\right)^{0.64\pm0.03}
\end{equation}

\begin{equation}
    \label{eq:20}
    \Delta\lambda_{H_\beta}(nm) \approx 4.5\left(\dfrac{n_e(cm^{-3})}{10^{17}}\right)^{0.71\pm0.03}
\end{equation}
The empirical formulae for H$_\beta$ is valid for electron densities in the range of $0.03$ to $8 \times 10^{17}$ cm$^{-3}$ and for H$_\alpha$ in the order of $0.01$ to $100 \times 10^{17}$cm$^{-3}$. 

The accuracy of electron density measurements of an LPP using recorded spectral lineshapes depends on the errors of the reference Stark-broadening parameters. \citet{Radziemski1983} measured the time evolution of electron density in an aerosol plasma using several lines of multiple elements and found that its values are significantly scattered, and this discrepancy is attributed to the limitation in the accuracy of the theory to determine the reference parameters.

Stark broadening becomes negligible when the electron density drops $\leq$ 10$^{16}$ cm$^{-3}$. Therefore, determining electron density via spectral linewidths measured in OES is limited to early times in the LPP with sufficiently high electron density.  It is also very challenging to measure other line-broadening mechanisms such as Doppler, van der Waals, etc. using emission spectroscopy because of their smaller contributions $(\sim$ 1-10 pm) relative to the instrumental profile of standard spectrometers. 

\subsubsection{Measurement of temperature} 
Eq.~\ref{eq:10} provided the relationship between emission intensity and excited level population, which in turn depends strongly on temperature through the Boltzmann relation (Eq.~\ref{eq:11}). But the definition of temperature in an LPP system is somewhat ambiguous. In fact, several temperature nomenclature terms exist in an LPP system, including excitation temperature, neutral, electron and ion temperature, molecular or gas (vibrational and rotational) temperature, kinetic, translational temperature, etc., and all these temperatures should be similar if the LPP system is in LTE. Optical spectroscopic tools are useful for measuring each of these temperature categories. For example, the Boltzmann analysis is used to measure excitation (atom, ion) and molecular temperature, and the Saha Boltzmann equation and line-to-continuum intensity ratio provide electron temperature. Boltzmann and Saha's methods are most commonly used by the LPP community, and therefore, a brief account of these is given below. 

\paragraph{\textbf {Boltzmann method} }
The Boltzmann method is used routinely for measuring the temperature of the plasma using the spectral lines of a similar charge state \cite{Aguilera2004, Zhang2014, Kunze2009}. By combining equations (\ref{eq:10}) and (\ref{eq:11}), the spectral emission coefficient of a transition from energy level $j$ to $i$ is written as \cite{Kunze2009}
\begin{equation}
\label{eq:21}
    I(\lambda)_{j \to i} = \dfrac{hc}{4\pi \lambda_{ji}}A_{ji}\dfrac{n_{tot}g_j}{U(T)}e^{-E_j/k_bT}\chi(\nu)
\end{equation}
Using Eq.~\ref{eq:21}, the intensity ratio of the spectral intensities of two lines (1 and 2) with the same ionization stage can provide the excitation temperature: 
\begin{equation}
    \label{eq:22}
    \dfrac{I_2}{I_1} = \dfrac{g_2}{g_1}\dfrac{A_2}{A_1}\dfrac{\lambda_1}{\lambda_2}e^{-\left(E_2-E_1\right)/{k_bT}}
\end{equation}
However, to obtain accurate temperature measurements, it is recommended to use several lines for constructing a Boltzmann plot with a large upper energy difference. The natural logarithm of Eq.~\ref{eq:21} gives
\begin{equation}
    \label{eq:23}
    \ln\left(\dfrac{I_{ji}\lambda_{ji}}{g_jA_{ji}}\right) = -\dfrac{1}{k_bT}E_j + \ln\left(\dfrac{hcn_{tot}}{4\pi U(T)}\right)
\end{equation}

\begin{figure}[t]
\includegraphics[width=0.75\linewidth]{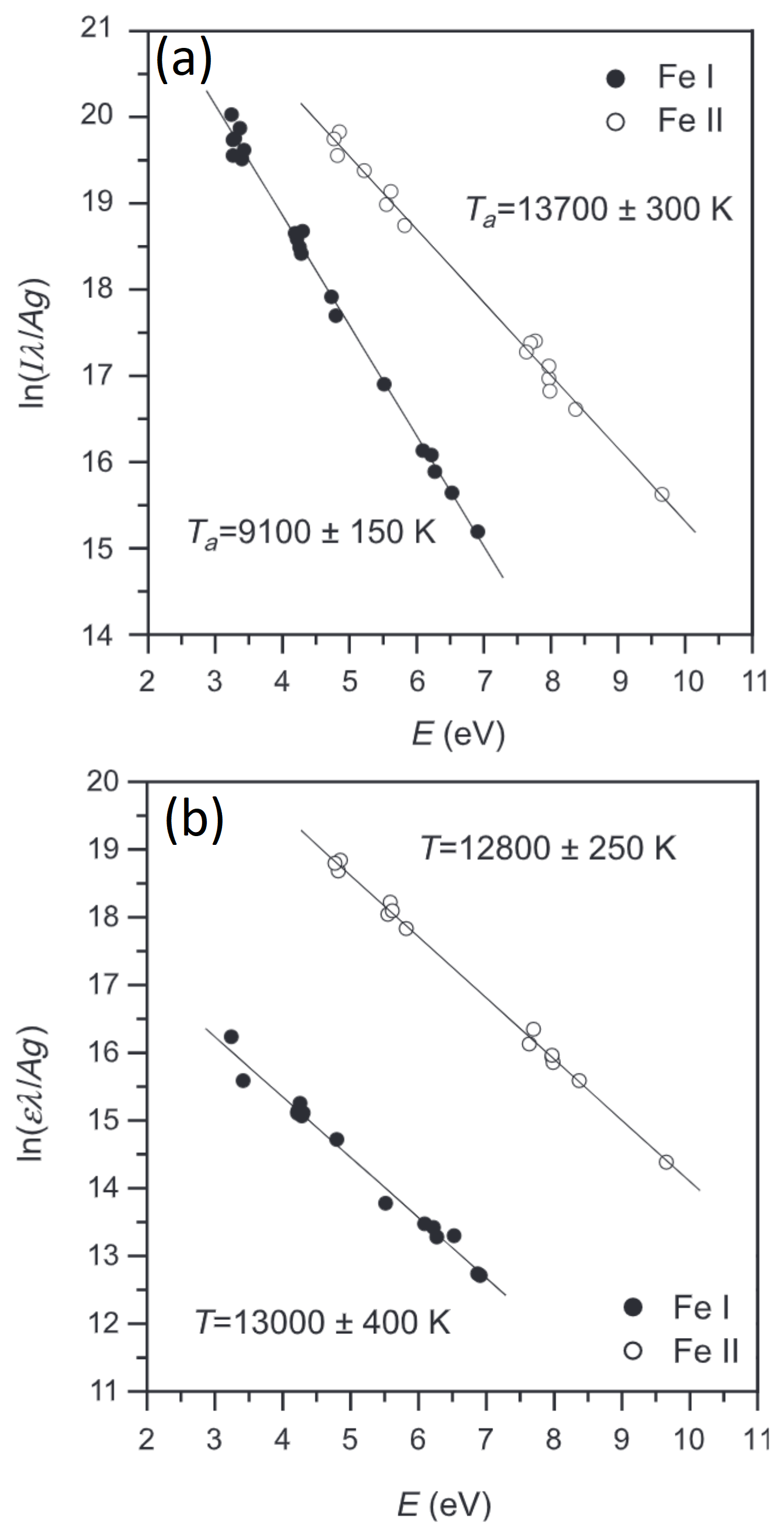}
\caption{\label{fig:16} Examples of Boltzmann plots obtained from Fe I lines and Fe II lines are given for (a) spatially integrated measurement and (b) spatially resolved measurement and Abel inversion. The temperature values deduced from the linear fit are also given. From \citealp{Aguilera2004}}. 
\end{figure}

Eq.~\ref{eq:23} gives a linear plot if $\ln(I\lambda/gA)$ is plotted against the energy levels E of several transitions, and the slope of the plot corresponds to temperature. Because a similar ionization stage is used for the Boltzmann method, the measured temperature is typically called the excitation temperature. If the spectral lines of a certain element with charge state $z=0$ (neutrals) are used, it can be referred to as neutral atom excitation temperature. Similarly, if the ions $(z \geq 1)$ are used for temperature measurement, it is referred to as ion temperature. An example of a Boltzmann plot is given in Fig.~\ref{fig:16}. \citet{Aguilera2004} used several Fe I and Fe II lines for constructing Boltzmann plots for an LPP system generated from a Fe–Ni alloy and noticed that excitation temperatures obtained through the spatially integrated measurement of neutrals (Fe I) and ions (Fe II) possess different values (See Fig.~\ref{fig:16}a). However, the measured excitation temperatures from neutral atoms and ions through a spatially resolved analysis together with an Abel inversion showed similar temperatures within the error (Fig.~\ref{fig:16}b). These results indicate the importance of spatially resolved analysis and Abel inversion to account for the inhomogeneity of the LPP \cite{Konjevic2010}.

\begin{figure}[t]
\includegraphics[width=0.75\linewidth]{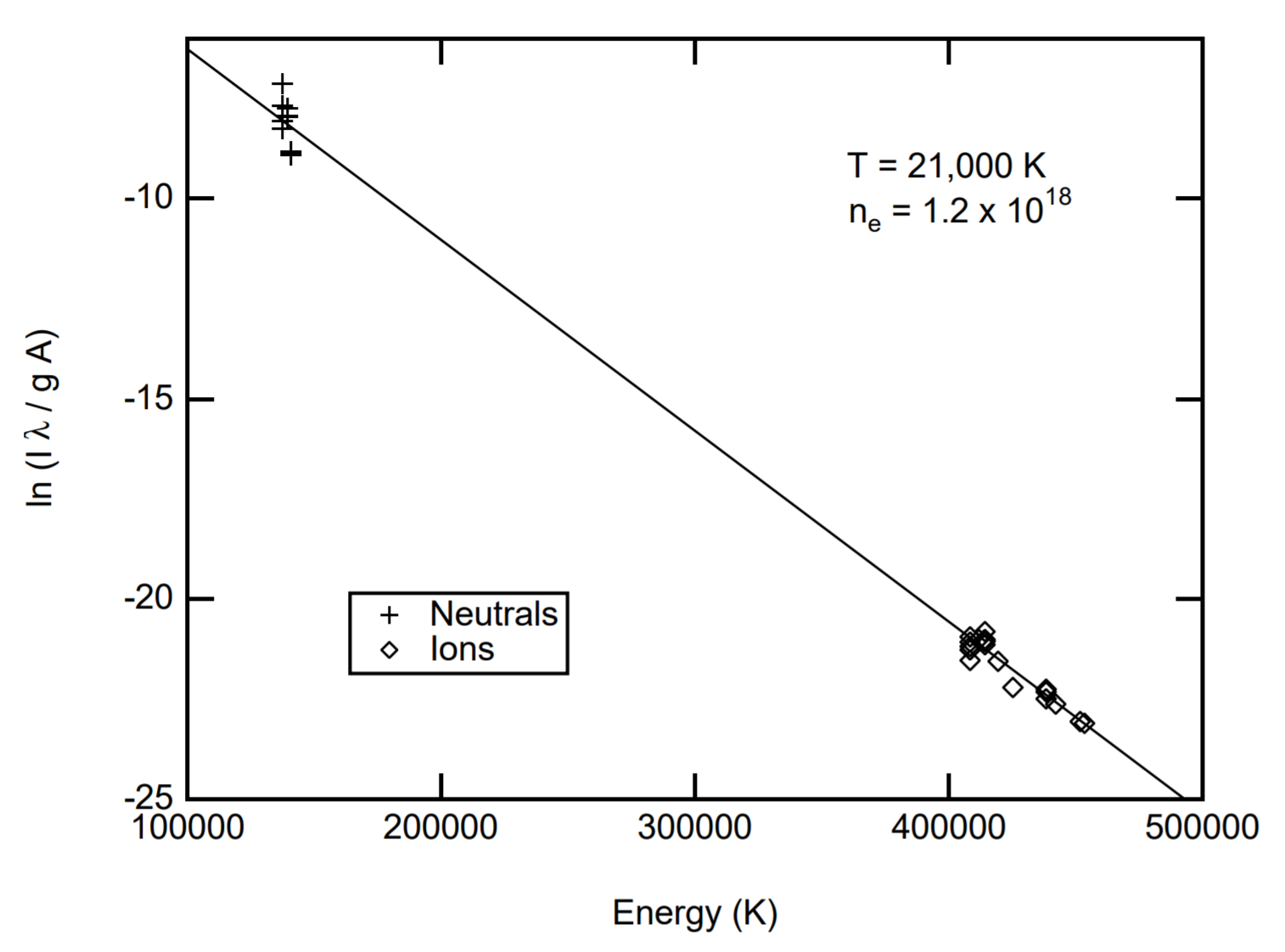}
\caption{\label{fig:17} The Saha–Boltzmann plot for N lines at a delay of 0.35 µs during LPP evolution. The LPP was generated using a frequency-doubled Nd:YAG laser in a metal aerosol. From \citealp{Yalcin1999}.}
\end{figure}

\paragraph{\textbf{Saha-Boltzmann method}} 
The electron temperature of an LPP system can be measured using the Saha equation by taking the intensity ratio of lines originating from successive ionization states. Compared to the Boltzmann method that employs line intensities of the same element and ionization state, improved sensitivity can be obtained using this method because the effective energy difference is now enhanced by the ionization energy, which is much larger than the thermal energy. Therefore, the Saha–Boltzmann analysis is more reliable than the simple Boltzmann analysis, providing a more accurate temperature \cite{Aguilera2004}. Under LTE, the intensity ratios of lines originated from successive ionization stages is given by Saha-Boltzmann \cite{Kunze2009}: 
\begin{equation}
\label{eq:24}
    {\scriptstyle\dfrac{I'}{I} = \dfrac{g'A'\lambda}{gA\lambda'}\left(\dfrac{2\pi m_e k_bT}{h^2}\right)^{3/2} \dfrac{2}{n_e}e^{-(E'-E+E_{IP}-\Delta E)/{k_{b}T}}}
\end{equation}
where the primed symbols represent the line of an atom with a higher ionization stage, $E_{IP}$ represents the ionization potential of a lesser ionization state, and $\Delta E$ is the correction to the ionization potential for interaction in the plasma, which is given by 
\begin{equation}
    \label{eq:25}
    \Delta E = 3z \dfrac{e^2}{4 \pi \epsilon_0}\left(\dfrac{4\pi n_e}{3}\right)^{1/3}
\end{equation}
where $z$ is the lower ionization state. Eq.~\ref{eq:24} is suited for temperature measurements by taking the ratio between two lines of successive ionization states. However, the knowledge of electron density $n_e$ is necessary to measure temperature. It is possible to perform the temperature measurement without the knowledge of $n_e$ if two sets of equations (two lines each of successive ionization states) are considered \cite{Hari-thesis}. 
\newpage
Similar to the Boltzmann plot (Eq.~\ref{eq:23}), a multi-line approach was suggested by \citet{Yalcin1999} based on the Saha and Boltzmann equations.  An example of a Saha-Boltzmann plot employing neutrals and singly ionized lines of N is shown in Fig.~\ref{fig:17} \cite{Yalcin1999}. Here, the x-coordinate values of the ions were modified by a correction term that depend on the temperature and the electron density which is given by

\begin{equation}
    \label{eq:26}
    \ln\left(\dfrac{I_{ji}\lambda_{ji}}{g_jA_{ji}}\right)^* =  \ln\left(\dfrac{I_{ji}\lambda_{ji}}{g_jA_{ji}}\right) -z\ln\left(2\left[\dfrac{2\pi m_ek_B}{ h^2}\right]^{3/2}\dfrac{T^{3/2}}{n_e}\right)
\end{equation}
where z=1 for singly ionized atoms. Apart from these, the excitation energy (y-axis in Fig.~\ref{fig:17}) of the ions in the 
Saha–Boltzmann plot should also be modified by adding the ionization energy of the lower ionization stages.  

Spectral simulation tools are useful for measuring the excitation or electronic temperature of an LPP system, and this is performed by comparing the simulation with the experimentally measured spectral features . Spectral simulation is possible by using a comprehensive collisional radiative (CR) model with atomic codes or level populations that are calculated according to the Boltzmann and Saha equations under the LTE assumption \cite{Hermann2017, NIST-LIBS}. This method is certainly useful for measuring temperatures of complex spectral features of high-Z plasmas \cite{2019-PCCP-Hari}. However, to obtain accurate characterization of the plasma, it is imperative to consider spatial inhomogeneity and radiative transport simulations in the modeling tool. 

\paragraph{\textbf{Molecular temperature} }
Similar to atomic transitions, each excited molecule emits a set of discrete frequencies determined by the energy levels and their populations.  For molecules, each electronic level ($T_e$) consists of a vibrational energy level manifold ($G_v$), and for each vibrational level, there is a rotational energy level manifold ($F_J$). Therefore, the energy corresponding to an optical transition ($\epsilon$) for a molecular transition is given by the difference between the energies of the electronic, vibration, and rotational levels in the upper and lower electronic states \cite{Herzberg1950}
\begin{equation}
    \label{eq:27}
    {\scriptstyle\epsilon = E(e',\nu',J')-E(e'',\nu'',J'') = T_e' + G_\nu'+F_J' - (T_e'' + G_\nu'' + F_J'')}
\end{equation}
where the upper-state and lower-state energy terms are represented by $'$ and $''$, respectively. Because each electronic level for a molecule contains vibrational and rotational manifolds, each molecular species has many more transitions than the corresponding atoms composing the molecule. 

The population of vibrational and rotational levels is related to its temperature (vibrational and rotational) through the Boltzmann equation. For example, in the case of vibrational transitions, the sums of the strengths of all bands with the same upper ($\nu$) or lower ($\nu'$) state are proportional to the number of molecules in the respective states, and the line intensities of various vibrational levels are related to temperature through the following equation (Herzberg, 1950b):
\begin{equation}
    \label{eq:28}
    \ln(\sum_{v'}(\lambda^4I_{vv'})) = C_1 - G(\nu)\dfrac{hc}{k_bT_{vib}}
\end{equation}
where $C_1$ is a constant and $G(\nu)$ is the term value corresponding to the vibrational level in the upper electronic state, which can be estimated using the following relation: 

\begin{eqnarray}
    \label{eq:29}
    G(\nu) = \dfrac{E_{vib}}{hc} = \omega_e \left(\nu + \dfrac{1}{2}\right) - \omega_e x_e \left(\nu + \dfrac{1}{2}\right)^2 + \nonumber\\
    \omega_e y_e \left(\nu+\dfrac{1}{2}\right)^3 + \omega_e z_e \left(\nu+\dfrac{1}{2}\right)^4
\end{eqnarray}
where $E_{vib}$ represents the molecule’s vibrational energy levels and $\omega_e,x_e,y_e,z_e$ are molecular constants. Similar to the Boltzmann plot generated for the atomic excitation temperature, the vibrational temperature can be obtained by plotting $\ln(\sum_{\nu'}(\lambda^4 I_{ \nu'}))$ vs $G(\nu)$, and the slope provides $1/k_bT_{vib}$. 

Typically, spectral fits are preferred for measuring vibrational and rotational temperatures rather than using Boltzmann plots because molecular spectral features contain crowded rotational and vibrational energy states compared to the atomic spectra. Spectral simulation allows the measurement of both rotational and vibrational temperatures, and they may show significant differences in the case of gaseous (low-temperature) plasmas \cite{laux2003optical, Zhu2017}. However, in the case of LPP, one frequently finds equal vibrational and rotational temperatures $(T_{vib} = T_{rot})$ \cite{Hornkohl1991, Parigger1995, 2018-POP-Hari}. 

\begin{figure}[t]
\includegraphics[width=0.9\linewidth]{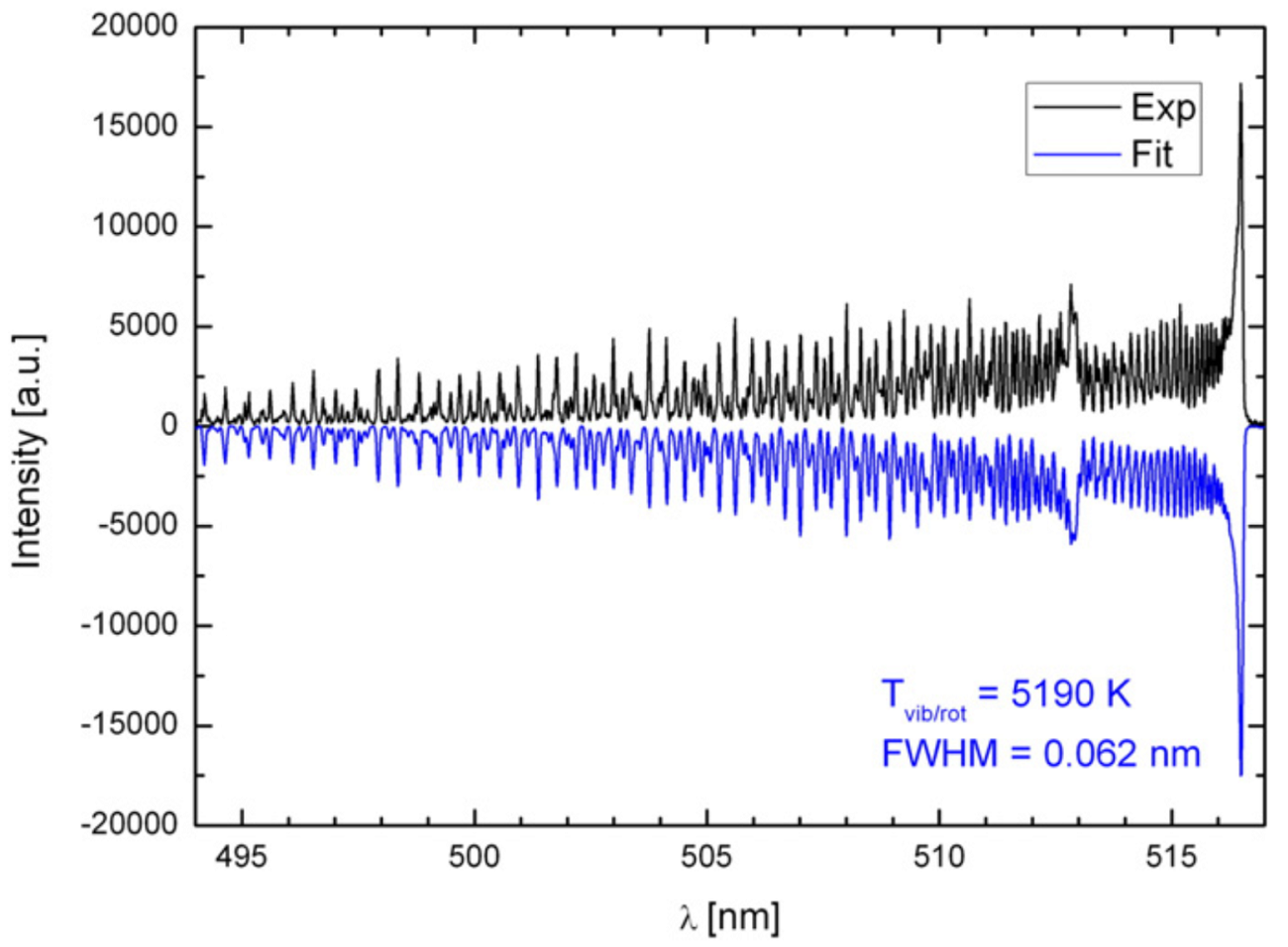}
\caption{\label{fig:18} Comparison of measured CN violet $\Delta v = 0$ emission band and spectral simulation. The measured emission spectrum is shown in black in the upper part of the figure, and the inverted simulated spectrum is shown in blue in the lower part. Adapted from \citealp{Trautner2017}.}
\end{figure}

There are several molecular simulation programs available (e.g., PGOPHER \cite{Western2017}, SPECAIR \cite{laux2003optical}, BESP \cite{Parigger2015}, and LIFBASE \cite{Luque1999}) that provide molecular spectral features by utilizing standard molecular constants. The PGOPHER simulation tool was used by several research groups for extracting rotational and vibrational temperatures from the LPP system \cite{Western2017}. \citet{Parigger2015} developed the BESP simulation tool. For laser-produced atmospheric air plasmas, SPECAIR (a commercial software) is routinely used and includes the most important radiating atoms, ions, and diatomic molecules present in air plasmas \cite{laux2003optical, Kimblin2017, 2018-POP-Hari}. A comparison between the experimental and simulated spectra of CN from an LPP source is given in Fig.~\ref{fig:18} \cite{Trautner2017}.

\citet{Lam2014} compared the atomic excitation and molecular rotational temperatures of a multi-component plasma and found a mismatch between them and explained due to the lack of thermodynamic equilibrium. A similar study involving laser-produced air plasma showed a good agreement between the atomic oxygen excitation temperature and the molecular temperatures for N$_{2}^+$ and CN \cite{2018-POP-Hari}. However, they observed a discontinuity in the temperature decay for N$_{2}^+$, CN, and OH molecular temperatures, explained by changes in plasma hydrodynamics and chemistry that lead to spatial inhomogeneity. 

Because the physical properties of the LPP system vary significantly with time and space, spatially and temporally resolved diagnostic measurement techniques are appropriate for retrieving accurate distributions of temperature and density \cite{Aguilera2004}. Even though spatially resolved collection of the emission is possible by 2D side-on imaging of the 3D expanding LPP system onto the slit of the spectrometer, it must be mentioned that it only provides the line-of-sight integrated measurements and it indicates that the collected data correspond to the weighted average values of the spectral properties. Considering the spatial inhomogeneity of the LPP system, this is one of the major hurdles in using optical spectroscopy for quantitative spectroscopic measurements \cite{Konjevic2010}. Abel inversion can be utilized to address these issues, and it converts the axially and laterally resolved data into radially resolved data, assuming cylindrical symmetry of the LPP \cite{Merk2013}. Therefore, line-of-sight measurements without performing Abel inversion may introduce systematic error in the measured temperature and density values. Other approaches to account for line-of-sight averaging through inhomogeneous LPPs may include the use of radiative transport models or simulations \cite{Gornushkin2001}.

\subsubsection{Optical time-of-flight measurements}

Temporally resolved spectral emissions from various species in an LPP are very important for understanding plume kinetics \cite{Irimiciuc2020, 2016-AC, Diaz2020}.  This is typically performed using a combination of the monochromator and a single channel detector such as PMT (see Fig.~\ref{fig:13}). Similar information about the kinetics of plume species can be obtained using a spectrograph-ICCD combination. However, this requires multiple laser shots by varying gate delay \cite{Camacho2015, Ursu2020}. Considering the transient nature of the LPP, such analysis is based on analyzing different plasmas generated by numerous laser pulses, and it would therefore be less accurate. Instead, the combination of monochromator-PMT provides the kinetic distribution of an emitting species in the plume using a single laser pulse. Compared to array detectors, the PMTs also provide broad spectral coverage with high quantum efficiency (QE) and excellent time response and sensitivity.   

\begin{figure}[t]
\includegraphics[width=65mm]{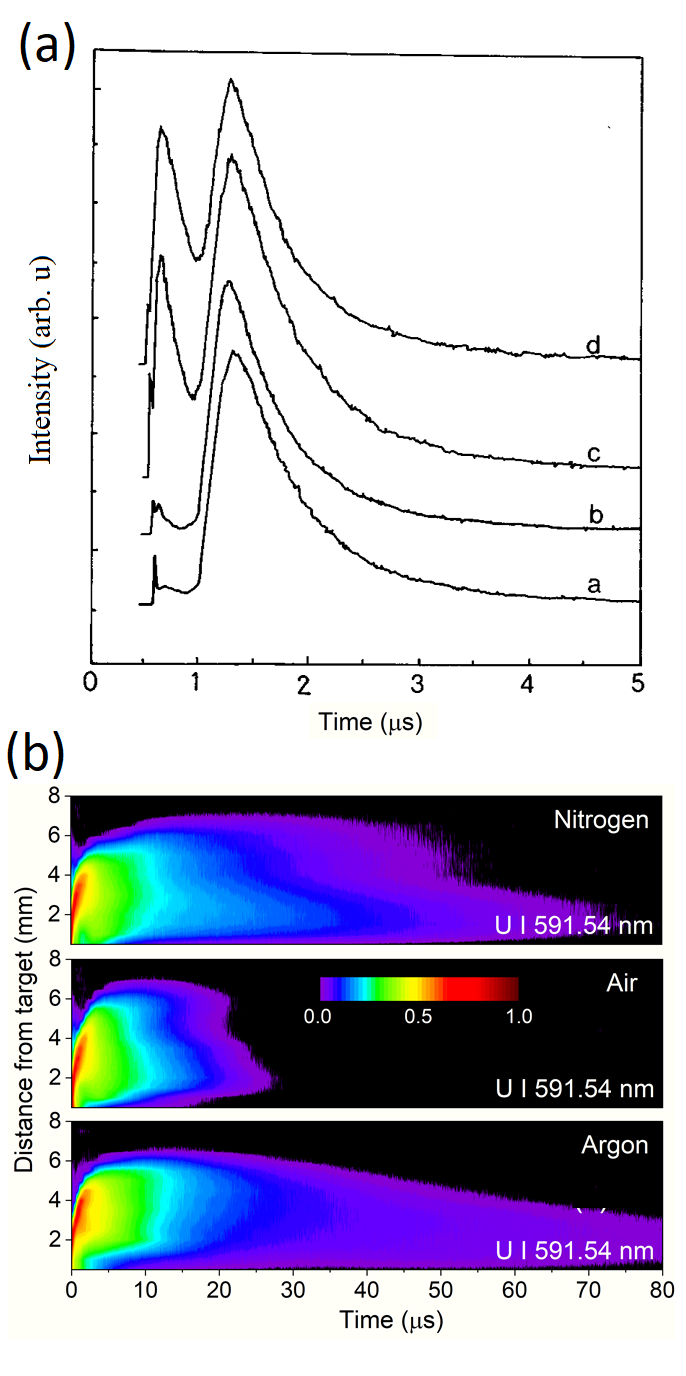}
\caption{\label{fig:19} (a) OTOF of C$_2$ emission at 516.5 nm recorded at 5 mm away from the target for various laser fluences (a) 12.7 J.cm$^{-2}$, (b) 26.7 J.cm$^{-2}$, (c) 28 J.cm$^{-2}$, and (d) 29.3 J.cm$^{-2}$. Adapted from \citealp{1996-JAP-HariC2}.  (b) Spatiotemporal emission contours for the U I 591.54 nm line under various background gases (nitrogen, air, and argon) recorded at 100 Torr pressure, showing the effect of plasma oxidation on emission persistence. Space- and time-resolved emission contours were generated by combining OTOF profiles recorded at various distances from the target. Adapted from \citealp{2019-POP-Patrick}.
}
\end{figure}

For spatially resolved studies, the plasma plume from different regions is imaged onto the monochromator slit, and the temporal evolution of the selected transition is collected using PMT, which is then fed to a digital oscilloscope for recording. This diagnostic method is often referred to as optical time of flight (OTOF) and provides the delay as well as the persistence of emitting species in the plume \cite{Sivakumar2014, Smijesh2014, Ying2015}. By selecting various species in the plume (ions, atoms, molecules, etc.), the kinetic distribution of the LPP can be mapped. An example of optical TOF from C$_2$ molecules from laser-produced graphite plasma at various laser fluences is given in Fig.~\ref{fig:19}a \cite{1996-JAP-HariC2}, and it shows the modifications in C$_2$ emission kinetics due to the changes in species formation mechanisms. 

The OTOF method is routinely used by the LPP community for studying plasma chemistry \cite{2019-POP-Patrick}, self-absorption \cite{Fu2019}, ion acceleration \cite{Thomas2020}, 2D mapping of various species in the plume \cite{2016-PSST-Khaled}, and optimizing parameters for PLD \cite{Druffner2005}. The application of OTOF for studying the LPP plume chemistry is shown in Fig.~\ref{fig:19}b, and it shows spatiotemporal contours of U I emission in the presence of various background gases (air, nitrogen, and argon), and the presence of reacting gas (air) reduces the emission persistence of U significantly \cite{2019-POP-Patrick}.  

\subsection{Absorption spectroscopy}
Absorption spectroscopy (AS) refers to the absorption of radiation as a function of wavelength or frequency when a probe beam passes through a medium and is a well-established technique for identifying and/or quantifying gas-phase atomic and molecular species. Unlike emission spectroscopy, AS is an active technique that necessitates the use of a light source (laser, arc lamp, etc.) for probing the absorption by the LPP species. AS can be performed across the entire electromagnetic spectrum if a suitable light source is available, and depending on the probing spectral region, there are a wide variety of experimental schemes (X-ray AS, UV-VIS AS, IR-AS, etc.); however, the fundamental principles are the same.  

In AS, the amount of light transmitted through the LPP is measured.For a spatially uniform system, the relationship between the transmitted and initial light intensity is given by:
\begin{equation}
    \label{eq:31}
    I(\nu , t) = I_0(\nu)e^{-A(\nu,t)}
\end{equation}
where $I(\nu, t)$ is the intensity of the probe laser at the optical frequency $\nu$ incident on the detector at time $t$, $I_o(\nu)$ is the intensity of the probe laser in the absence of absorption, and $A(\nu,t)$ is the absorbance. Based on Eq.~\ref{eq:31}, the absorbance can be expressed as $A(\nu,t) = -\ln[I(\nu,t)/I_0(\nu)]$  (note that an equivalent expression may be defined using base-10 rather than base-e, and conventions may differ among sub-fields). The absorbance may also be expressed as the product of an absorption coefficient $\alpha(\nu,t)$ and an optical path length $L : A(\nu,t) =\alpha(\nu,t)\cdot L$, which is again valid for a spatially uniform system along the measurement path. The absorption coefficient is the product of the absorption cross-section with the difference in population density between lower and upper states of the probed transition and may be expressed as:
\begin{equation}
    \label{eq:32}
    \alpha_{ij}(\nu) =\tilde{\sigma_0} \cdot g_{i}f_{ij} \cdot \left[\dfrac{n_i}{g_i}-\dfrac{n_j}{g_j}\right] \cdot \chi(\nu)
\end{equation}
where $\tilde{\sigma_0} = e^2/4\epsilon_0 m_e c$ is a constant equal to $2.654 \times 10^{-6} $m$^2$ s$^{-1}$, $f_{ij}$ is the transition oscillator strength, and $n_i$ is the number density in the lower level (m$^{-3})$. The population difference term in Eq.~\ref{eq:32} accounts for stimulated emission, but for thermal distributions following Boltzmann statistics at temperatures typically probed by AS in LPPs, this is often negligible for optical transitions at visible wavelengths. Thus, it is often safe to assume $n_j \ll n_i$, in which case Eqns.~(\ref{eq:31}) and (\ref{eq:32}) may be simplified to: 
\begin{equation}
    \label{eq:33}
    A(\nu,t) = \alpha_{ij}(\nu,t)\cdot L = \tilde{\sigma_0}\cdot f_{ij}\cdot \chi(\nu,t)\cdot n_i(t)\cdot L
\end{equation}
Absorbance is a dimensionless quantity, while absorption coefficient represents the optical attenuation per unit length of the medium. The change in absorbance with respect to frequency provides the absorption spectrum, which varies in time in the LPP due to differences in both the lineshape and the atomic number density in the energy levels being probed.

Similar expressions may be derived without assumptions of spatially uniform conditions using radiative transport theory \cite{apruzese2002physics}. As a simple example, in Eq.~(\ref{eq:33}) the product of the absorption coefficient with length may be replaced by an integral along the measurement path:
\begin{equation}
    \label{eq:34}
    {\scriptstyle A(\nu,t) = \int \alpha_{ij}(\nu,t,x)\cdot dx   = \tilde{\sigma_0}\cdot f_{ij}\cdot \int \chi(\nu,t,x)\cdot n_i(t,x)\cdot dx }
\end{equation}
If the conditions in the LPP at a given time can be approximated as uniform in temperature, pressure, and electron density such that the lineshape function $\chi(\nu,t,x)$ does not vary with position, the spatial integral in Eq.~(\ref{eq:34}) only includes the number density over the measurement path and the resulting integral $\int  n_i (t,x)\cdot dx$ is known as the column density. Furthermore, the column density is reduced to $n_i(t)\cdot L$ for spatially uniform conditions along the measurement path. 

The basic experimental set up for performing AS of LPPs includes (1) a pulsed laser to generate an LPP; (2) a light source for probing the plasma; and (3) a detector for analyzing the transmitted/absorbed intensity. Both broadband and narrowband light sources can be used for performing AS \cite{koch2002narrow}; however, the characteristics of the probe light source (e.g., laser, arc lamps, frequency combs, etc.) may affect the properties of the collected data, as well as the analysis method \cite{kautz2021optical,merten2022laser}.  Unlike emission analysis, AS requires a certain experimental orientation for the probe beam. For example, the probe beam should be directed through the plasma parallel to the sample surface, and it may be useful to keep a smaller probe beam size for reducing the interaction region within the LPP. The following subsections provide details of experimental schemes and analysis methods used for AS employing tunable lasers (laser absorption spectroscopy, or LAS), broadband arc sources, and frequency combs. 

\subsubsection{Laser absorption spectroscopy}

\begin{figure*}[t] 
\includegraphics[width=0.8\linewidth]{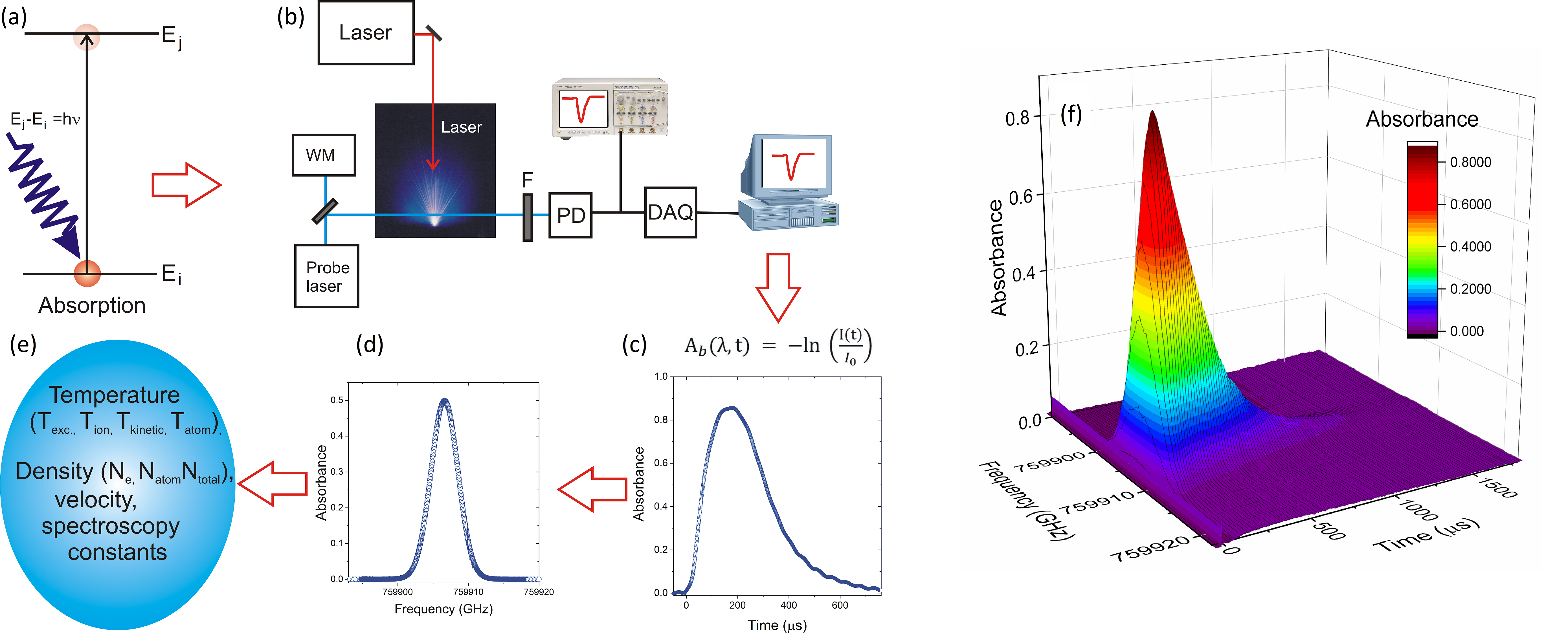}
\caption{\label{fig:20} (a) Absorption process between two electronic levels in an atomic system, (b) schematic of the laser absorption spectroscopy experimental scheme for analyzing an LPP, (c) time evolution of absorbance, (d) absorption spectrum, (e) typical fundamental properties of the plasma that can be obtained from absorption spectroscopy, and (f) the time-resolved absorption spectrum obtaining using tunable laser are given. The selected transition is Al I 394.4 nm. A ns laser was used to produce plasma from an Inconel alloy sample. From \citealp{2021-PSST-Hari}.}
\end{figure*}

 A schematic of the absorption spectroscopy setup for LPP analysis that employs a tunable laser source is given in Fig.~\ref{fig:20}. Examples of time-resolved absorbance, absorption spectrum, and potential LPP parameters that can be gathered using tunable LAS are also given in this figure. For recording the absorption spectrum using tunable CW laser sources, the probe laser wavelength is scanned across the selected transition, and for each ablation event the time-resolved intensity transmitted through the LPP is recorded using a photodiode and digitizer. The wavelength step size and total scan range may be adjusted depending on the linewidth of the transition being probed. Typical spectral coverage of AS using narrow linewidth lasers is limited to the tunability of the laser, which is often $\approx$ 60 GHz ($<$ 0.1 nm) or lower  for continuous high-resolution scans. An example of time-resolved absorption spectra obtained from an LPP source using a CW tunable laser  is given in Fig.~\ref{fig:20}f \cite{2021-PSST-Hari}.

For LAS data collection and analysis, an oscilloscope or analog-to-digital converter (ADC) is routinely used. Considering the dynamical nature of LPP expansion and large dynamic range of measured signals, a time response $\lesssim 1\mu$s and 16-bit ADC resolution is desirable. Signals may be averaged over multiple ablation shots to reduce noise if desired. The wavelength or frequency of the probe laser at each step may be measured to high accuracy using a wavemeter or determined using other methods such as comparison with a reference absorption cell \cite{cervelli1998situ}, hollow-cathode lamp absorption or opto-galvanic signal \cite{barbieri1990optogalvanic}, or measuring transmission fringes through an etalon with a known free-spectral range \cite{2017-SR-Mark}. Typical tunable lasers such as dye or Ti:Sapphire or external cavity diode lasers (ECDL) provide linewidths in the range of $\sim $ 0.1-10 MHz, which is small compared to major line-broadening mechanisms in an LPP (Stark, Doppler, van der Waals). Lineshape measurements of atomic transitions in the LPP require multiple wavelengths to be measured across the absorption profile. If Lorentzian broadening mechanisms are dominant, determination of an accurate spectral baseline may require that the profile be measured over a total range that is significantly broader than the linewidth.   

  Spontaneous emission from the LPP reaching the detector can cause errors in quantitative absorbance measurements \cite{merten2022laser}, and it can be reduced using filters, gratings, prisms, and by limiting the detector field of view.  Ultimately, it is not possible to completely remove the resonant spontaneous emission at the wavelength of the transition probed by the LAS. However, if it can be assured that the probe intensity on the detector is much higher than the collected spontaneous emission intensity, which is usually practical for laser-based absorption light sources, then the spontaneous emission contribution to the signal can be neglected.  This approximation becomes more valid at later times in the LPP evolution probed by LAS. Late time analysis of plasmas with limited spectral broadening and negligible instrumental broadening of LAS enables high-resolution spectroscopic applications such as isotopic splitting and hyperfine structure analysis \cite{smith1999measurement,quentmeier2001measurement, miyabe2013absorption, bushaw2009isotope, hull2021isotopic}.

Given the requirement to measure a large number of wavelengths in sequence for separate LPP ablation events, combined with potential averaging over multiple ablation shots, tunable CW measurements of absorption lines are often time-consuming, especially for low-repetition rate (10-20 Hz) ablation lasers. However, despite the potentially long acquisition times, the measurements provide unparalleled spectral resolution and high-speed temporal information over the full lifetime of the LPP, along with high SNR. Although it is possible to scan a tunable laser at high speed to record an absorption spectrum during a single ablation event, changes in atomic number density and temperature during the scan make spectral analysis problematic, especially at earlier times of LPP evolution \cite{Liu2002LAS}. 

Nonlinearities of Beer’s Law arising from the logarithmic relationship between absorbance and measured intensities may be important for transitions with high peak absorbance. For laser sources with a linewidth much less than the transition linewidth, the linearity of Beer’s law is preserved to high absorbance values. In these cases, the practical limit to the maximum measurable absorbance is typically dictated by the dynamic range of the measurement system and the ability to eliminate all sources of stray light and detector offsets. Detector offsets may often be removed via:
\begin{equation}
    \label{eq:36}
    A(\nu,t) = - \ln \dfrac{I(t,\nu)-I_b}{I_0(\nu)-I_b}
\end{equation}
where $I_b$ is the detector signal in the absence of the light source used for the absorbance measurement. As Eq.~(\ref{eq:36}) makes clear, when the absorbance is high such that $I(t,\nu)\to 0$, any errors in determining $I_b$ will lead to corresponding significant errors in the calculated absorbance. 

Using LAS, it is often possible to measure peak absorbance values up to $\sim 3 - 5$ \cite{2017-SR-Mark, 2021-SCAB-Nicole}. Absorption lines with higher absorbances near the peak may still be measured accurately away from the peak center, and in some cases may be used for spectral fitting if care is taken to exclude the inaccurate points near the peak center \cite{2017-SR-Mark}. However, this situation will naturally lead to higher uncertainty in the fit parameters because the true peak center and maximum amplitude are not measured. For absorption measurements where the measurement resolution is broader than the actual linewidth of the measured lines, Beer’s Law nonlinearities become more problematic \cite{griffiths2007fourier}. In these cases, it may be appropriate to measure only weak absorption lines or to account properly for an instrument lineshape function in the spectral fitting routine \cite{koch2002narrow}. 

 Potential sources of noise in LAS measurements include laser amplitude noise due to power fluctuations, detector noise, pointing instabilities of the probe beam, frequency fluctuations, and density fluctuations in the plasma plume \cite{demtroder2015laser}.  Assuming low-noise and stable CW laser sources are used, the dominant source of noise in LPP measurements is often “flicker noise” arising from variations in LPP conditions between each ablation event. Because the LAS measurement is performed serially at each wavelength, each measured wavelength may experience a plasma with slightly different physical conditions. Thus, for each LPP, the probe beam may encounter different atomic number density, temperature, scattering, and/or beam steering, any of which may lead to amplitude variations in the spectral absorbance profile that are indistinguishable from noise. Averaging over multiple ablation shots may help reduce the amplitude of flicker noise. However, the methodology used for signal averaging (transmitted probe beam intensity vs. absorbance) can influence the errors in the measured absorbance signal due to its nonlinear exponential relation with intensity. Various approaches have been developed to reduce the effects of flicker noise in LAS measurements, including measurement of differential absorption relative to a second co-aligned laser with a wavelength that is not resonant with the probed transition \cite{taylor2014differential}. When other noise sources are removed, the ultimate sensitivity of the LAS method is determined by photon shot noise.  

\begin{figure}[t]
\includegraphics[width=0.3\textwidth]{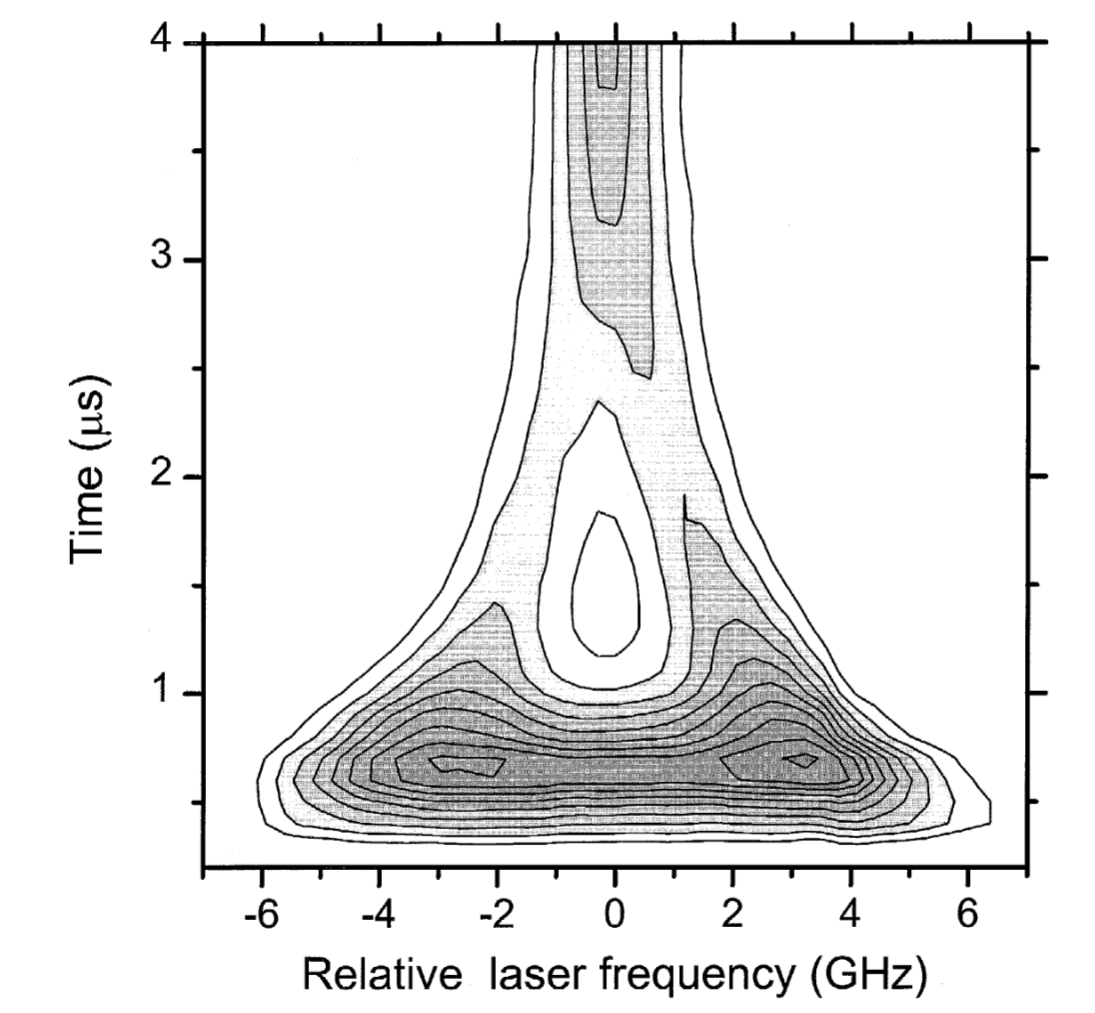}
\caption{\label{fig:25} Contour map of the time-resolved optical spectra of ablated Sr atoms in the presence of 2.5 Torr He cover gas with a probe beam at 3 mm height. Absorbance units with the strongest absorption are indicated by the darkest shading.  Adapted from \citealp{bushaw1998investigation}.}
\end{figure}

Absorption spectral features may be distorted at early times of plasma evolution due to inhomogeneities in the plasma. As an extreme example, at low background pressure conditions and at early times during the LPP evolution, Doppler splitting may be observed in the spectral absorption profiles of lines because of the counter-propagating velocity distributions of atoms or ions along the probe laser line of sight \cite{bushaw1998investigation, Miyabe2012}. Fig.~\ref{fig:25} shows contour maps of the time-resolved absorption of neutral Sr atoms formed in the ablation of CaCO$_3$, which shows Doppler splitting \cite{bushaw1998investigation}. Although most distinct at low pressures and giving rise to a dual-peaked absorption profile, similar effects of non-thermal atomic/ionic distributions may be present at early times for higher pressures as well, which may distort the absorption lineshape from a simple Voigt profile.

Other factors may also influence the observed lineshape in LAS measurements and lead to errors in determined plasma parameters if not properly accounted for. For example, high laser intensity used for probing the plasma may induce power broadening or optical pumping effects in the probed spectral line \cite{vitanov2001power}, although it has been shown that a Doppler temperature may still be obtained by using a correction factor for power broadening \cite{matsui2006influence}. Hyperfine structure in absorption lines may often be observed in high-resolution measurements \cite{2020-SCAB-Hari}, but even if not fully resolved it should be accounted for in spectral fitting to avoid an overestimation of the true spectral linewidths \cite{2021-SCAB-Nicole}. Similar considerations apply for elements with multiple isotopes such as Rb, even if the isotope shifts are too small to resolve experimentally \cite{King1999}.

The absorption spectra obtained using narrowband CW tunable lasers provide limited spectral bandwidth; however, they give high spectral resolution, time-resolved absorbance during the entire lifetime of the plasma, and very high signal-to-noise ratio (SNR) for each measurement because of the high source intensity. So far, the selection of an atomic transition for performing LAS of LPP is constrained to  the availability of reliable lasers with  narrow linewidth and a large tuning range. Recent advancements in diode and Ti:Sapphire laser technology can provide the capability to perform LAS of any atomic transition in the UV-VIS-IR spectral range.   

\subsubsection{Absorption spectroscopy using broadband sources}
\begin{figure*}[t]
\includegraphics[width=0.7\linewidth]{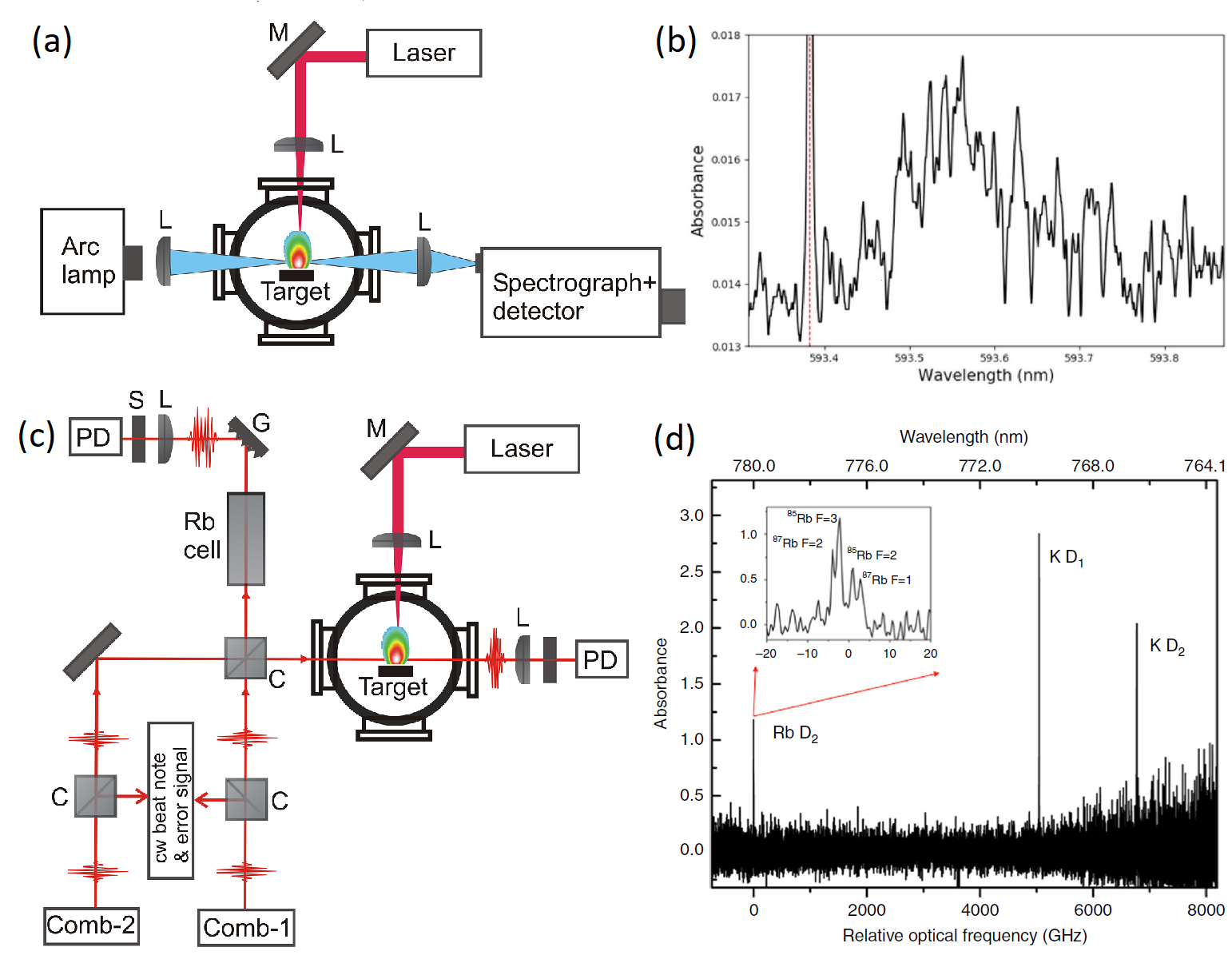}
\caption{\label{fig:22} (a) Schematic of the AS setup using a BB arc lamp. (b) The absorption spectrum from an LPP recorded using an arc lamp is given. From \citealp{weerakkody2021quantitative}. The absorption spectrum corresponds to the UO 593.55 nm band from a U metal plasma, and the peak marked with a red dotted line corresponds to the UI 593.38 nm atomic transition. (c) Schematic of AS employing frequency combs. (d) Broadband single-shot absorption measurement of the laser-induced plasma using frequency combs. The inset shows a zoomed-in view of the $5S_{1/2} - 5P_{3/2} Rb~D_2$ line. From \citealp{2018-Nature}. }
\end{figure*}

AS employing broadband light sources such as arc lamps, laser plasma continuum, etc. provide wide spectral bandwidth; however, a spectrograph is used for measuring the absorption spectrum (Fig.~\ref{fig:22}a). In this scenario, similar to emission spectroscopy, the spectral resolution available is constrained by the detection system. Information about the LPP at a certain time after ablation can be collected from a single shot, and the temporal dynamics are generated by delaying the probe source with respect to the LPP source or by using time-gated detection. Although broadband measurements are made on a single shot, improvements in SNR may require averaging over multiple shots. The use of broadband absorption methods where multiple wavelengths are measured simultaneously may also be advantageous for reducing some effects of flicker noise because the LPP variations affect all wavelengths uniformly.    

\citet{weerakkody2021quantitative} used a xenon flash lamp as a broadband light source for performing absorption spectroscopy of LPPs. The flashlamp provided a broadband light source with a duration of $\sim$ 10 µs, which was collected, directed through the LPP, and then dispersed using a custom 1.5 m CT spectrometer with gratings used in high orders similar to an echelle configuration. An example of a uranium oxide (UO) absorption spectrum obtained from an LPP source using an arc lamp is given in Fig.~\ref{fig:22}b. Because time resolution was provided by the pulsed light source, a non-intensified CCD detector was used for detecting the transmitted light, and absorption spectra at different times after LPP generation were probed. An optical parametric oscillator (OPO) based pseudo-continuum source was used for performing absorption spectroscopy in conjunction with a high-resolution echelle spectrograph with ICCD \cite{merten2018massing}.

Several authors used the early-time Bremsstrahlung emission from an LPP as a broadband pulsed light source for performing AS. For example, \citet{ribiere2010analysis} performed near-UV absorption spectroscopy of various elements including Al, Mg, Ni, Cu, and Si by employing the continuum emission from an LPP as a source to probe a second time-delayed LPP. In a similar experiment scheme, called dual plasma absorption spectroscopy (DP-AS), the broad VUV continuum emission from a high-Z LPP was used to probe VUV photo-absorption spectral features from a second LPP  \cite{carillon1970extreme, carroll1977doubly, costello1991xuv, neogi2001vacuum}.  Similar to a DP-AS scheme, an extreme ultraviolet (XUV) continuum generated from a first LPP was used to measure the XUV absorption spectra of Th \cite{meighan2000application}. A confocal configuration of the double-pulse measurement was performed to study Fraunhofer-type absorption lines in a cooling LPP \cite{nagli2012fraunhofer}.\

Compared to using arc lamps and tunable laser sources as probes, the AS of LPPs using frequency combs is a  relatively new approach. The limitations of LAS (narrow spectral range) and AS employing broadband arc lamp light sources (spectral resolution) can be overcome with the use of frequency combs, which provide both high spectral resolution and broadband detection \cite{2018-Nature}. A frequency comb is a stabilized mode-locked laser, which provides a set of narrow-linewidth frequency modes \cite{cundiff2003colloquium}. Dual-comb spectroscopy (DCS) uses multi-heterodyne optical interference between two frequency combs with slightly different repetition rates and provides a representation of the optical spectrum in the radio-frequency (RF) domain, with a one-to-one mapping of RF frequency to optical frequency \cite{coddington2016dual}. A schematic of a DCS AS setup is given in Fig.~\ref{fig:22}c.

A benefit of DCS for the measurement of absorption spectra in LPPs is that all wavelengths are measured simultaneously.  As a result, flicker noise in DCS does not lead to spectral noise as it does in scanning CW LAS.. However, the trade-off for measurement of all wavelengths simultaneously in DCS is that the average power on the photodiode must be reduced well below what is typically used for CW LAS measurements. As a result, the noise levels are typically higher for DCS relative to CW LAS but may be reduced by averaging over multiple LPP events. Another trade-off with DCS arises between time resolution and frequency resolution. Because DCS measurements are based on a Fourier transformation of an interferogram in the time domain to generate the spectral measurement, the time and frequency resolutions are linked \cite{2021-SCAB-Weeks}. In  \citet{2018-Nature}, an absorption spectrum of Rb and K was measured over an optical bandwidth of 13 nm with sub-GHz resolution (Fig.~\ref{fig:22}d).  The spectral resolution was sufficient to measure the hyperfine and isotope splitting of the Rb D2 line. DCS AS was used for time-resolved plasma characterization and for  resolving the closely spaced absorption lines of higher-Z materials \cite{2019-OL-Zhang, 2021-SCAB-Weeks}.   

\subsubsection{Temperature and density measurement}
Most of the data analysis methodologies presented for emission spectroscopy (Section \ref{sec:Optical_spectroscopy}.A) are also applicable to AS analysis. For example, by monitoring the Stark width or shift of the absorption line, the electron density can be inferred. Similarly, the Boltzmann plot can be constructed using absorption lines for inferring excitation temperature. In the last three decades, several groups have used the AS of LPPs to measure lower state populations, transition linewidths in the plasma, and absorption-based kinetic and excitation temperatures \cite{mitzner1993time, duffey1995absorption, cervelli1998situ, miyabe2013absorption, merten2018massing}.  

In the following discussions, we will use the simplified forms of the equations without explicitly including the spatial integration along the line of sight, but it should be understood that these conditions do not exist at all times in the LPP. For late times of LPP evolution that are usually probed by AS, it is often acceptable to assume spatially uniform conditions. Similar equations can be written in units of frequency, wavelength, or wavenumber, with corresponding changes in units for spectral integrations. Finally, it is also possible to write the above equations using Einstein A and B coefficients instead of oscillator strength, and conventions differ between various sub-fields. 

\begin{figure}[t]
\includegraphics[width=0.4\textwidth]{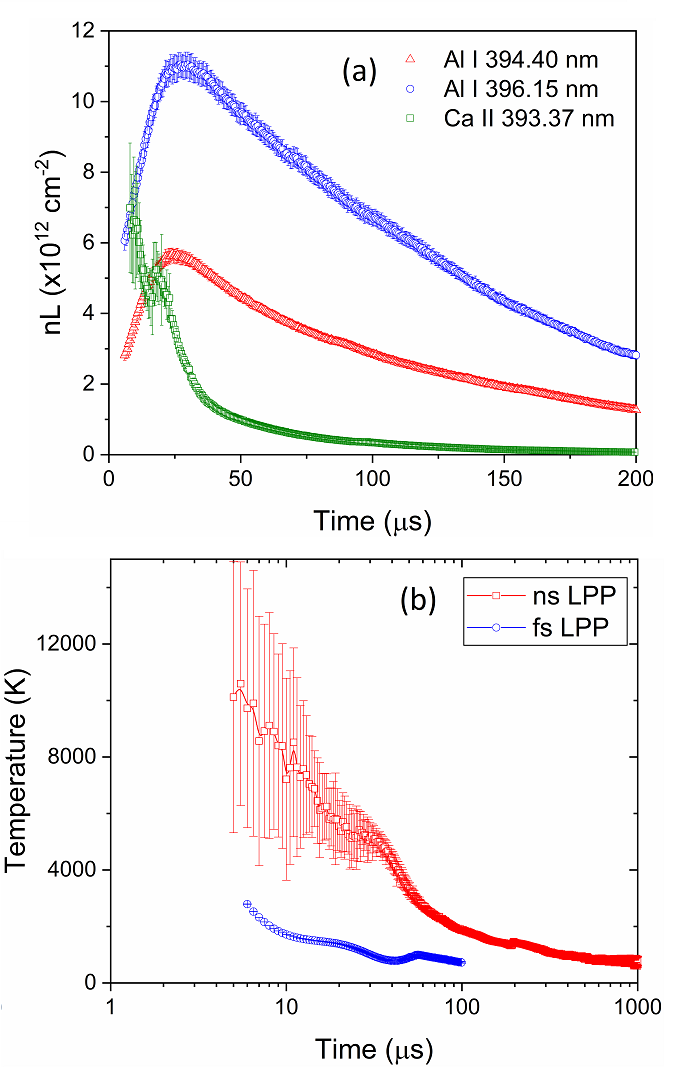}
\caption{\label{fig:23} Column density (nL) as a function of time calculated from fits to LAS absorbance spectra for two Al transitions at 394.4 nm and 496.15 nm and a Ca II transition at 393.37 nm. 266 nm, 6 ns pulses were used for producing LPP on a glass target, and LAS spectra were obtained using a CW tunable laser. Adapted from \citealp{2021-SCAB-Nicole}. Comparison of time evolution of kinetic temperatures measured using LAS of ns and fs LPPs. An Inconel sample containing a small amount of Al $(< 0.4$ at$\%)$ was used. The large error bars at early times of ns LPP are a result of uncertainty in the spectral line fitting due to low absorbance. From \citealp{2021-PSST-Hari}.
}
\end{figure}

For LAS measurements with a tunable CW laser source, the experimental results provide an absorbance spectrum of an optical transition for multiple times in the LPP evolution. As shown by Eq.~(\ref{eq:33}), in spatially uniform conditions, each absorbance spectrum may be represented by $A_{ij}(\nu,t) = \tilde{\sigma_0}\cdot f_{ij}\cdot \chi(\nu,t)\cdot n_i(t)\cdot L$. At each time delay after ablation, the absorbance spectrum may be fit using an appropriate lineshape function $\chi(\nu,t)$, and the spectrally integrated area of the absorbance peak is proportional to the column density in the lower level of the probed transition. An example of time-resolved, path-integrated atom density for two Al I transitions (394.4 nm and 396.15 nm) and ion density of a Ca II transition (393.37 nm) are given in Fig.~\ref{fig:23}a \cite{2021-SCAB-Nicole}. This measurement was performed using a tunable laser as a probe, and a spectral fit was performed for the experimental spectra at each time delay to determine the column density via the peak areas. Several authors used LAS for measuring path-integrated (column) number densities of a selected transition and performing quasi-three-dimensional spatial mapping \cite{duffey1995absorption, AlWazzan1996, Martin1998, Mazumder2002, Gordillo-Jap-2001}.  

The column density measured by LAS is determined by the number density in the lower energy level probed by the measurement and not the total atomic number density. This is true even for ground-state resonance transitions. Assuming a Boltzmann distribution, the atomic populations in the energy levels are given by Eq.~(\ref{eq:11}), from which it is apparent that the ground state number density is reduced according to the partition function at the given temperature. In tunable LAS measurements of resonance lines, this effect can often be observed via a reduced column density at early times in LPP evolution due to the high excitation temperatures, as shown in Fig.~\ref{fig:23}a for Al I transitions.

The fitting of experimental absorbance lineshapes may also be used to determine various physical properties of the LPP. As described in Section \ref{sec:Optical_spectroscopy}.A, the spectral lineshape is usually modeled as a Voigt profile, wherein the width of the Lorentzian component is determined by collisional broadening (Stark, resonance, van der Waals) and the width of the Gaussian component by Doppler broadening. For LAS with CW lasers, the instrumental broadening is negligible; however, it may be important for broadband absorption methods. Because LAS probes the electronic lower energy state of an atomic population, analysis of plasma evolution is possible at lower temperatures and later times than what is accessible by emission measurements.  At late times, Stark broadening becomes negligible, and the linewidth is dominated by Doppler broadening and possibly van der Waals broadening at higher ambient pressures. The kinetic (Doppler, translational) temperature can be calculated from the Gaussian linewidth according to the relation \cite{demtroder2015laser}

\begin{equation}
    \label{eq:40}
    w_g = 7.16 \times 10^{-7}\lambda_0 \left(\dfrac{T}{m}\right)^{1/2}
\end{equation}
where $w_g$ is the Gaussian FWHM of the line in nm, $\lambda_0$ is the center wavelength of the transition in nm, T is the kinetic temperature in K, and m is the mass of the species in amu.  

Examples of kinetic temperatures of ns and fs LPPs measured using LAS are given in Fig.\ref{fig:23}b \cite{2021-PSST-Hari}. The large errors in the measured temperature at early times of ns LPP evolution are due to low absorbance signals and therefore the uncertainty in the spectral line fitting. \citet{2021-SCAB-Nicole} used LAS for measuring the time evolution of kinetic temperature by employing Doppler-broadened Al transitions in a ns LPP and comparing to excitation temperature determined from OES, thereby finding a discontinuity between temporal decay of the kinetic and excitation temperature, which can be explained by the inhomogeneous nature of the LPP. Therefore, even if the plasma is in LTE, differences in various temperatures (neutral vs. ion vs. molecular vs. kinetic) can be expected because of the inhomogeneous properties of LPP systems, similar to observations using emission spectroscopy.

\begin{figure}[t]
\includegraphics[width=0.45\textwidth]{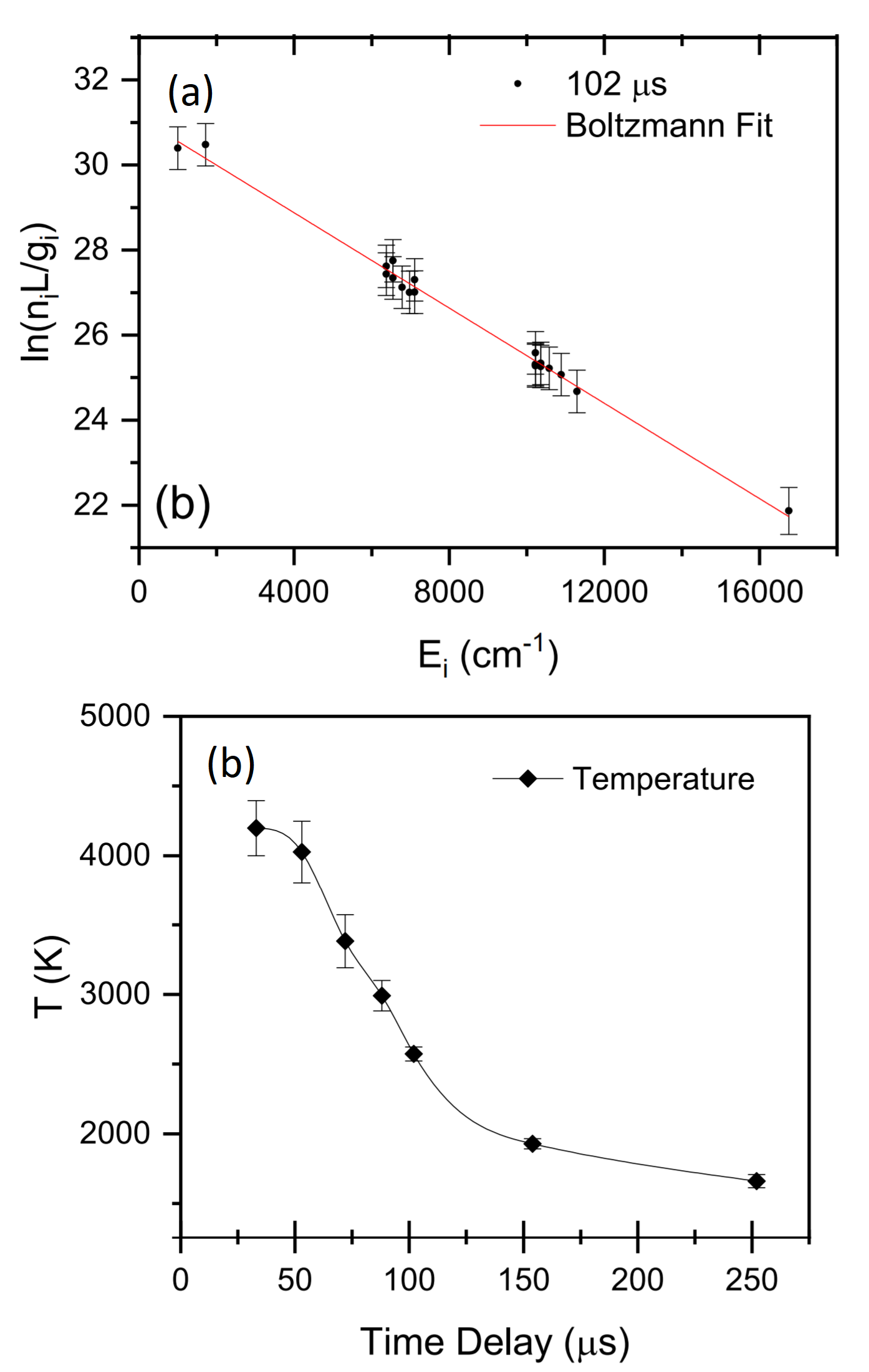}
\caption{\label{fig:24} (a) An example of a Boltzmann plot generated from DCS absorption spectroscopy of Gd plasma and using Gd lines is given. (b) The measured excitation temperature evolution is given. From \citealp{2021-SCAB-Weeks}}. 
\end{figure}

The absorption spectrum obtained using broadband light sources (arc, frequency comb) can be used for excitation temperature measurement by employing Boltzmann plot methods. Using the column density measured from absorption $(n_i \times L)$ combined with the Boltzmann relation given in Eq.~(\ref{eq:11}), the electronic excitation temperature can be measured under the assumption of LTE using the following relation  
\begin{equation}
    \label{eq:39}
    \ln\left(\dfrac{n_i\cdot L}{g_i}\right) = \ln \left(\dfrac{n_{tot}\cdot L}{Z(T)}\right) - \dfrac{E_i}{k_bT}
\end{equation}
Thus, the slope of a linear fit of $\ln\left({n_i\cdot L}/{g_i}\right)$ vs $E_i$ gives the excitation temperature, while the y-intercept is related to the total column density divided by the partition function.   Boltzmann plots from multi-line absorption spectra used to determine the absorption excitation temperature and column density are given in Refs. \cite{duffey1995absorption, 2019-OL-Zhang, weerakkody2021quantitative}. An example of a Boltzmann plot using Gd absorption lines measured using DCS, and the resulting excitation temperatures are given in Fig.~\ref{fig:24} \cite{2021-SCAB-Weeks}.

In principle, it should be possible to determine Stark broadening at early times of plasma evolution via an increase in the Lorentzian linewidth \cite{bratescu2002electron, 2021-PRE-Hari}; however, these measurements are complicated due to depletion of low-energy-level populations, leading to a low absorption strength. The high spatial inhomogeneity at early times and non-thermal velocity distributions may also lead to deviations in the observed lineshape from a simple Voigt profile. Furthermore, the large density gradients present in the early times of plasma expansion may distort the line shapes arising partly from optical refraction \cite{El-Astal-JAP1996}.  The measurement of Lorentzian width at late times of plasma evolution at various pressure levels provides van der Waals broadening \cite{2021-PRE-Hari, taylor2014differential}. 

\citet{Merten2018Pseudo} used an OPO-based pseudo-continuum source for performing absorption spectroscopy and reported Li isotopic shifts as well as the time evolution of the kinetic temperature of a ns LPP. They used Abel inversion methods to determine total number density and mass of atomic Ti in LPPs. \citet{weerakkody2020time, weerakkody2021quantitative} used broadband, high-resolution absorption spectra of U I and U II to determine absorption excitation temperature via a Boltzmann plot analysis. Molecular absorption spectra from SiO, BeH, UO, and ZrN were reported, and when possible were compared with simulated spectra modeled using PGOPHER to estimate molecular temperatures.

\subsubsection{Kinetics of LPP using absorption spectroscopy}
Several authors used AS for investigating the LPP spatiotemporal dynamics \cite{geohegan1989characterization, yang1999spatially, Miyabe2012, Cheung-JAP1991, Krstulovic-SCAB-2008, Harilal2022SpatiotemporalEO}. These methods utilize a combination of temporal dynamics of the observed absorption signal at the probe laser positions, along with variations in the absorption spectral profile with time. As was shown in Fig.~\ref{fig:25}, at early times of plasma evolution a dual-peaked absorption spectra may be observed, especially at low pressure, which may be used to estimate the velocity distributions of atoms and ions in the expanding LPP \cite{bushaw1998investigation}. \citet{Miyabe2012} used similar methods to investigate plume dynamics of neutral and ionized Ce. 

\begin{figure}[t]
\includegraphics[width=0.4\textwidth]{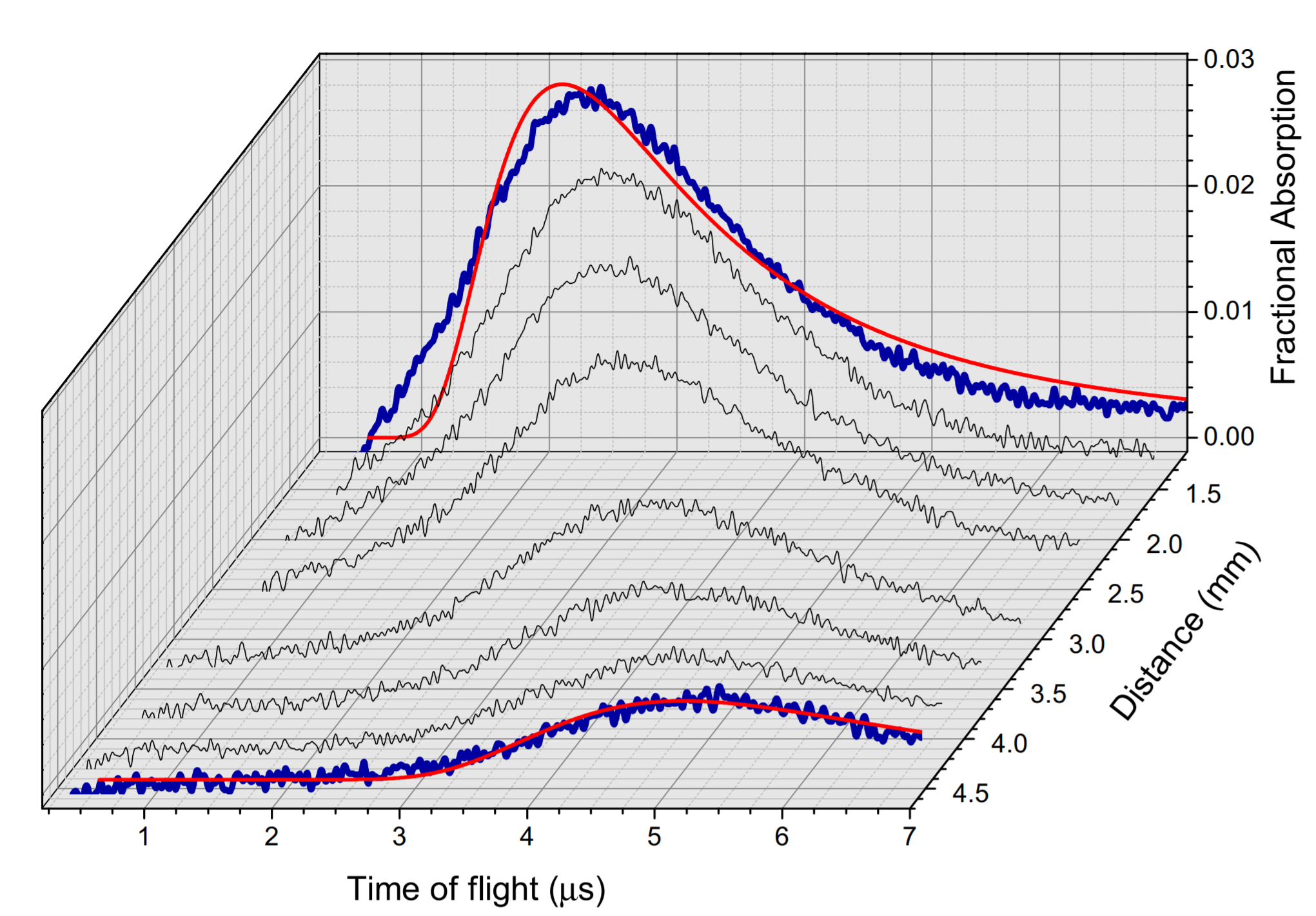}
\caption{\label{fig:26} TOF absorption signals (blue and black curves) of BaH molecules from an LPP are recorded at different distances from the target surface. The smooth red lines represent the best fit to the data. From \citealp{tarallo2016bah}}. 
\end{figure}

\citet{tarallo2016bah} used LAS to measure plume dynamics and kinetic temperature of molecular BaH using Doppler-shifted absorption profiles and inclusion of nonlinear absorption saturation effects. Examples of the time-of-flight absorption signals recorded at various distances from the target for BAH molecules are given in Fig.~\ref{fig:26}. \citet{Harilal2022SpatiotemporalEO} used a combination of absorption and emission TOF profiles for comparing spatiotemporal evolution of excited and ground state species in a laser ablation plume. They also reported that the LAS TOF profiles are useful for studying shock propagation into ambient gas medium.  \citet{geohegan1996laser} used  a $\sim$ 1.5 µs Xe broadband source for recording the absorption spectrum and for studying kinetics of atoms and ions from an LPP.  

\subsection{Laser-induced fluorescence }
Laser-induced fluorescence (LIF) combines the principles of both absorption and emission. In LIF, the probe laser is tuned to match an atomic or molecular transition, and the spontaneous emission resulting from the decay of the resonantly excited atoms is measured. The following subsections provide details of LIF instrumentation for LPP analysis, and methods employed for plasma characterization.   

\subsubsection{Instrumentation and analysis considerations}
\begin{figure}[t]
\includegraphics[width=\linewidth]{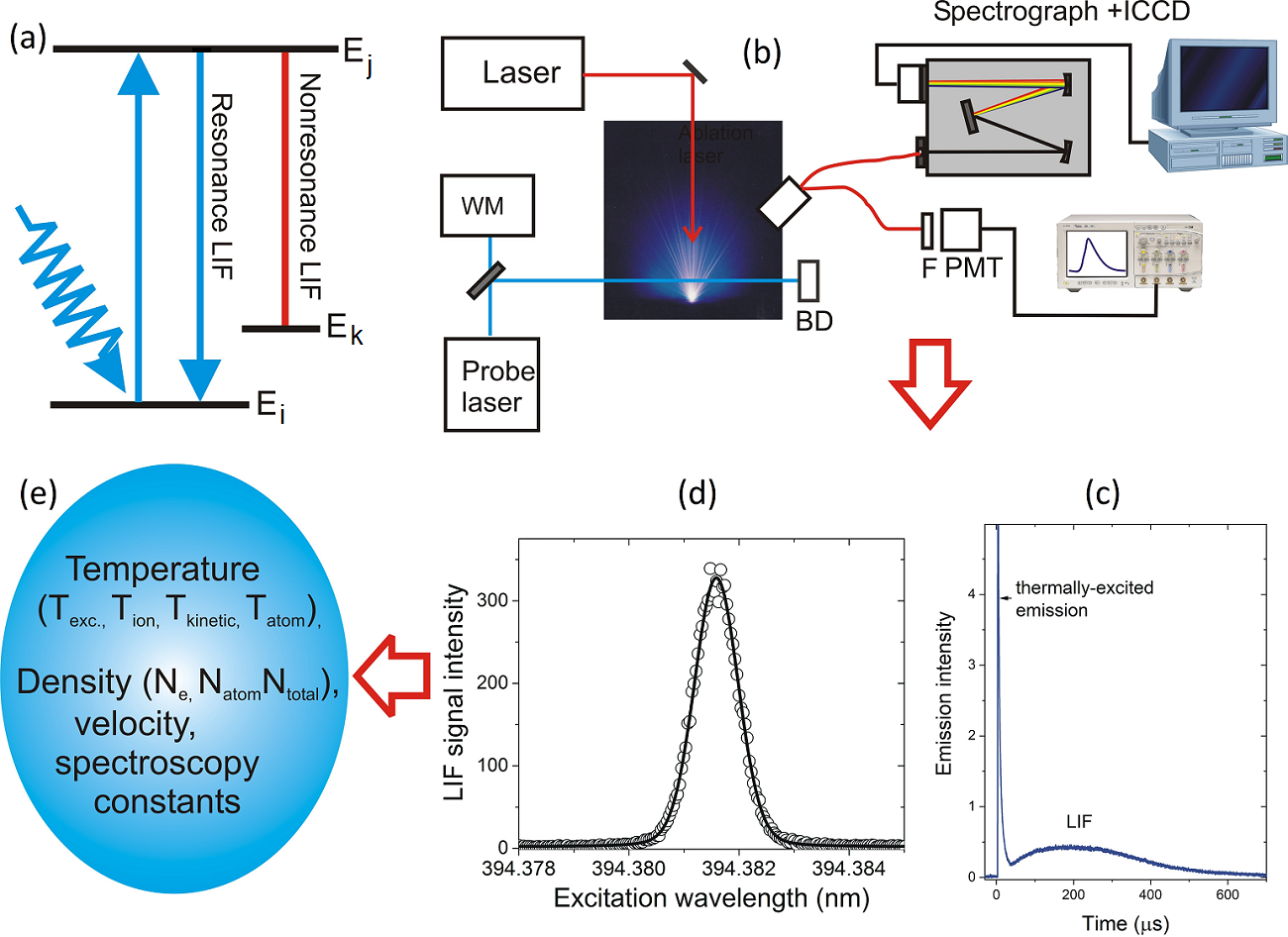}
\caption{\label{fig:27} (a) Schematic of LIF excitation and emission; (b) typical LIF of LPP experimental scheme; (c) time-resolved LIF emission from LPP using CW excitation, where the early time peak seen in the temporal profile is due to thermally excited emission; (d) excitation spectrum; and (e) potential properties of the plasma that can be gathered using LIF.}
\end{figure}

Since LIF combines the principles of absorption and emission spectroscopy, the instrumentation requirement is also a combination of LAS (for excitation) and OES (light collection and analysis).   A schematic of the LIF pumping and detection schemes, an example of a time-resolved LIF emission, and an excitation spectrum are given in Fig.~\ref{fig:27}, along with potential plasma properties that can be measured using LIF of LPP. If the resonantly pumped transition shares the upper energy level with more than one transition, LIF can be monitored by selecting any of these transitions. If the LIF emission is monitored at the same wavelength of probe radiation, it is commonly called resonance LIF. Similarly, if other transitions that are coupled to the same upper level are used for LIF monitoring, it is called directly coupled LIF or non-resonance LIF (see Fig.~\ref{fig:27}a). The use of a directly coupled LIF transition is useful in avoiding detection of the spurious scattering signals from the probe laser. Unlike LAS, the LIF pumping scheme has fewer geometric limitations. So, LIF can be used for standoff analysis \cite{2018-OL-Hari, Kautz2021LaserinducedFO}.  

 The general parameters of LIF spectroscopy for probing and sensing are the excitation spectrum, emission spectrum, and fluorescence lifetime or decay rate \cite{Stchur2001}. When the probe laser is tuned across the spectral range of an absorption line, the fluorescence intensity is monitored as a function of probe wavelength, and the spectrum obtained is called the excitation spectrum. Under conditions of low optical density and low probe intensity, the LIF excitation spectrum is directly proportional to the absorption spectrum obtained from LAS. For excitation, lasers with a wide wavelength tuning range and narrow linewidth are preferred.

Both pulsed lasers \cite{nilsen2005plasma, gormushkin1997determination} and CW lasers \cite{smith2002pu} can be used for LIF excitation, provided that the spectral bandwidth of the laser is smaller than the linewidth of the optical transition for efficient excitation of the targeted transition. However, the LIF probe laser properties influence the detection scheme, and the use of high laser intensities may affect the recorded lineshape. At low laser intensities, the LIF signal varies linearly, where the de-excitation rate of an atom is faster than the excitation rate. In this scenario, most of the atoms are in the lower/ground state, and the probe laser causes negligible depopulation. As the laser intensity of the LIF probe laser increases, the LIF signal becomes nonlinear with respect to pump laser intensity due to depopulation or bleaching of the lower-level population, leading to saturation of the detected LIF signal \cite{nakata1999correction}. The saturation effect during LIF experiments will lead to spectral broadening because saturation is strongest in the line center compared to the wings. Detailed description of saturation and power broadening is discussed by Demtroder \cite{demtroder2014laser}. Compared to the pulsed LIF excitation scheme, the laser-intensity saturation effects are expected to be minimal with the use of a CW excitation source. The other advantage of using CW lasers as an excitation source is that they continuously excite the lower-state population during the entire lifetime of the LPP plume, which may be monitored via time-resolved detection of the continuous LIF emission signal \cite{2021-PSST-Hari}.  

\begin{figure}[t]
\includegraphics[width=\linewidth]{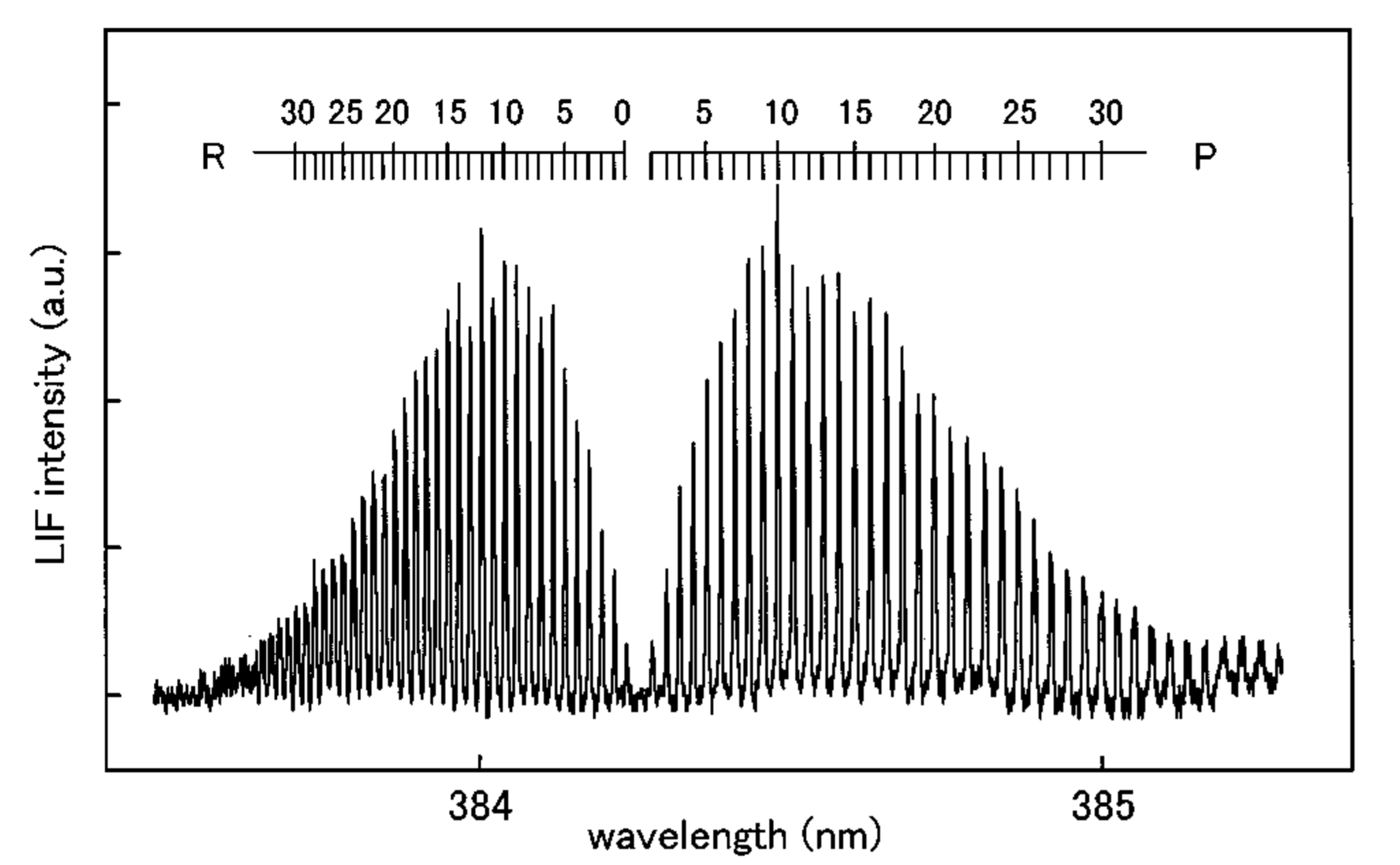}
\caption{\label{fig:28} The LIF spectra of the SiO$^+$ 0-0 band recorded at 80 mTorr O$_2$ pressure. The plasma was produced using a 1064 nm laser beam, and LIF was performed by scanning a pulse dye laser. From \citealp{matsuo1997formation}.
}
\end{figure}

Any traditional detectors with appropriate spectral filters can be used for measuring the LIF signal. If the wavelength of the LIF emission is different from the excitation wavelength (directly coupled LIF), the scattering of the LIF probe laser can be easily eliminated by filters or monochromators. For detection, PMTs are preferred for measurements requiring high time resolution or small signal intensities. The potential sources of noise or interfering signal associated with LIF measurements include thermally excited emissions from the plasma, stray laser light, inherent noises from the detector (dark current, shot noise), and electromagnetic pickup noises coming into detection lines. For LPPs, the thermally excited emission is typically the dominant signal interfering with the detection of the LIF emission signal. However, the thermally-excited emission and LIF emission may often be separated using gated detectors due to their different time-dependence (See Fig.~\ref{fig:27}c). LIF also provides benefits for detection of small emission signals by virtue of being a “background-free” detection method, in contrast to AS where small signal variations must be detected on a large background $I_0$. The spectral resolution available in LIF diagnostics of the LPP is dictated by the linewidth of the probe laser; therefore, LIF can be utilized for recording high-resolution spectral features. An example of the LIF spectrum is given in Fig.~\ref{fig:28}, where a pulsed dye laser is used to record the LIF spectrum of SiO$^+$ from a laser-produced Si plasma generated in an 80 mTorr oxygen ambient \cite{matsuo1997formation}.

\begin{figure}[t]
\includegraphics[width=\linewidth]{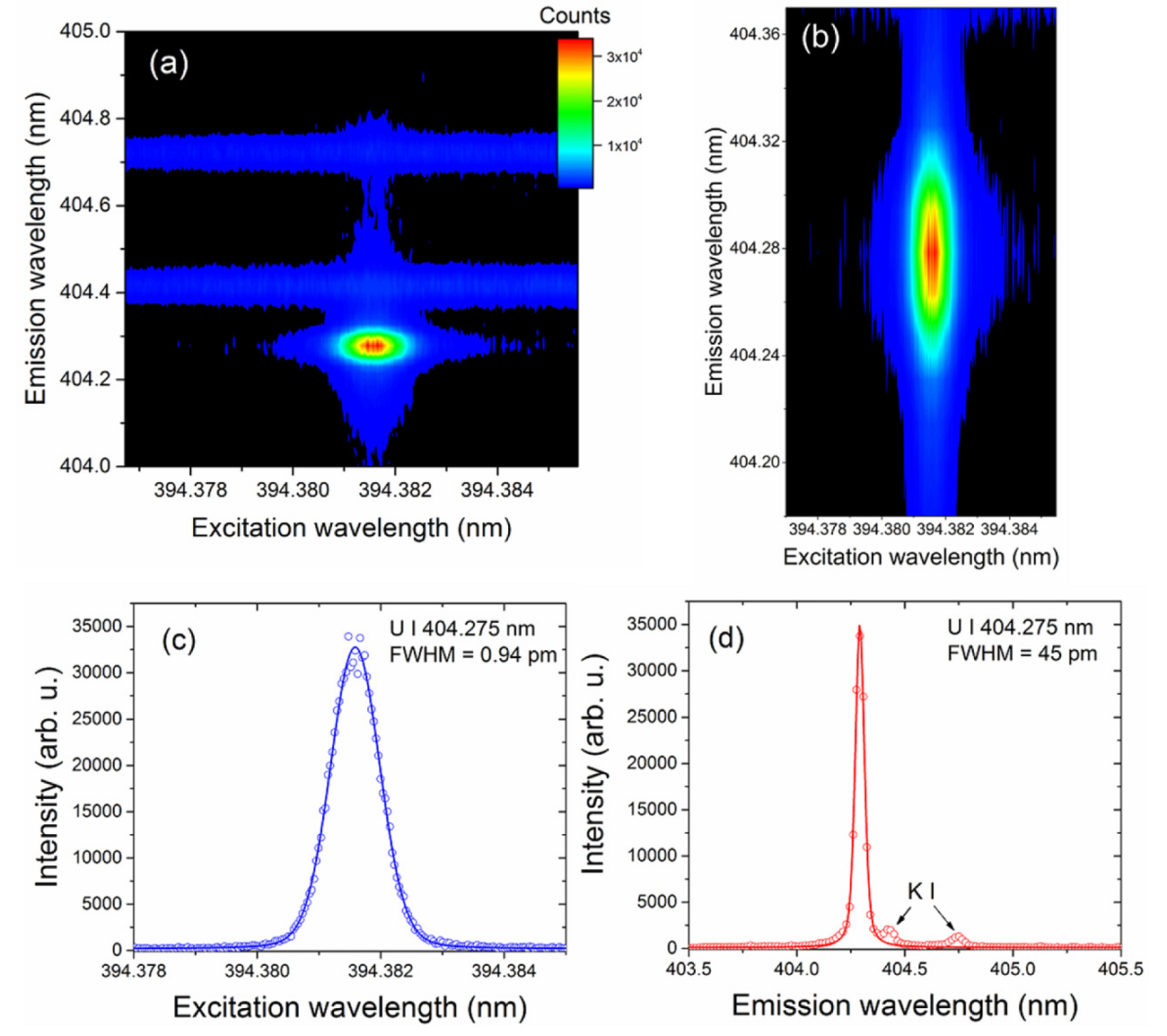}
\caption{\label{fig:29} (a) 2D-FS of U I  and weak thermal emission from K I lines at  45 Torr N$_2$ pressure from a ns-laser-produced plasma. (b) Zoomed-in image of 2D-FS of UI transition. The emission and excitation spectra are given in (c) and (d). From \citealp{2017-OE-Hari-LIF}. }
\end{figure}

Simultaneous emission and excitation spectral features can be measured using 2D-fluorescence spectroscopy (2D-FS), where a spectrograph - ICCD combination is used. This technique can also be used for reducing effects of shot-to-shot (flicker) noise in an LPP \cite{2017-SR-Mark}. For recording the 2D-FS, the LIF emission spectra are collected for each wavelength step during the excitation laser scan using a spectrograph-ICCD detector system, giving a map of emission intensity versus both excitation and emission wavelengths. An example of 2D-FS is given in Fig.~\ref{fig:29} where non-resonant LIF emissions from U I 404.275 nm was recorded while the excitation laser beam was tuned across the U I 394.3816 nm transition \cite{2017-OE-Hari-LIF}. The measurement was performed on a ns LPP from a glass target containing a trace amount of U at 45 Torr N$_2$ ambient pressure. A one-dimensional cross-section of the 2D-FS map at a fixed emission wavelength gives the corresponding excitation spectrum (Fig.~\ref{fig:29}c). Similarly, a one-dimensional cross-section at a fixed fluorescence excitation wavelength provides the emission spectrum (Fig.~\ref{fig:29}d). The weak emissions seen at 404.4 and 404.72 nm are from K I, and the 2D-FS spectrum shows a constant emission intensity for these transitions versus LIF probe wavelength, indicating that the K I transitions were not pumped.

 Pulsed LIF of LPPs is routinely used to boost the emission signal, which is important for analytical applications such as trace detection \cite{hilbk2001analysis, telle2001sensitive, laville2009laser, kang2017ultrasensitive, loudyi2009improving}. Previous studies showed that LIF of LPP improved detection limits and reduced matrix effects \cite{kwong1979trace}. LIF of LPPs is also useful for detecting isotopes \cite{smith1998laser, 2017-SR-Mark}, analyzing hyperfine structures \cite{2020-SCAB-Hari}, and reducing self-reversal effects in the spectral profiles \cite{Kautz2021LaserinducedFO}. Although the LIF provides high sensitivity for trace element detection, it is challenging to use LIF for quantitative measurements of bulk elements because of the inherent nonlinearity of LIF signals with atomic number density under conditions of high absorbance \cite{2020-SCAB-Hari}. For a low-intensity LIF probe laser, the detected LIF signal $S_F(\lambda_{ex},t)$ can be expressed in a simplified form as \cite{burns2011diode}:
\begin{equation}
    \label{eq:41}
    S_F(\lambda_{ex},t) = S_0 \cdot I_0 \cdot [1 - e^{-A(\lambda_{ex},t)}]
\end{equation}
where $\lambda_{ex}$ is the wavelength of the laser exciting the LIF transition, t is time, $S_0$ is a constant factor incorporating collection area/efficiency and fluorescence quantum yield, $I_0$ is the incident laser intensity, and $A(\lambda_{ex},t)$ is the absorbance of the LIF probe laser. Under conditions of high absorbance, the LIF intensity is nonlinear with the atomic number density and approaches a maximum value, which may distort the shape of the excitation spectrum in a manner similar to self-absorption effects observed in emission spectroscopy.

\subsubsection{Temperature and density measurement}

LIF is a very well-known technique in combustion research; fusion devices (e.g., Tokamak); and ICR plasmas for sensitive and spatially resolved analysis of particle behavior, density, and velocity distributions of atoms, ions, and molecules. In theory, the LIF of LPP can provide all plasma properties gathered using the LAS technique. However, challenges exist to perform accurate measurement of fundamental properties of a high-density plasma source such as LPP using LIF compared to LAS.  
For example, if LIF photons are reabsorbed by the plasma, it may distort the excitation spectra. Understanding the collisional quenching of LIF signal is also very important. Hence, a detailed understanding of signal-generation and radiation transport processes, which requires solving the rate equations combining both absorption and collisional decay, is essential for using LIF for accurate plume characterization.   

Several authors use the LAS technique for calibration of LIF signatures to measure the absolute density of particles in an LPP \cite{niemax1990optical, lui2008detection, nicolodelli2018determination}. \citet{Martin1998} measured Mg atom density from an LPP derived from a planar LIF image, and the calibration of the number densities was performed using the resonance broadening of the absorption line shapes. \citet{dutouquet2001laser} used LIF for measuring ground-state number densities of atoms and molecules from LPPs generated from Al, C, and Ti targets in N$_2$ or O$_2$ low-pressure atmospheres, and the absolute calibration of ground-state densities was performed using additional absorption measurements. \citet{orsel2016laser} used LIF for studying oxidation processes in YBiO$_3$ LPPs. They recorded Y, YO, and Bi spatiotemporal distributions in the plasma using LIF, and calibration of absolute density was performed using simultaneous measurement of AS (Fig.~\ref{fig:28}). 2D LIF imaging is also used by several authors for studying the dynamics of the LPP in the presence of an ambient \cite{nakata1999correction}, and details about LIF imaging are given in Section~\ref{sec:imaging}B.

\begin{figure}[t]
\includegraphics[width=\linewidth]{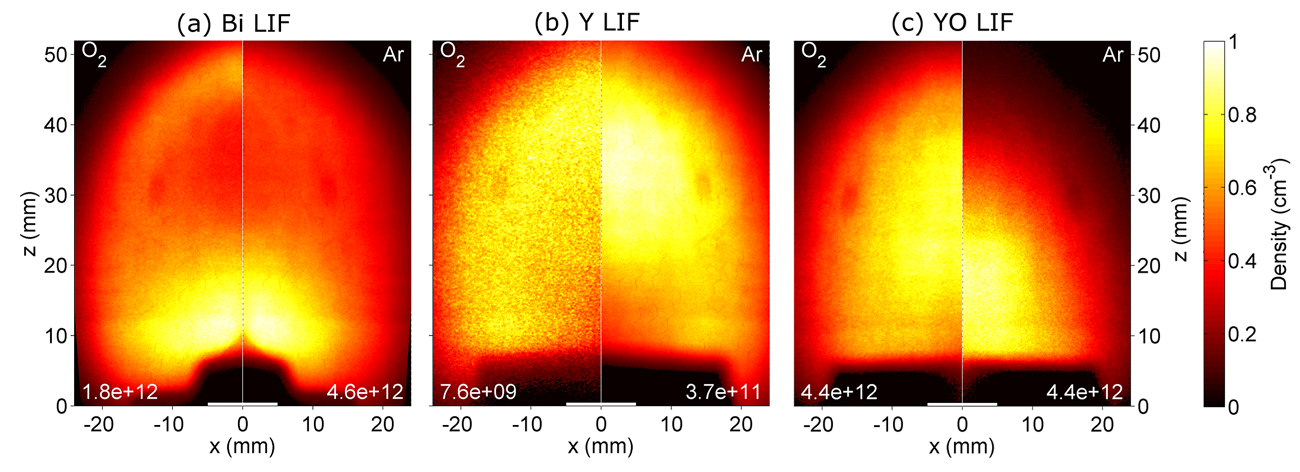}
\caption{\label{fig:30} Density distributions of Bi, Y, and YO from a YBiO$_3$ LPP. Each picture contains an LIF measurement in a 75 mTorr ambient environment: (left) $O_2$ and (right) Ar. All pictures are normalized, with the normalization factor shown at the bottom of each image. The densities shown are measured at 35 µs delay after ablation. From \citealp{orsel2016laser}.}
\end{figure}

\subsection{Local Thermodynamic Equilibrium}

For the temperature and density measurement employing spectroscopic tools, it is commonly assumed that the plasma is in LTE (except the electron density measurement using Stark broadening). Under thermal equilibrium (TE), the population distribution of atoms or ions at various energy levels in a plasma is related to the temperature and the density through a Boltzmann distribution, the population distributions of atoms of adjacent ions states are described by a Saha-Boltzmann relation and the intensity of radiation is given by the Planck function. However, it is challenging to establish TE in LPPs. So, the measurement of various temperatures in the LPP is carried out under the assumption of LTE. 

The existence of LTE in a transient system such as  LPP requires that the electron-atom and electron-ion collisional processes occur sufficiently fast and dominate the radiative processes.  The radiation processes in a LTE plasma are not following a Planck function; instead, it depends on local plasma conditions along with population distributions and atomic transition probabilities. LTE is a reasonably valid assumption for LPPs with relatively high densities and low plasma temperatures. In these scenarios, the population distributions at any time during plasma evolution are determined by the local plasma conditions set by Boltzmann and Saha relations. Even if there is a departure from LTE, a part of the energy levels favor collisional transitions over radiative transitions and those plasmas are considered to in partial LTE. 

There exist several articles discussing the existence of LTE in an LPP \cite{Cristoforetti2010, Zhang2014}, and still, it is a debated topic specifically when and where the LPP is considered to be in LTE or partial LTE.  One of the most commonly utilized methods for validating the existence of LTE in LPP is the McWhirter criterion \cite{McWhirter1969}, which states that the minimum density for LTE should be
\begin{equation}
    \label{eq:30}
    n_e(cm^{-3}) \geq 1.4\times 10^{16}T^{1/2}(\Delta E)^3
\end{equation}
where $T$ and $\Delta E$ (the energy difference between upper and lower energy levels) are in eV. But it must be pointed out that the McWhirter criterion, which is derived for homogeneous and optically thin plasma, may be a necessary condition but not a sufficient condition to ensure LTE for a transient and inhomogeneous plasma like an LPP.  In other words, the minimum electron requirement according to the McWhirter criterion warrants that the LTE conditions may exist in the plasma, but not with certainty.  According to \citet{Cristoforetti2010}, the mere use of the McWhirter criterion alone to assess the existence of the LTE in laser-induced plasmas should not be encouraged. Considering the existence of large gradients in fundamental parameters of the LPP with space and time, a significant deviation from LTE can be expected if the measurements were performed in a spatially and temporally integrated manner. Instead, the assumption of partial LTE may be valid for spatially and temporally resolved spectral analysis. 

The LTE is not a valid assumption for plasmas with low densities and/or high temperatures, which are typically called non-equilibrium plasmas or non-LTE plasmas. Examples of non-equilibrium plasmas include inertial and magnetic confinement fusion plasmas, as well as astrophysical and processing plasmas. The collision-radiative (CR) model is typically used to describe non-LTE plasmas, where the local population distribution is described by balancing local collisional processes and non-local radiative processes. The CR model uses multi-level atomic rate equations and radiation transport equations for calculating atomic-level population \cite{Chung2005, Ralchenko2016}.

\section {\label{sec:imaging}Passive and active Imaging tools} 

Passive and active imaging tools are widely used for LPP characterization. Fast gated photography employing ns/ps short gated cameras (ICCD, streak etc.) is a valuable tool studying the hydrodynamic expansion features, plasma morphology and excited species distribution in an LPP system. Similar features can also be recorded using LIF and absorption imaging, however they are recording the lower/ground state species. In this section, experimental details and measurement examples of passive and active imaging tools are given. 

\subsection{Fast photography}
 Time-gated  or  ungated  cameras  are  widely  used  for two-dimensional  (2D)  imaging  and  are  perhaps  one  of the  most  direct  characterization  tools  for  LPPs. The LPP community extensively uses ICCD cameras to capture plasma images in the optical spectral region \cite{Siegel2004, 2003-JAP-Hari}. For shorter wavelengths (X-ray, EUV, and VUV), vacuum-compatible cameras or microchannel plates coupled to a CCD are used in conjunction with a pinhole camera \cite{2001-JPD-Atwee}. In this scenario, the position of the pinhole with respect to the source and camera determines the magnification. Because the dynamics of the LPP change significantly during the earliest times of its expansion, higher magnification is favored for capturing the internal structure and/or instabilities. Moreover, 2D snapshots of the 3D plasma evolution, the plume morphology, and species distribution can be gathered from imaging diagnostics \cite{bai2015morphology}.

Most of the reports in plasma imaging are focused in the visible spectral range because of the availability of large aperture and ICCDs with high quantum efficiencies operating in the VIS spectral region \cite{amoruso2006propagation, gurlui2008experimental, irimiciuc2018influence, min2018investigation}. The ICCD is usually positioned orthogonal to the plume expansion direction. Appropriate notch filters in front of the camera reject scattered laser light and avoid laser-borne damage to the ICCD. By selecting the gate width, delay, and magnification, the spatiotemporal evolution of the supersonic plasma expansion can be recorded with precision. Because of the electronic delay in fast-gated devices, an advanced triggering approach must be adopted to record the early stages of plasma evolution. The entire plume expansion dynamic can be reconstructed by delaying the camera gate with respect to the arrival of the laser pulse; however, ablation events from several consecutive laser pulses are necessary to capture the complete temporal sequence. 

\begin{figure}[t]
\includegraphics[width=\linewidth]{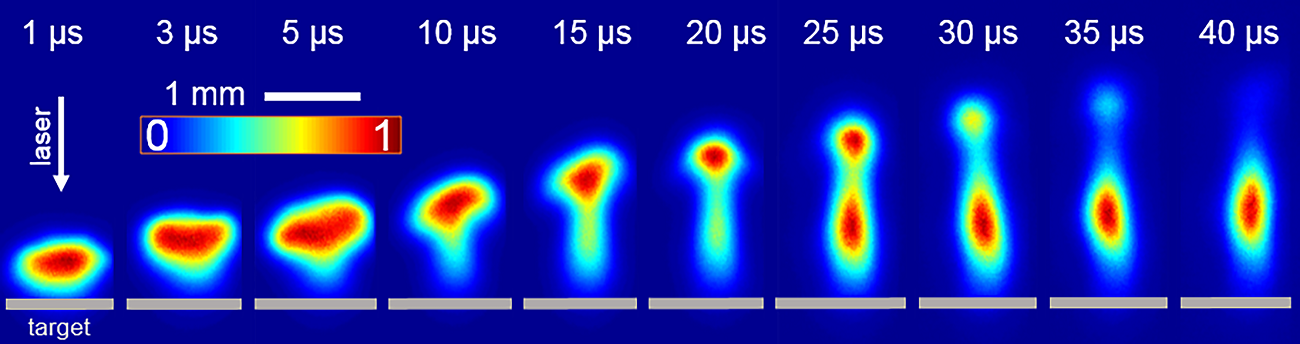}
\caption{\label{fig:31} Time-resolved spectrally integrated  2D images of U plasma in 700 Torr air. Gate delay times are indicated above each plasma image. Each image is from a single laser shot and was normalized to its maximum intensity. The direction of the incident laser is indicated with a white arrow, and the target position is shown with a grey rectangle in each image. From \citealp{2020-PCCP-Liz}. }
\end{figure}

Several authors used spectrally integrated, fast-gated imaging with ICCD for analyzing plume morphology, velocity \cite{lafane2010laser}, species kinetics \cite{2014-JAP-Prasoon}, shock propagation and confinement in a background gas medium \cite{2003-JAP-Hari}, plume splitting \cite{mahmood2010plume}, internal structures, and instabilities \cite{focsa2017plume}. An example of the time-resolved, spectrally integrated (350-800 nm) 2D expansion features of an ultrafast LPP from a U target in air is given in Fig.~\ref{fig:31} \cite{2020-JAAS-Liz}. Here, plume splitting and late time nanoparticle thermal emissions are observed. 2D gated images of expanding LPPs are also used for matrix correction in LIBS \cite{zhang2020plasma} and measurement of residual pressure in sealed containers \cite{yuan2018laser}.

Monochromatic imaging is performed by positioning narrow bandpass filters corresponding to various emission lines in front of the camera, and this method is useful for recording the spatial distribution of species (atoms, ions, molecules, etc.) in the plasma \cite{2016-JAP-Anoop, 2019-JAAS-Liz}. It is appropriate to select an isolated strong-line emission as well as a bandpass filter that transmits only the selected line for avoiding spectral interference. Instead of a narrow band line filter, the use of an acousto-optic tunable filter (ATOF) provides filter transmission tunability \cite{chen2015tracing}. Several groups use monochromatic imaging for the optimization of the growth of films using PLD \cite{bator2013oxidation} to understand the plasma oxidation \cite{2019-JAAS-Liz} and nanoparticle generation \cite{anoop2014spectrally}. \citet{bai2015morphology} studied the emission morphology of various species in the plume as well as background gas excitation using monochromatic imaging and reconstructed the plume emission morphology by combining emission features from various species. Studies showing such a spatiotemporal species’ emission distribution and its dynamics play an essential role in optimizing LPP properties for various applications.  

For imaging the X-ray, soft X-ray, and EUV emission, metallic filters are positioned in front of the pinhole with an appropriate thickness, which limits the observed radiation to the spectral region of interest \cite{2001-JPD-Atwee}. The transmission details of high-energy radiation through various metallic filters are tabulated \cite{henke1993x}.  

Streak cameras, on the other hand, provide simultaneous spatial and temporal features of LPPs and capture the entire spatiotemporal dynamics of LPPs in a single laser shot \cite{rabasovic2014time}. In this device, a streak tube is positioned in front of the image intensifier to sweep the electrons generated on a photocathode radially. Streak cameras provide a high dynamic range, excellent time resolution ranging from several femtoseconds to picoseconds, and ultrahigh sensitivity across wavelengths from X-rays to NIR. By combining the streak camera with a spectrograph, time-resolved, single-shot spectroscopic analysis of the LPP is carried out \cite{rabasovic2019laser}. 

\subsection{LIF and absorption imaging}
Fast-gated imaging employing ICCD records the emission from the excited atoms, ions, and molecules. However, a plasma plume contains both excited- and ground-state species. The hydrodynamics of species in the lower state can be monitored using LIF or absorption imaging. Planar LIF imaging is a well-established method in combustion measurements and analysis of engine gases and steady-state plumes, especially for monitoring various molecular species of gas such as acetone, OH, CH, and NO and their instantaneous distribution and number density \cite{patnaik2017recent}. However, the number of studies on the application of LIF imaging to LPPs to monitor atomic and molecular species in the ground state is limited.  

\begin{figure}[t]
\includegraphics[width=\linewidth]{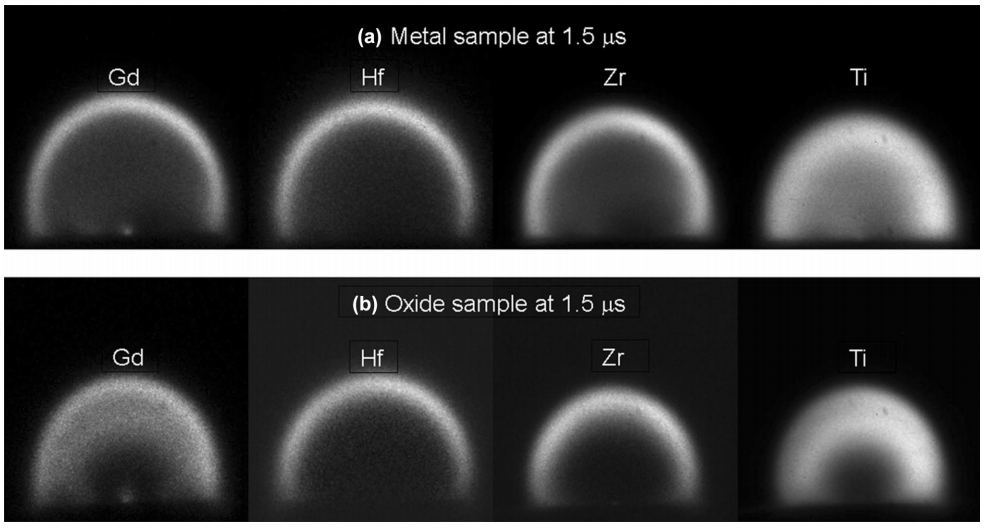}
\caption{\label{fig:32} (a) LIF images for Gd, Hf, Zr, and Ti species observed with a metal and (b) mixed oxide samples at 1.5 µs delay at an ablation energy of 0.5 mJ. The plasmas were produced using 532 nm pulses from a frequency-doubled Nd:YAG laser, and 6 Torr He gas was used as a background medium. From \citealp{miyabe2020development}.}
\end{figure}

For performing LIF imaging, a tunable second laser is used to preferentially excite selected ground-state atomic or molecular transitions. Because LIF emission depends on a selected transition, it inherently provides the distribution of individual species in the plasma, similar to monochromatic self-emission imaging described earlier. Several groups have used LIF for investigating the dynamical behaviour of LPP expansion \cite{nakata1996two, sasaki2002dynamics, miyabe2015ablation}. \citet{miyabe2015ablation} studied the dynamics of an LPP in the presence of ambient gas and noticed that a significant portion of ground-state atoms and ions accumulate in the contact region between the plasma and ambient gas.  Examples of  LIF images obtained from LPPs generated on various metal and metal oxide targets are given in Fig.~\ref{fig:32} \cite{miyabe2020development}. 

Resonant absorption properties of the LPP can also be utilized for species-specific imaging similar to monochromatic self-emission imaging and LIF imaging. Several authors performed absorption imaging of laser ablation plumes using CW probe laser when it is in resonance with a selected atomic or molecular transition \cite{bulleid2013characterization, skoff2011diffusion}.   Gilgenbach \emph{et al.} used a pulsed dye laser for resonance absorption photography (DLRAP) to study the hydrodynamics of excimer laser ablation processing of polymers and metals both in vacuum and background gas environments \cite{gilgenbach1991dynamics, ventzek1992schlieren}. For this, they used a collimated dye laser beam, which was absorbed by atomic or molecular species in the plume and cast a shadow on a photographic film. This method combines the features of absorption spectroscopy and shadowgraphy.

\section {\label{sec:optical probing}Optical Probing} 

Optical probing methods involve the use of an external electromagnetic source,typically a laser or an arc lamp, for measuring the properties of the plasma, and they are routinely used by the LPP community for accessing the density changes in the plasma by measuring the probe beam's phase (interferometry), displacement (shadowgraphy \cite{SettlesBook}), refraction angle (Schlieren imaging \cite{SettlesBook}, Moiré deflectometry, angular refractive refractometer \cite{follett2016plasma}), or measuring the scattered signal from the plasma species \cite{froula2006thomson}. Among these, shadowgraphy and Schlieren imaging are the oldest and remain a traditional method for imaging shock waves in large-scale experiments and are widely used for understanding LPP shock wave generation and propagation into an ambient gas medium, as well as for monitoring internal structures and material ejection. Although both shadowgraphy and Schlieren imaging provide qualitative pictures of plasma density variation and shocks, these techniques are not precise enough to extract plasma density. The interferometry diagnostic tool is one of the most common methods for measuring plasma density \cite{schittenhelm1998two, 2012-MST-Hough, cao2018dynamics}.    However, at higher densities typically seen in HEDP plasmas, the interference fringe become closer and are eventually unresolvable. The angular refractive refractometer is a useful tool for obtaining the complete density profile in long scale-length LPPs where interferometry does not work \cite{follett2016plasma}.

There are three methods involved in the light-scattering techniques, {\it viz.}  Rayleigh, Raman, and Thomson.  Rayleigh scattering is elastic scattering from the particles and molecules and is regularly used by the combustion community \cite{glumac2005temporal}. Raman scattering corresponds to inelastic scattering from molecules. Thomson scattering corresponds to the elastic scattering of optical photons from the free electrons in the plasma. Thomson scattering is one of the most accurate methods for measuring the temperature and density of an LPP with no presumptions about the thermodynamic equilibrium or symmetry. With the first experiments done soon after the invention of the laser \cite{kunze1964measurement}, Thomson scattering (TS) has now developed into an established temporal and spatially resolved measurement technique for electron density, electron temperature, ion temperature, and electron and ion distribution functions of plasmas generated by both low-and high-intensity lasers \cite{glenzer2009x, ross2010thomson, nedanovska2011comparison, Yiming2021TS}.

For all optical probing tools, one of the important considerations is the critical density ($n_c$) of the plasma corresponding to the probe laser beam wavelength, which is given by $    n_c \approx 1.1\times10^{21} \lambda^{-2}$   where $\lambda$   is the probe beam wavelength in µm and $n_c$ is in cm$^{-3}$. When the electron density of the plasma $n_e < n_c$, the plasma is underdense and the probe will propagate through the plasma. However, when $n_e > n_c$, the plasma is overdense and opaque for the probe beam. For example, the critical density of the plasmas for the Nd:YAG laser wavelength and its harmonics (1064 nm, 532 nm, and 266 nm) are $\approx9.7\times10^{20}$ cm$^{-3}$, $\approx3.9\times10^{21} $cm$^{-3}$, and $\approx1.6\times10^{22} $cm$^{-3}$, respectively and therefore represent the upper limits of measurable density if one of these wavelengths are used for optical probing. Shorter probing wavelengths can penetrate higher-density regions of the plasma compared to longer wavelengths. 

In this section, the details of shadowgraphy, Schlieren, interferometry, and Thomson scattering are discussed. 

\subsection{Shadowgraphy}
In the most general terms and in the present context, a shadowgram is the shadow of a plasma  on a photographic screen or on a CCD/CMOS camera. This shadowgram represents the second spatial derivative of the refractive index ($\partial ^2\mu/\partial x^2$), which will reveal the inhomogeneities in the medium of interest in the optical path \cite{SettlesBook}. Because $\partial ^2\mu /\partial x^2$ is much larger than $\partial \mu/\partial x$ in scenarios such as shock waves, turbulence, etc., the shadowgram imaging tool is ideally suited for recording sharp refractive index gradients in the expanding plume boundary. For example, shock waves produce a strong, higher derivative of the refractive index and appear as sharp lines in a shadowgram. 

Light with planar and spherical wavefronts can be a source for generating the shadow. Laser beams traditionally provide extremely high-quality collimated beams and are therefore a  suitable light source for recording a shadowgram. They are also useful for avoiding chromatic aberration effects and provide large photon flux to overcome thermal emission from LPPs. Considering the transient nature of laser ablation plumes, pulsed lasers with shorter pulse widths are preferred illumination sources, especially for capturing the early dynamics of the plume.  

\begin{figure}[t]
\includegraphics[width=\linewidth]{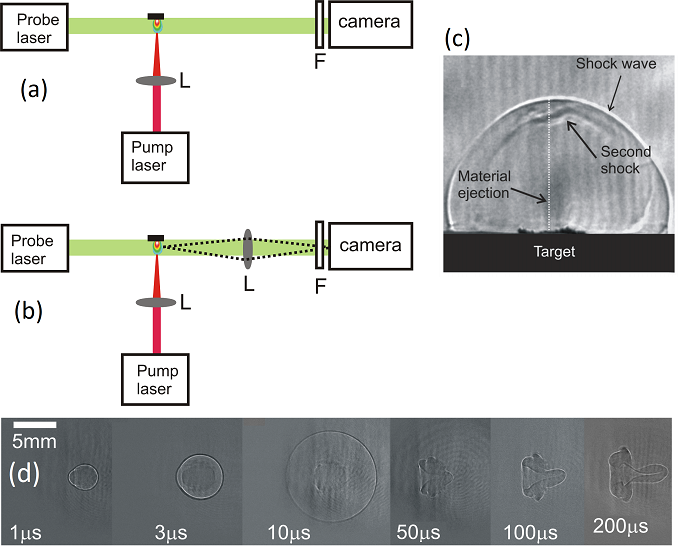}
\caption{\label{fig:8} Experimental schematic for (a) direct shadowgraphy and (b) focused shadowgraphy set up. The pump laser is used to generate plasma, and a transverse probe laser is used for generating shadowgram images. (c) A typical LPP shadowgram image recorded using focused shadowgraphy (ns LPP, Al plasma in air, 150 ns after onset). Both primary and secondary shock waves are clearly visible along with material ejection From \citealp{2012-POP-Hari}. (d) Shadowgrams taken at different times during the evolution of laser-detonated air plasma. From \citealp{2015-POP-Hari}.}
\end{figure}

Two types of shadowgraphy are routinely used for studying LPP shock expansion: \emph{viz.} direct shadowgraphy and focused shadowgraphy. In direct shadowgraphy (Fig.~\ref{fig:8}a), a laser light source is used for casting a shadow of the LPP directly onto the detector. In focused shadowgraphy (Fig.~\ref{fig:8}b), a relay lens is used to cast the shadow of the plume onto the detector. The advantage of focused shadowgraphy is that one can vary the magnification of the image at the detector plane; however, the aperture size of the lens or mirror used for focusing may limit the field of view. Considering small sizes of LPPs in the air at atmospheric pressure ($\sim$ 1-3 mm), the focused shadowgraphy is superior for visualizing the shock waves, the internal structures of the plume, and for capturing turbulent mixing. 

Both continuous wave (CW) or pulsed laser beams can be used for performing shadowgraphy of laser ablation plumes. In the former case, the time resolution is provided by the detector (e.g., ICCD), while in the latter, the pulse width of the laser determine the time resolution with the use of a regular CCD or CMOS camera. The dynamics of shock expansion during LPP evolution are typically captured using shadowgraphy by delaying the probe laser with respect to the plasma production laser. An example of a shadowgram image recorded during LPP expansion from a metal target at 1-atmosphere pressure using focused shadowgraphy is given in Fig.~\ref{fig:8}c, which shows primary and secondary shock waves along with material ejection \cite{2012-POP-Hari}. Fig.~\ref{fig:8}d gives the time sequence of laser-detonated air breakdown, highlighting features such as shock expansion, shock decoupling from the plasma, and turbulent mixing \cite{2015-POP-Hari}. The high repeatability of the LPPs, which enable consistent comparison of shadowgrams from different ablation events, allows movies to be created that show dynamics. The quality of the shadowgram depends strongly on the illumination beam transverse profile, camera specifications (pixel density, dynamic range), and optics. The presence of dust particles on the camera or optics may generate interference fringes in the shadowgram, and the shadowgram image quality can be improved by background subtraction.

There are numerous reports on using shadowgrams for studying the propagation of shock waves during laser-plasma generation in gases, liquids, and solids in a background medium \cite{breitling1999shadowgraphic, sobral2000temporal, schoonderbeek2005shadowgraphic, thiyagarajan2008experimental, gravel2009study, 2014-POP-Alex}. The shadowgram technique is capable of monitoring laser-supported detonation (LSD) waves, primary and secondary shock waves, mass ejecta, and turbulent mixing as shown in Figs.~\ref{fig:8}c and \ref{fig:8}d.  Shadowgraphy is also used for elucidating the role of the cavitation bubble during nanoparticle formation \cite{chen2017probing}. Previous studies employing $\approx$10 ns duration laser pulses  as a probe showed the presence of darkened regions in the shadowgrams without any structures at the earliest times of plasma evolution and described due to high densities of the plasma and/or the expansion of the plasma during the probe laser pulse duration \cite{2015-POP-Hari}. It indicates the necessity of shorter pulse (fs or ps) lasers for tracking early time information of the plasma, which reduces the effects of large-density gradients seen during the early times of LPP evolution \cite{prasad2010chirped}. Shorter wavelength lasers will be useful for negating the deflection effects at the critical density surface. \citet{key1978pulsed} unveiled early time dynamics of high-density LPP using X-ray shadowgraphy. 

The shadowgraphy imaging technique is not typically used for quantification of fundamental plume properties because of the challenges associated with beam diffraction, small angle deflection, and solving the Poisson equation on large data arrays. However, \citet{gopal2007quantitative} demonstrated that the shadowgraphic technique could be used for measuring 2D density profiles of laser breakdown of air by relating the transverse variation of the optical path of the sample to the shadowgram. They compared the measured density values to simultaneously measured density by employing Nomarski interferometry and found that the shadowgraphic technique provided better sensitivity. \citet{kasim2017quantitative} retrieved quantitative information from shadowgrams, based on computational geometry with measurement uncertainties less than 10\%.

Traditionally, the second harmonic radiation  from a Nd:YAG laser ($\lambda$ = 532 nm) is used for probing laser plasmas, and recording is performed using a conventional CCD or CMOS detector. The drawbacks of using such a system are the inability to penetrate at high densities due to plasma critical density at 532 nm and inflexibility of capturing the time sequence of events on a single plasma. Due to critical density effects, the visible lasers are capable of probing only densities much less than the solid density hence limiting penetration and the phenomena that can be measured. So the development of compact and tabletop sources emitting in EUV or soft X-ray spectral region such as high order harmonic radiation, discharge-driven x-ray lasers {\it etc.} could be very impactful for studying high density plasma regime  \cite{hammarsten2004soft, key1978pulsed}.

Several groups recently used high-speed cameras and time-stretch imaging techniques for capturing time sequence of shadowgram images during a single plasma event. For example, laser diode illumination in conjunction with a high-speed camera was used to demonstrate a sequence of shock wave propagation and their interplay with cavitation structures in transparent media \cite{agrevz2020high}. In another study, time-stretch imaging was used to record the time lapse of shock wave propagation during a single-shot LPP, enabling its full dynamics to be monitored \cite{hanzard2018real}. Multiple shadowgram/Schlieren images were captured by splitting the beams into four and probing the plasma at various delays by using a regular CMOS digital single-lens reflex (DSLR) camera as the detector \cite{collins2021direct}. In high-speed laser stroboscopic videography, a high-repetition-rate probe laser in conjunction with a high-speed camera is used for capturing the time sequence of images during a single-shot laser ablation \cite{tanabe2015bubble}.  

\subsection{Schlieren imaging}
The Schlieren method is very closely related to shadowgraphy, with subtle differences. Shadowgram provides a shadow of an object and not a focused image, while Schlieren gives an optical image  that  bears an optical conjugate relationship with an object. Secondly, the Schlieren image requires spatial filtering of refracted light, which is typically done with the use of a knife-edge or a pinhole at the Fourier plane. Thirdly, the Schlieren image displays the deflection angle, while the shadowgram gives the ray displacement resulting from the deflection. Finally, Schlieren images provide the first spatial derivative of the refractive index ($\partial \mu / \partial x $), while the shadowgram corresponds to the second derivative. Therefore, the Schlieren tool is better suited for recording weaker disturbances because of higher sensitivity compared to the shadowgram \cite{SettlesBook, traldi2018schlieren}. Both laser and arc lamps are used for Schlieren illumination \cite{hosokai2006observation, hammarsten2004soft, gottfried2014influence}.  

\begin{figure}[t]
\includegraphics[width=\linewidth]{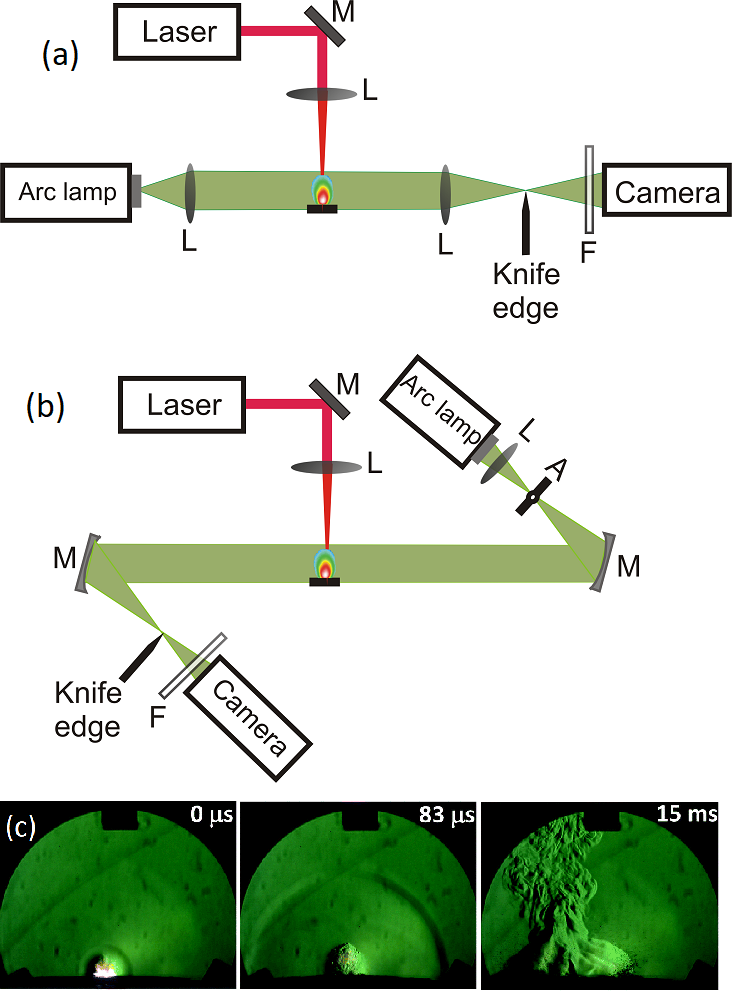}
\caption{\label{fig:9} (a) Schematic of the lens-type Schlieren system with an arc lamp. (b) Folded (Z-shaped) Schlieren setup using mirrors (F, filter; L, lens; A, aperture; M, mirror). (c) Schlieren images of laser-created plasma from RDX. A Nd:YAG laser operating at 1064 nm was used for ablation, and a 200 W HgXe lamp served as the illumination source. . From \citealp{gottfried2014influence}.}
\end{figure}

The Schlieren photography setup is similar to a focused shadowgraphy setup combined with a knife edge for blocking half of the spatial frequencies at the Fourier plane. Lenses or mirrors can be used for setting up Schlieren imaging. Lens-type Schlieren instruments can be set up on a straight line and therefore relatively easy to align (Fig.~\ref{fig:9}a), while mirror-type instruments (z-shaped) are inherently folded (Fig.~\ref{fig:9}b). The LPP is placed in between the lenses or folding mirrors. The positive and negative  µ gradients generated by the plasma refract the probe light rays upwards and downwards, respectively. A sharp opaque object such as a knife-edge is used to block about half of the light beam at the geometrical focus. The type and position of the spatial filter control the intensity distribution of the Schlieren image. The Schlieren image is captured using a CCD/CMOS camera where the features of the plasma are provided as light and dark zones against a uniform background.  For  good-quality Schlieren images, the optics should be free from spherical aberration, chromatic aberration, coma, astigmatism etc. The z-shaped mirror-based systems shown in (Fig.~\ref{fig:9}b) provide a larger field of view and can be aligned free from coma. The use of large f-number ($\geq f/6$) mirrors is recommended for minimizing astigmatism.

Similar to shadowgraphy, Schlieren photography is often used only for understanding the fluid dynamics of the LPP, and it is challenging to gather quantitative information. Several researchers used the Schlieren method to track shock waves and mass ejecta from the laser ablation zone \cite{camacho2002temporally, gottfried2014influence}. \citet{gottfried2014influence} used a mirror-based Schlieren system employing a high-speed camera to study the role of exothermic reactions in an LPP, and  Fig.~\ref{fig:9}c gives the time-snapped images of laser-shocked RDX, highlighting the shock wave expansion and deflagrating particles. The detailed late time features observed attest to the superior sensitivity of the Schlieren technique, which is not possible with simpler shadowgraphic techniques. \citet{vogel2006sensitive} demonstrated a very sensitive white-light Schlieren system that provided visualization of complex ablation plumes with high resolution, a large dynamic range, and color information. This is achieved by a modified Hoffmann modulation contrast technique.

\subsection{Laser Interferometry}
The use of interferometry in science and technology, including analysis of plasmas, became widespread after the invention of lasers. In fact, the early work of the use of laser interferometry for inferring the electron density of different types of plasmas can be traced back to immediately after the advent of lasers in the 1960s \cite{ashby1963measurement}. In interferometry, the electromagnetic waves are superimposed to generate an interferogram, and it is normally performed by amplitude splitting a light source into two beams and recombining after they have traversed different optical paths with path difference that are shorter than the coherence length of the source. The resulting interference pattern provides the phase or optical path differences between the two beams. There are several experimental configurations in interferometry, and they can be broadly classified as double path, common path, and polarized light interferometers \cite{Hariharanbook}. In double path interferometry, the two beams travel in different paths (e.g., Michelson and Mach-Zehnder interferometers); in common-path interferometry, both beams travel in the same path (e.g., Sagnac and Shearing interferometers). In the case of polarized light interferometer (e.g., Nomarski interferometer), a polarizing beam splitter is used.  

\begin{figure}[t]
\includegraphics[width=\linewidth]{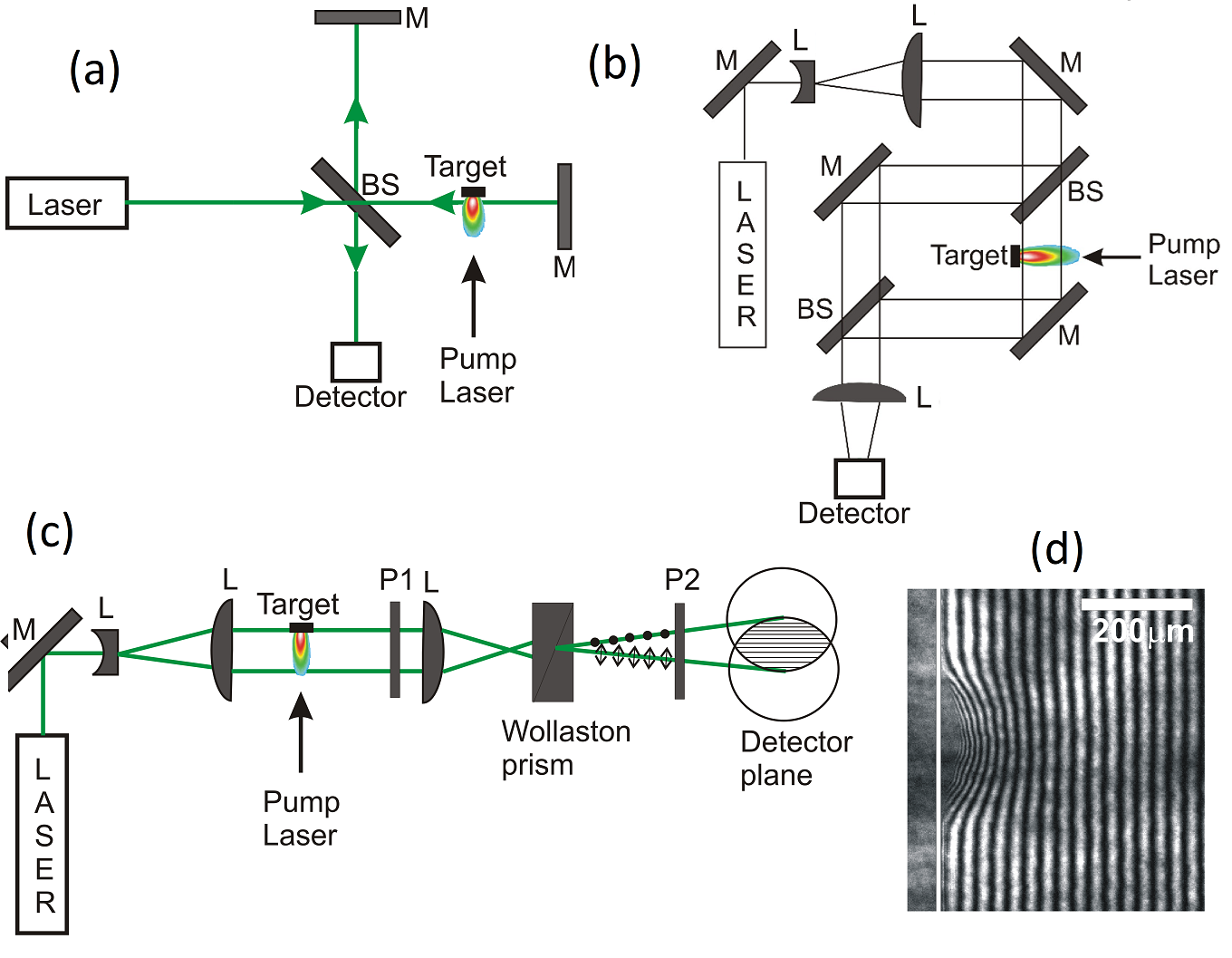}
\caption{\label{fig:10} Various laser interferometry configurations for measuring LPP electron density are given. (a) Michelson interferometer; (b) Mach-Zehnder interferometer; and (c) Nomarski interferometer. An example of fringe shifts due to refractive index variation caused by a ns LPP recorded using a Nomarski interferometer is given in (d). (L - lens; P$_1$, P$_2$ - polarizers; M - mirror; BS – beam sampler)
}
\end{figure}

To measure the physical properties of the LPP using interferometry, the LPP is positioned in one of the arms of the interferometer (Fig.\ref{fig:10}a), and the refractive index changes due to the presence of the plasma are manifested as fringe shifts in the interferogram. The refractive index contribution by free electrons in the plasma is given by    
\begin{equation}
    \label{eq:8}
    \mu = \left(1 - {\omega^2_p}/{\omega^2}\right)^{{1/2}} =  \left(1 - {n_e}/{n_c}\right)^{{1/2}}
\end{equation}
where $ \omega $  is the frequency of the probe beam and $ \omega_p $ is the plasma frequency. Because the critical density defines the change-over from being underdense to overdense for a given plasma, the probe beam wavelength governs the maximum electron density that can be probed using an interferometer.  As Eq.~\ref{eq:8} shows, the plasma index of refraction is proportional to the square root of the density of free electrons in  plasmas, provided contributions from bound electrons are negligible. 

For an underdense plasma, Eq.~\ref{eq:8} is simplified as  $ \mu \approx 1 - {n_e}/{2{n_c}}$. The corresponding phase shift due to the presence of a homogeneous plasma with length L can be written as \cite{da1995electron, Hutchinson2005}  
 \begin{equation}
    \label{eq:9}
   \Delta \phi = \dfrac{2\pi}{\lambda}\left(1- \mu \right)L \approx \dfrac {2\pi L}{\lambda} \dfrac{n_e}{2n_c}  
\end{equation}
where $ \lambda $ is the probe laser wavelength. Therefore, the fringe shifts induced by the presence of plasma in an interferometer arm is given by $N_{fringe} \approx \left({L}/{\lambda}\right) \left({n_e}/{2n_c}\right) $. The LPP is not a homogeneous plasma, and the density changes along the line of sight. The average density along the line of sight is thus given by $ \left<n_e\right> =  \left({1}/{L}\right) \int n_e dl  $.  

  Any interferometric configurations can be used for the mapping electron density of LPP, and the most common configurations are Michelson, Mach-Zehnder, and Nomarski. The schematics of these widely used interferometry configurations for LPP characterization and an example of an interferogram recorded during LPP evolution is given in Fig.~\ref{fig:10}. For any interferogram, the shift in the fringe pattern is measured using a fast photodiode,  photo-multiplier tube, or ideally a CCD/CMOS camera. The photodiodes and PMTs provide 1D measurement of average density along the probe beam path, while CCD/CMOS cameras as detectors are useful for obtaining 2D map of the density. 

In Michelson Interferometer, the probe laser is split into two beams with nearly equal amplitudes using a beam sampler (BS) (Fig.~\ref{fig:10}a). These two beams are reflected back with the help of two mirrors (M) and recombined to form an interference pattern. For measuring the refractive index of the plasma, the target is placed in one of the arms in such a way that the probe beam grazes the sample surface. Several groups used a Michelson interferometer for measuring free electron density in an LPP system \cite{walkup1986studies, 1997-SCAB-Geetha}.  Compared to other interferometry schemes, the probe laser in a Michelson interferometer passes through the plasma system twice, complicating alignment and data analysis.  

A Mach-Zehnder interferometer (Fig.~\ref{fig:10}b) uses well-separated beam paths before interfering, and this therefore simplifies the analysis of the recorded fringes because the beam passes through the plasma only once. This type of interferometer is used extensively by the LPP community for electron density measurements \cite{doyle1999three, mao2000initiation, lemos2013plasma}. A Mach-Zehnder interferometer can be effortlessly transformed into a shadowgraphy experimental set up simply by blocking the reference beam path \cite{mao2000initiation}.   

The Nomarski interferometer is a polarization-based interferometer, and a Wollaston prism is typically used for generating two orthogonally polarized diverging beams \cite{benattar1979polarized, 2007-JAP-Tao, 2009-ASS-Hough, borner2012development}. It is preferable to use a Wollaston prism with a small angle beam separation ($\leq 10 $ mrad) so that the overlapping beam can be easily imaged onto a detector. A schematic of the Nomarski interferometer setup is given in Fig.~\ref{fig:10}c.  A polarizer P1 assures that the polarization of the incoming beam is at 45$^{\circ}$. After the polarizer, a positive lens is used to focus the beam so that the beam goes into the Wollaston prism with a spherical wavefront. As the beam passes through the Wollaston prism, two orthogonally polarized diverging beams are generated with an angular separation. A second polarizer P2 positioned before the detector and orientated orthogonally to P1 assures equal intensity and polarization for both beams arriving at the detector and therefore producing high-visibility fringes. The fringe separation can be easily adjusted by varying the spacing between the imaging lens and Wollaston prism. Because the interferometer assures an equal optical path length between the two beams, it is ideally suited for very short pulse (picoseconds) illumination with its inherently low temporal coherence. The Nomarski interferometer produces two partially overlapped images, and therefore, two images of the targets will be visible in the detector plane.  Shadowgrams of the LPP can also be recorded by controlling the polarization of the beam. The other advantages of using Nomarski interferometry for LPP diagnostics are its relative simplicity, compactness, ease of alignment, and fringe stability. A study comparing the gas density measurements using Mach-Zehnder and Nomarski interferometers showed that the latter provided more accurate results \cite{liu2021application}.

To obtain the density information from the recorded interferograms, certain assumption on the spatial density distribution is to be made.  Since LPPs expands orthogonal to the target surface, typically, the assumption of axial symmetry is considered for fringe analysis. By assuming axial symmetry of the LPP, the Abel inversion technique can be used for obtaining radially dependent density from line-integrated measurement. There are freely available software packages for fringe analysis through Abel inversion (e.g., IDEA - Interferometric Data Evaluation Algorithms) \cite{hipp2004digital}. Algebraic reconstruction and multi-angle tomography techniques are useful for reconstructing  of non-symmetric refractive index fields \cite{Sweeney:73, zhou2019optical}.  

Most of the reported work on LPP electron density measurements using interferometers relied on the approximation that the refractive index in plasmas is contributed solely by free electrons. However, both free electrons and bound electrons in LPPs can contribute fringe shifts, although in opposite directions. So, it is important to select the wavelength of the probe beam such that it is far from any absorption resonances in the plume to avoid contributions to the refractive index from bound electrons. For example, interferometric measurements of Al plasmas using an X-ray laser showed that the bound electrons could have a dominant effect, with the index of refraction greater than one \cite{filevich2005observation, nilsen2005plasma}.  

Considering the transient nature of LPP, high temporal precision is essential, and the time resolution of the interferometry system depends on both the duration of the probe laser pulse and/or gating resolution of the detector. Interferograms are susceptible to fringe blurring if the gradients in electron density are significant during the probe laser pulse. Therefore, using a shorter pulse laser as the probe is preferred to overcome the loss in fringe visibility caused by density variations taking place over the duration of the probe laser pulse. 

The sensitivity of interferometric measurements depends on the configuration and selection of the probe laser wavelength. In addition to these, the beam quality and mechanical stability of the set up may influence the sensitivity of the measurement.  In laser interferometry, the probe laser wavelength sets both the upper and lower limits of electron density measurement in an LPP system. Here the upper limit is governed by the penetration of the beam through the plasma due to critical density effects ($n_c \propto 1/\lambda^2$) and the lower limit (sensitivity)  is dictated by the minimum fringe shift one can measure ($N_{fringe} \propto 1/\lambda$). Hence, for a similar electron density plasma, the fringe shift will be 2$\times$ higher for 1064 nm wavelength probe beam instead of 532 nm.  So the selection of the probe laser is very important for any interferometric analysis of the LPP. Typical phase sensitivity of a two-arm interferometer is $\sim$ 0.1 rad \cite{brandi2019optical} which corresponds to an electron density ~ 7 $\times$ 10$^{17}$ cm$^{-3}$ for a 532 nm probe beam and 100 $\mu m$ plasma. Lasers with shorter wavelengths are also necessary for measuring higher densities because of beam penetration limitation due to critical density \cite{da1995electron}.

Although the upper limit of the measurable electron density with interferometry is the critical density, the refraction and opacity effects may limit the measurement when the plasma density approaches a fraction of the critical density. The change in interferometer contrast due to beam deflection in plasmas with density gradients reduces the fidelity of interferometer measurements. Therefore, information on the effect and extent of refraction is an important prerequisite for accurate analysis of the data \cite{lisitsyn1998effect}. The fringe contrast is also governed by the time resolution of the system as well as probe beam attenuation. For example, at high plasma temperatures and densities, the absorption  through  inverse-Bremsstrahlung  can attenuate the probe beam obscuring part of the interferogram. Shortening the probe laser wavelength will help address these issues. Fringe reconstruction is also a challenge in many cases, especially where the plasma emits significant light that can contaminate the interferogram. So, the intensity of the probe beam should be appropriate to overcome absorption losses and optical noise from intense plasma self-emission. 

The other interferometric configurations of interest that are not discussed here for LPP diagnostics include  folded wave interferometer \cite{martin1980folded}, second harmonic interferometer (also known as dispersion interferometer) \cite{brandi2019optical} and self-mixing interferometry \cite{donadello2020time}. Techniques other than interferometry also can be used to study phase changes in the probe beam when it passes through the LPP.  For example, \citet{plateau2010wavefront} demonstrated that electron density of an LPP can be measured using direct wavefront analysis using a wavefront sensor and it  offers improved phase sensitivity in addition to greater ease of operation in comparison with a folded interferometry setup. Dark field photography is a useful diagnostic of electron density measurement method where visible interferometry does not have the sensitivity \cite{stamper1981dark}. 

\subsection{Thomson scattering}

\begin{figure}[t]
\includegraphics[width=0.45\textwidth]{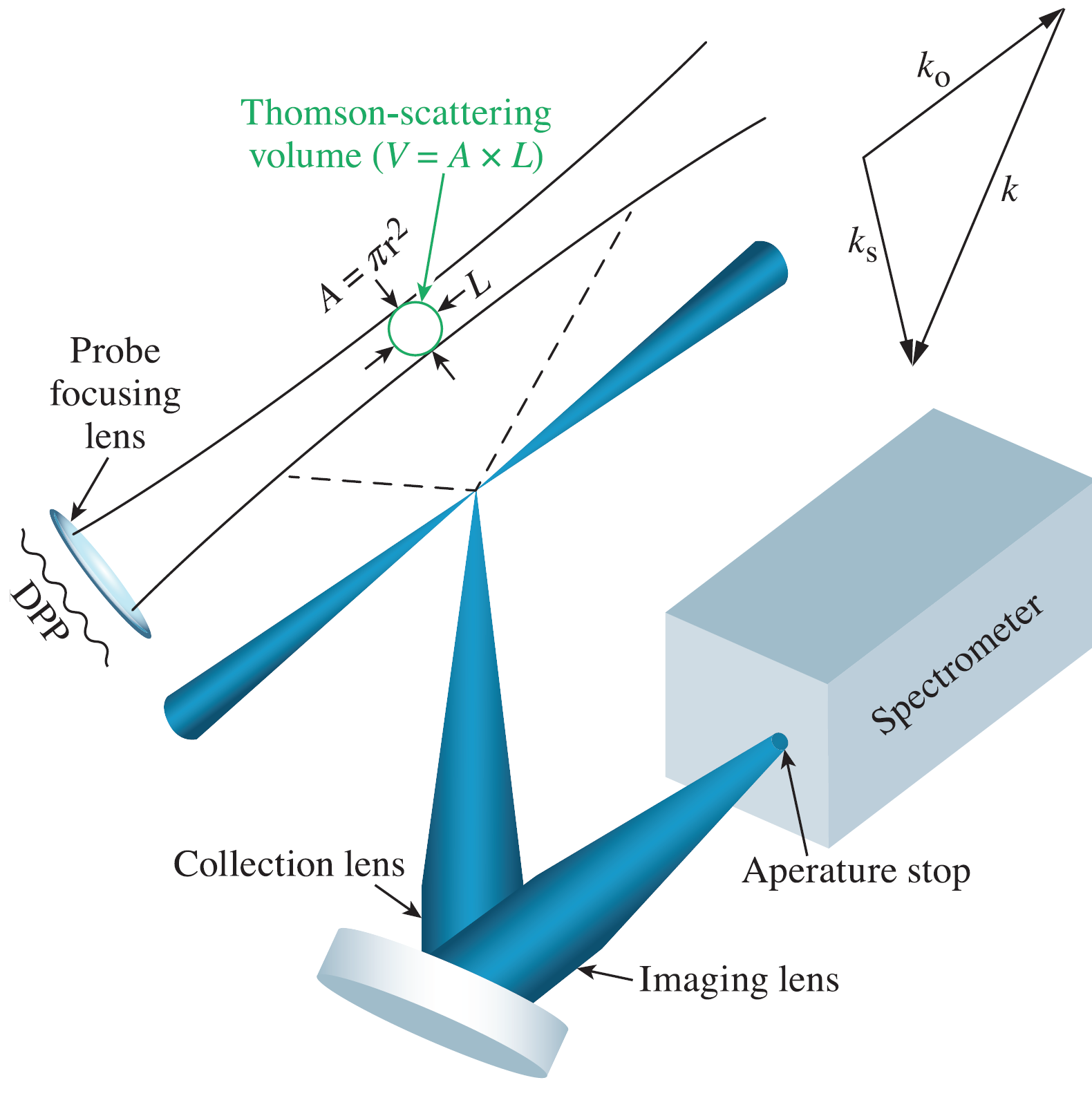}
\caption{\label{fig:3} A typical Thomson-scattering system is shown where the probe laser beam is propagated through a plasma before being focused on an area (A) at the Thomson-scattering volume. An aperture stop is imaged into the plasma to define the Thomson-scattering volume along the propagation of the probe beam (L). (DPP: distributed phase plate).}
\end{figure}

Thomson scattering provides a direct observation of electron motion in a plasma by encoding the electron velocities on the frequency spectrum of the scattered light. By propagating a beam of photons ($\omega_0, \bm{k}_0$) through a plasma and isolating the Thomson-scattering volume collected into a spectrometer (Fig.~\ref{fig:3}), a spatially resolved measurement of the plasma conditions can be determined from the scattered frequency spectrum ($\omega_s, \bm{k}_s$) \cite{Froula2011}. The scattered-power spectrum observed by the detector is given by
\begin{equation}
    \label{eq:1}
    \dfrac{dP_s}{d\omega_s} = \dfrac{P_i r_{0}^{2} L d\Omega}{2\pi}\left(1 + \dfrac{2\omega}{\omega_0}\right)n_eS(k,\omega)
\end{equation}
where $r_0^2=7.95\times10^{-26}$ cm$^2$ is the classical electron radius, $L$ is the length of the scattering volume along the probe beam, $\bm{k = k_s - k_0}$, $\omega = \omega_s - \omega_0$, $d\Omega$ is the solid angle of the collected scattered photons, and $P_i$ is the average incident laser power. The density fluctuations of the plasma around its average density dictates the primary shape of the scattered spectrum through the dynamic structure factor. For a collisionless plasma with no magnetic fields affecting the motion of the particles, the dynamic structure factor is,
\begin{equation}
    \label{eq:2}
    S(\bm{k},\omega)=\dfrac{2\pi}{k} \left|1-\dfrac{\chi_e}{\epsilon} \right|^2 f_e\left( \dfrac{\omega}{k}\right) + \sum_j\dfrac{2\pi}{k}\dfrac{Z_j^2 n_j}{n_e} \left|\dfrac{\chi_e}{\epsilon}\right|^2 f_j\left( \dfrac{\omega}{k}\right)
\end{equation}
where $f_e$ and $f_j$ are the normalized one-dimensional electron and ion velocity distribution functions, respectively, projected along the scattering vector $(\bm{k})$, $Z_j$ is the average charge of the $j^{th}$ ion species, $n_e=\sum_j n_j Z_j$, and $n_j$ is the density of $j^{th}$ ion species. The longitudinal dielectric function is
\begin{equation}
    \label{eq:3}
    \epsilon = 1 + \chi_e + \Sigma_j\chi_j
\end{equation}
where the kinetic plasma susceptibilities are given by
\begin{equation}
    \label{eq:4}
    \chi_e(\bm{k},\omega) = \dfrac{4 \pi e^2 n_e}{m_e k^2} \int_{-\infty}^{\infty} d\bm{v}\dfrac{\bm{k}\cdot\partial f_e/\partial \bm{v}}{\omega - \bm{k}\cdot\bm{v}}
\end{equation}
\begin{equation}
    \label{eq:5}
    \chi_j(\bm{k},\omega) = \dfrac{4 \pi Z_j^2 e^2 n_j}{m_i k^2} \int_{-\infty}^{\infty} d\bm{v}\dfrac{\bm{k}\cdot{\partial f_i/\partial \bm{v}}}{\omega - \bm{k}\cdot\bm{v}}
\end{equation}

\begin{figure}[t]
\includegraphics[width=0.4\textwidth]{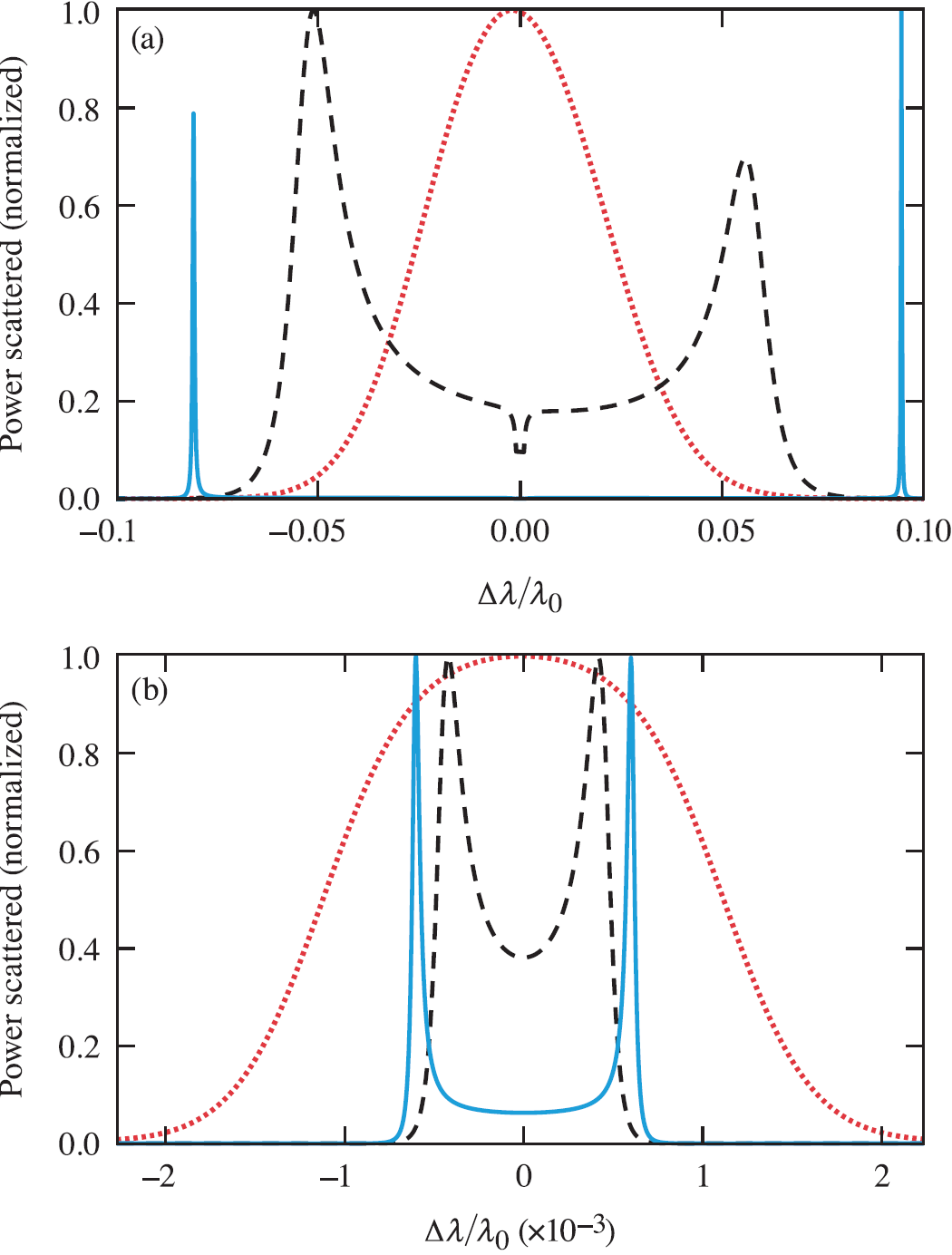}
\caption{\label{fig:4} (a) High-frequency spectrum calculated from Eq.~\ref{eq:1} in the heavily damped noncollective regime, $\alpha = 0.25$ (red dotted curve); mildly damped collective regime, $\alpha = 2.0$ (black dashed curve); and weakly damped collective regime, $\alpha= 4.0$ (blue solid curve). The temperature was maintained at $k_b T_e = 100$ eV, and the density was scaled $n_e = 1 \times 10^{17} \mbox{cm}^{-3}$ (red), $n_e = 6 \times 10^{18} \mbox{cm}^{-3}$ (black), $n_e = 2.5 \times 10^{19} \mbox{cm}^{-3}$ (blue). The low-frequency spectrum has been suppressed. (b) The low-frequency spectrum calculated from Eq.~\ref{eq:1} in the heavily damped noncollective regime, ${ZT_e}/{T_i} = 0.5$ (red dotted curve); mildly damped collective regime ${ZT_e}/{T_i} =3.5$ (black dashed curve); and weakly damped collective regime ${ZT_e}/{T_i} =10$ (blue solid curve). The scattering parameter $\alpha =$ 2 and ${T_e}/{T_i} =0.5$  were held constant. For all calculations, the angle between the incident and scattered light was held constant $(\theta = 90^\circ)$.}
\end{figure}

The scattering spectrum can be used to measure the electron distribution function, which is most evident in the high-frequency non-collective Thomson-scattering regime. Here, the collective motion of the electrons is screened, and the power scattered at a particular frequency is proportional to the number of electrons with a velocity that Doppler shifts the frequency of the probe laser to the measured frequency (Fig.~\ref{fig:4}a). In this regime, where the scattering parameter $\alpha \equiv {1}/{k\lambda_{De}} << 1$ $(\lambda_{De}^2={k_b T_e}/{4\pi e^2 n_e}$ is the electron Debye length), Eq.~(\ref{eq:1}) is reduced to light that is scattered from an ensemble of uncorrelated electrons,
\begin{equation}
    \label{eq:6}
    \dfrac{dP_s}{d\omega_s} = \dfrac{P_ir_0^2Ld\Omega}{2\pi}\left(1 + \dfrac{2\omega}{\omega_0}\right)n_ef_e\left(\dfrac{\omega}{k}\right)
\end{equation}
From here it is evident that the noncollective spectrum provides a direct measurement of the electron distribution function, but in practice, the small scattering cross-section of the electron and small number of electrons at high velocities leads to low SNR, typically limiting this technique to measuring electrons in the bulk of the distribution function.

It is also possible to measure the electron distribution function in the regime where the high-frequency scattering spectrum is governed by the collective electron motion introduced by weaker screening of the density fluctuations \cite{MilderPRL, MilderPOP}. In this collective regime, the thermal particle motion drives a rich spectrum of fluctuations that, when probed, can present themselves in the scattering spectrum as peaks shifted around the incident frequency of the laser (Fig.~\ref{fig:4}). As charged particles propagate through the plasma at velocities greater than the thermal velocity, the surrounding electrons are not able to screen the perturbation, which leaves electrostatic fluctuations in their wake. The amplitude of each fluctuation is determined by the balance of its damping rate by the rate at which it is driven by the plasma particles.

The high-frequency electron plasma wave fluctuations start to play an important role in the scattering spectrum when $\alpha \sim 1$, but when the fluctuations are more weakly damped ($\alpha > 2)$, the resonant features have separated clearly from the noncollective scattering spectrum (Fig.~\ref{fig:4}a). For the low-frequency fluctuations (Fig.~\ref{fig:4}b), there are similar regimes but related to ion motion. The transition between the collective and noncollective regime in a collisionless plasma is governed by the balance between electron screening and ion Landau damping, $\left(\beta \equiv \sqrt{\frac{ZT_e}{T_i(1+k^2\lambda_{de}^2)}}\right)$. When the electrons perfectly screen the ions (e.g., at low electron temperatures or high electron densities), the spectrum represents the ion distribution function ($\beta<<1$). As the electron screening breaks down, damping of the ion perturbations governs the collective motion. Collective low-frequency motion occurs from the inability of the electrons to perfectly screen the ion motion due to the electrons thermal motion ($\beta>1$).

The frequency of these resonant peaks can be approximately determined by solving for the natural modes of the plasma through finding the real roots of the dielectric function Eq.~(\ref{eq:3}), which is where one can see the power of collective Thomson scattering in determining the plasma conditions. Assuming a Maxwellian electron distribution function and weakly damped fluctuations, the dispersion relation for the ion-acoustic waves is evident, $\omega_{iaw} \simeq k\sqrt{\left[\dfrac{(Zk_bT_e+3k_bT_i )}{m_i}\right]}$, in the low-frequency spectrum and the electron plasma wave dispersion, $\omega_{epw}^2 = \omega_{pe}^2+{3k_bT_e k^2}/{m_e}$ , in the high-frequency spectrum, where $\omega_{pe}^2 = {4\pi n_e e^2}/{m_e}$  is the electron plasma frequency. Light that is Thomson scattered from electrons participating in the collective motion and imaged into the detector plane generates constructive interference. The frequency spectrum can be directly related to the plasma conditions through the plasma dispersion relations; note that measuring the difference between frequency of the laser and the peak features in the spectrum $(\omega = \omega_s - \omega_0 = \Delta\omega)$ is a measure of the plasma conditions through the associated dispersion relations $({\Delta\omega}/{\omega_0} \approx {\Delta\lambda}/{\lambda_0})$.

Collective Thomson scattering is a powerful diagnostic regime used to overcome background radiation because of the need to only resolve the frequencies of the spectral peaks. This is in contrast with the noncollective regime, where the shape of the scattering spectrum is used to infer the plasma conditions, therefore challenging one to understand the background radiation spectrum and the wavelength sensitivity of the diagnostic. In practice, modern collective Thomson-scattering systems can resolve the complete spectrum, providing detailed measurements of the electron distribution functions \cite{MilderPOP}, electron temperatures, ion temperatures \cite{glenzer1996observation, froula2002observation}, plasma flow velocities, and electron densities \cite{froula2006thomson, ross2010thomson}.

\subsubsection{Laser beam propagation}
The small electron scattering cross-section is one of the most challenging aspects of Thomson scattering. Integrating Eq.~\ref{eq:1} over frequency provides the total power scattered, $\dfrac{P_s}{P_i} \simeq \dfrac{8\pi}{3}n_e r_0^2 Ld\Omega \sim 10^{-12}$, for typical parameters $(n_e = 10^{19}~\mbox{cm}^{-3}$, L = 50 µm, $d\Omega = 10^{-4})$. To overcome this small cross-section, lasers are used to deliver sufficient power to the Thomson-scattering volume, but the laser power must be balanced against laser–plasma instabilities that can prevent the laser beam from reaching the Thomson-scattering volume. One of the most limiting instabilities is ponderomotively driven self-focusing. For a laser beam with a Gaussian spatial profile, the self-focusing power threshold is $P_c[W] = 1.65 \times 10^{24} \left[{T_e [eV ]}/\left({n_e[\mbox{cm}^{-3}]\lambda_0^2[\mu \mbox{m}}]\right)\right]$, where $\lambda_0$ is the wavelength of the probe laser.

By limiting the power of the laser to the critical power for self-focusing, the maximum power scattered is given by
\begin{equation}
\label{eq:7}
    P_s^{max}(W) = \lambda_0^{-2}[\mu \mbox{m}]T_e[eV]L[cm]d\Omega 
\end{equation}
To demonstrate how restrictive this condition is on the parameter space accessible by Thomson scattering, the SNR can be calculated by assuming Poison statistics, SNR $\approx \sqrt{P_s^{max}\Delta t\dfrac{\lambda_0}{h c}}$ where $h c$ is Planck’s constant times the speed of light \cite{hansen2019mitigation}. For typical conditions ($T_e$ = 100~eV, L = $10^{-2}$ cm, $\Delta t$ = 50~ps, $d\Omega = 10^{-4}, \lambda_0 = 0.5~\mu\mbox{m}) $  spread evenly over 100 resolution units in an ideal system, the SNR $\sim 10$. From here, it is evident that Thomson scattering requires high-electron temperatures, long integration times $(\Delta t)$, large Thomson-scattering volumes along the axis of the probe beam $(L)$, or large solid angle collection optics $(d\Omega)$ to increase the SNR, but each of these parameters has significant constraints within the experimental design. This is a fundamental limitation due to the number of scattered photons, which cannot be overcome by improved diagnostics. 

Intuitively one would expect higher laser powers or higher densities to improve the SNR, but once the laser power has reached the critical power for self-focusing the beam will not propagate well to the Thomson-scattering volume. Increasing the density does not help because the increased signal that results from the higher density is directly compensated by the need to reduce the laser power to remain below the critical power for self-focusing. One way to overcome self-focusing, typically at the cost of increasing the Thomson-scattering volume, is to use a random phase plate \cite{kessler1993phase}. A random phase plate introduces spatial phase modulation across the laser beam prior to the focusing lens. This phase increases the diameter of the laser spot by distributing the laser power into many speckles, which increases the self-focusing threshold by a factor of $\sim$100 \cite{hansen2019mitigation}.

\subsubsection{Thomson scattering from a Maxwellian plasma}
\begin{figure}[t]
\includegraphics[width=0.4\textwidth]{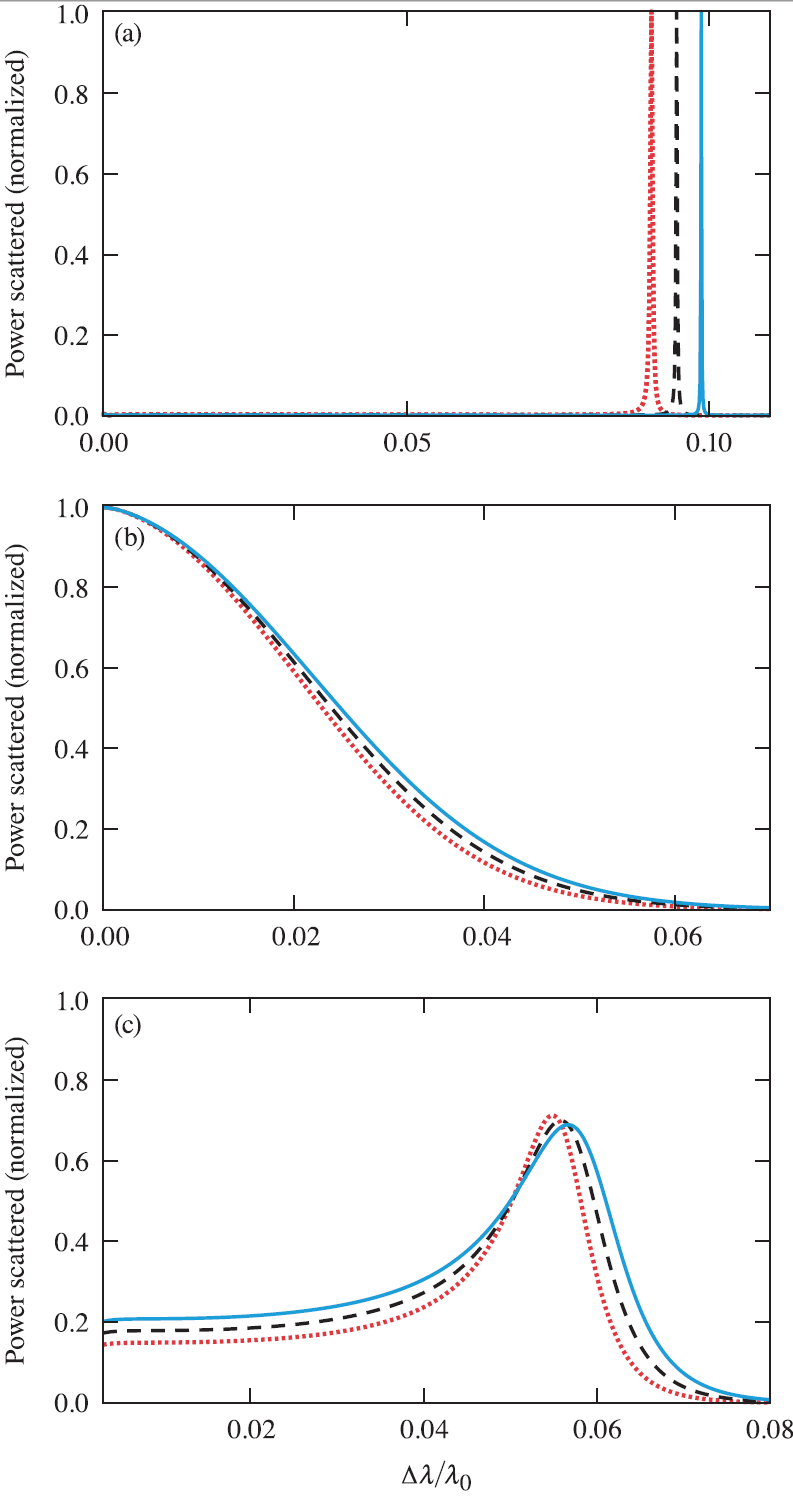}
\caption{\label{fig:5} The sensitivity of the spectrum shown in Fig.~\ref{fig:4} to (a) electron density in the weakly damped regime, (b) electron temperature in the strongly damped regime, and (c) mildly damped regime. The parameters were varied around the central value (black dashed curve) by $+10\%$ (blue solid curve) and $- 10\%$ (red dotted curve).}
\end{figure}

Figure~\ref{fig:4} shows the high-frequency and low-frequency parts of the Thomson-scattering spectrum calculated using Eq.~\ref{eq:1}, assuming Maxwellian ion and electron distribution functions. To measure these spectra, a typical Thomson-scattering instrument uses two spectrometers to independently resolve the high-frequency and low-frequency regimes \cite{ross2011ultraviolet, katz2012reflective}. The high-frequency spectrum requires lower dispersion to spread the $\Delta\lambda/\lambda_0\sim0.1$ spectrum over a detector with approximately 200 resolution units. This can be achieved with a $1/3$-meter spectrometer with a 150 grooves/mm grating. Resolving the low-frequency spectrum requires a high-dispersion system that can resolve the separation between the ion-acoustic peaks $\Delta\lambda/\lambda_0\sim10^{-3}$ over at least 20 resolution units. This can be achieved with a 1-meter spectrometer with a 2,400 grooves/mm grating. Often, the spectrometers are coupled to optical streak cameras to measure the evolution of the plasma conditions. In these systems, the temporal resolution is determined by the pulse-front tilt introduced by the spectrometers, which is typically on the order of 100 ps \cite{visco2008temporal}. By trading unrealized spectral resolution for improved time resolution, the temporal resolution can be optimized to the Heisenberg limit \cite{katz2016pulse, davies2019picosecond}.
\paragraph {High-frequency fluctuations – electron plasma waves}
Figure~\ref{fig:5} shows the sensitivity of the high-frequency spectrum to the plasma conditions in three different scattering regimes. In the weakly damped regime, the scattering features are very narrow, and the sensitivity of the frequency of their peaks provides an accurate measure of the electron density. In this regime, the width of these features is typically dominated by instrument broadening and density gradients within the Thomson-scattering volume \cite{follett2016plasma}. Reducing the scattering parameter such that the waves are heavily damped allows their width to be increased significantly beyond typical broadening due to gradients, and the shape becomes an accurate measurement of the electron temperature while the peak location remains a measure of the electron density. Further reducing the scattering parameter results in a regime where the electron perturbations are screened by the faster moving electrons and  a noncollective spectrum is evident in the scattering spectrum, which represents the shape of the electron distribution function.
\begin{figure}[t]
\includegraphics[width=0.4\textwidth]{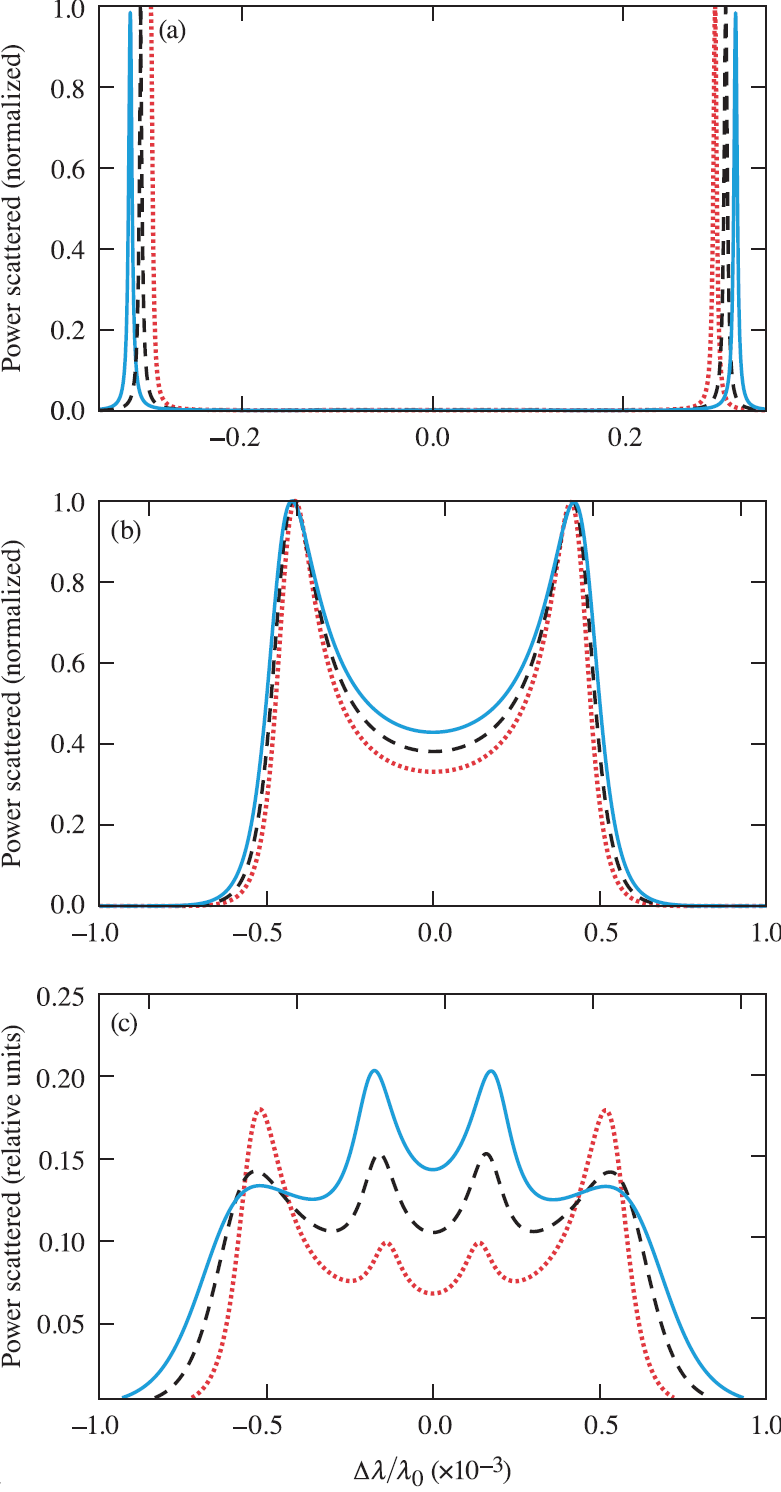}
\caption{\label{fig:6} Low-frequency spectrum (a) for a single-species nitrogen plasma where $Z k_b T_e$ = 630 eV (red dotted curve), $Z k_b T_e$ = 700 eV (black dahsed curve), and $Z k_b T_e$ = 770 eV (blue solid curve), where $T_i$ = 20~eV. (b) In the mildly damped regime, the width of the ion feature can be used to measure the ion temperature; $k_b T_i$ = 18 eV (red dotted curve), $k_b T_i$ = 20 eV (black dashed curve), $k_b T_i$ = 22 eV (blue solid), and $ Zk_b T_e$ = 700 eV. (c) Introducing $5\%$ nitrogen (Z = 7) to a hydrogen (Z = 1) plasma provides two low-frequency modes, and their relative amplitudes provide an accurate measure of the ion temperature,  ${T_e}/{T_i}$ = 5  (red dotted curve), ${T_e}/{T_i}$ = 3.3 (black dashed curve), and ${T_e}/{T_i}$ = 2.5 (blue solid curve); $T_e$ = 100 eV was held constant. For all calculations, $\alpha$ = 2.}
\end{figure}
\paragraph{Low-frequency fluctuations – ion-acoustic waves}
Figure~\ref{fig:6}a shows the sensitivity of the low-frequency spectrum in the collective regime to the product $ZT_e$. In this weakly damped regime, the scattering features are very narrow, and the sensitivity of their peak location in frequency provides an accurate measure of $ZT_e$, provided $ZT_e \gg 3T_i$. When this condition is not met, it is convenient to work in the mildly collective regime where the shape of the ion-acoustic peaks can be resolved, providing a measure of the ion temperature (Fig.~\ref{fig:6}b). Another technique that is often used to measure the ion temperature in low-Z plasmas is to introduce a small fraction of higher-Z atoms \cite{glenzer1996observation, froula2002observation}. When the ratio of atomic number to the average ionization ($A/Z$) is sufficiently different between the two species, additional low-frequency modes are resolvable in the scattering spectrum (Fig.~\ref{fig:6}c) \cite{williams1995frequency}. From the relative amplitudes of these two modes, an accurate measure of the ion temperature can be obtained \cite{froula2006thomson, Froula2006POP}.

\begin{figure}[t]
\includegraphics[width=\linewidth]{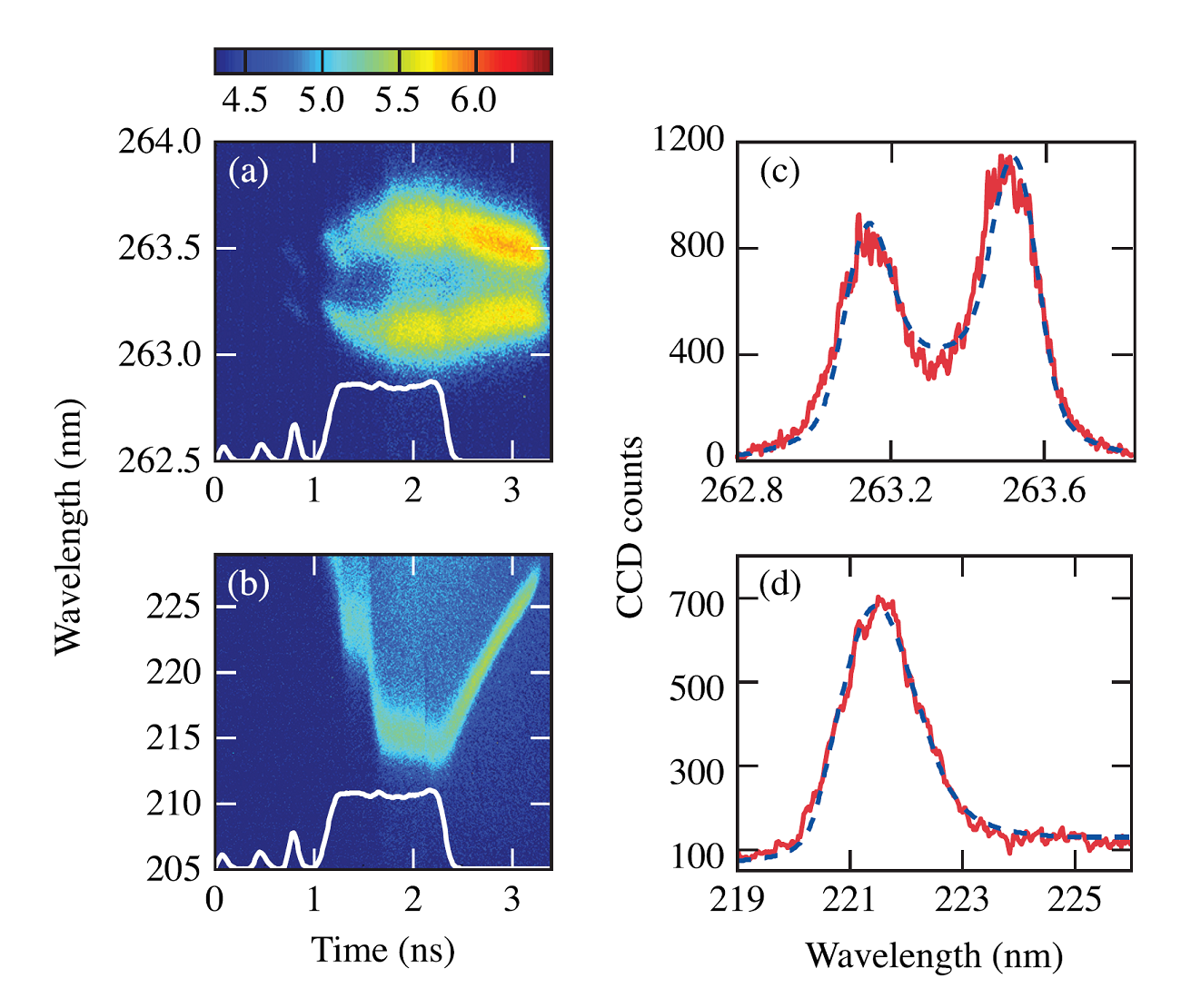}
\caption{\label{fig:7} Collective Thomson-scattering spectrum simultaneously recorded to reveal (a) the low-frequency ion-acoustic wave spectrum (${ZT_e}/{T_i} \sim$ 3), and (b) the high-frequency $(\alpha \sim$ 2.3) electron plasma wave spectrum (blue shifted peak only). The drive-laser pulse shape is overlayed. The Thomson-scattering spectrum at 2.8 ns for the (c) ion-acoustic waves and (d) electron plasma waves. The solid red curves are the measured spectra, and the dashed blue curves are the best-fit spectrum ($k_bT_e$=0.9 KeV, $k_b T_i$=0.8 KeV, $n_e = 4.4 \times 10^{20}~\mbox{cm}^{-3}$).}
\end{figure}

Figure~\ref{fig:7} shows an example of a Thomson-scattering spectrum measured from a multi-species CH plasma where both the electron-plasma and ion-acoustic features were resolved \cite{follett2016plasma}. The low-frequency spectrum shows the ion-acoustic wave features separating in frequency as the plasma heats (0.5–2 ns) and then coming back together, indicating cooling ($>$2.5 ns) after the heating beams turn off. The high-frequency spectrum shows the blue-shifted electron plasma feature, the increasing density in the Thomson-scattering volume as the plasma is formed at early times ($<$ 1.5 ns), and the relatively constant density (1.5 ns-2.5 ns) before decompressing once the drive lasers turn off ($>$ 2.5 ns). 

The ultraviolet Thomson-scattering probe beam ($\lambda_0$ = 263.25 nm) had a best-focus diameter of $\sim$ 70 µm at the scattering volume \cite{mackinnon2004implementation}. The scattered light was collected from a 50~µm $ \times$ 50 µm $ \times$ 70 µm volume located 400 µm from the initial target surface in the coronal plasma surrounding a direct-drive fusion capsule driven by 60 351-nm beams at the OMEGA laser \cite{boehly1995upgrade}. The geometry was configured to probe wave vectors perpendicular to the target normal. The angle between the probe beam and the collection optic was $120^\circ$. The spectral resolutions of the ion-acoustic wave and electron-plasma wave systems were 0.05 nm and 0.5 nm, respectively. 

In summary, Thomson scattering provides a window into the motion of the electrons by encoding their velocity onto the scattered spectrum and measuring this spectrum is a powerful way to determine the spatially- and temporally-resolved plasma conditions. Measurements in the noncollective scattering regime show spectrum that directly represent the electron or ion velocity distribution functions and are used to measure the electron or ion temperatures. Measurements in the collective regime allow the frequencies of the resonant plasma waves to be measured allowing the electron temperature, ion temperature, and electron density to be determined.

\section {\label{sec:Other-methods}Summary of other optical methods}
In addition to regularly used optical diagnostic tools discussed in the previous sections, there are other useful techniques for measuring physical properties of an LPP.  A brief account to Moiré deflectometry and velocimetry is given in this section. Other techniques of interest that are not discussed here are optical polarimetry \cite{Davies2014}, angular refractive refractometer \cite{follett2016plasma}, THz spectroscopy \cite{Herzer2018, Jamison2003}, Zeeman splitting \cite{Zeeman1984}, dark-field photography \cite{stamper1981dark}, and direct wavefront analysis \cite{plateau2010wavefront}.     

\subsection{Moiré deflectometry}
\begin{figure}[t]
\includegraphics[width=0.85\linewidth]{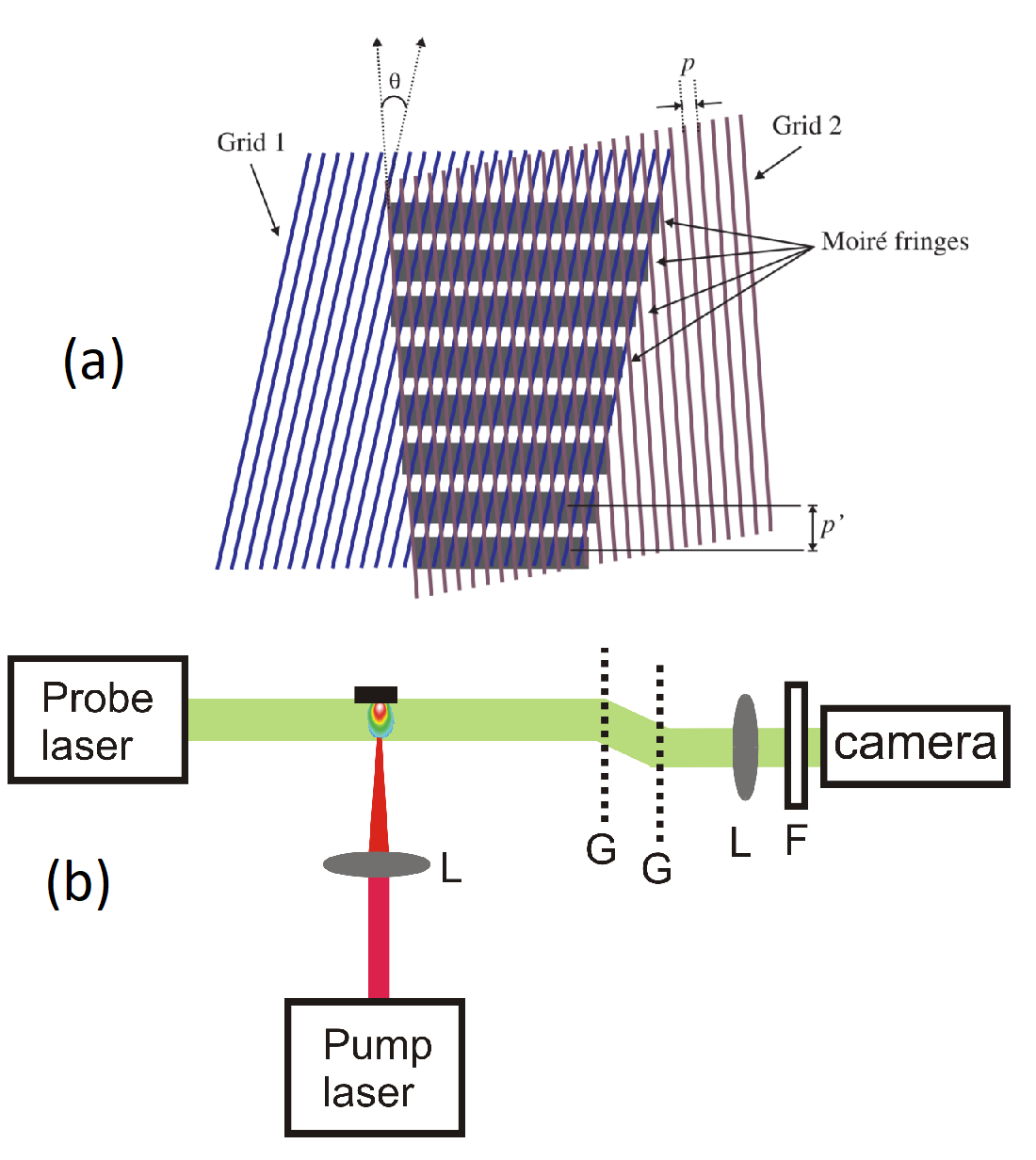}
\caption{\label{fig:11} Typical Moiré fringes produced by overlaying two rotationally offset Ronchi gratings. Adapted from \citealp{ruiz2007comparison}. (b) Experimental arrangement for measuring electron density of an LPP. (G – Ronchi grating, L – lens; F - filter).
}
\end{figure}

Moire deflectometry is a modified version of the Schlieren imaging technique; however, it provides quantitative information about the plasma electron density. Moiré deflectometry uses the deflection of a collimated beam as it passes through the plasma medium where the deflection is proportional to transverse gradients in the object’s refractive index \cite{kafri1980noncoherent}. A pair of Ronchi gratings is used in the Moiré deflectometry setup, and the Moiré pattern corresponds to a series of straight parallel equidistant fringes separated by $p'= p/ {\theta}$, where $p$ is the ruling pitch and $\theta$ is the angular separation between two gratings. Fig.\ref{fig:11} shows the Moiré pattern produced by two identical Ronchi gratings along with a typical Moiré deflectometry arrangement for the LPP density measurement. The Ronchi rulings are  transverse to the light path, parallel, and separated by a distance, $D = Nd$, where $N$ is an integer value and $d$ is called the Talbot spacing \cite{ruiz2007comparison}. When the probe beam passes through the plasma, it deflects, consequently distorting the Moiré fringe pattern. The resulting Moiré deflectogram is recorded using a 2D array detector placed behind the second Ronchi grating. A high-contrast fringe pattern is possible only for small offset angles $\theta$ and at limited distances between the rulings, usually of a few Talbot spacings.   The quality of the fringe pattern is influenced by the spectral width, divergence, and the diameter (with respect to grating pitch) of the probe beam, as well as the quality of the gratings and grating inter-distance (Talbot order). The relation between the angular refraction ($ \alpha $) of the probe beam  and the plasma electron density is given by \cite{Pia-RSI2018}
\begin{equation}
    { \alpha (x \cdot y) =  \frac{\lambda^2 r_0}{2\pi} \dfrac{\partial}{\partial x}\int n_e (x,y,z) dz}   
\end{equation}
where r$_0$ is the classical electron radius. 

Published work on Moiré deflectometry for LPP electron density is, however, limited  \cite{zakharenkov1990laser, decker1998electron}. A soft X-ray Moiré deflectometer was used for the measurement of high-density LPPs \cite{ress1994measurement}. Talbot-Lau based Moiré deflectometry with X-ray backlighter was used for measuring electron density of high energy density plasmas \cite{Pia-JAP2013}. 

\citet{ruiz2007comparison} compared the sensitivities of Moiré deflectometry and Nomarski interferometry by measuring the electron densities of transient z-pinch plasmas and found that Nomarski interferometry was more suitable for plasmas with low-density gradients, while Moiré deflectometry provided more accurate results for measuring plasmas with large density gradients. They also found that Moiré deflectometry provides higher electron density sensitivity ($\sim 10^{16}$ cm$^{-3}$) and hence it is better suited for studying low-density coronal plasma compared to interferometry. 

\subsection{Optical velocimetry}
When an intense laser interacts with a target, the generated plasma expands while initiating a shock wave that travels toward the target that compresses and heats the target material. Precision velocity measurements are important in the study of shocks with impulse excitation and associated phase transitions. Doppler velocimetry methods, such as a Velocity Interferometer System for Any Reflector (VISAR) \cite{miller2007streaked, batani2016diagnostics} and Photonic Doppler velocimetry (PDV) \cite{dolan2010accuracy, dolan2020extreme} are useful interferometric tools for measuring such compression events. VISAR was developed in the early 1970s \cite{barker1972laser} for measuring shock dynamics; however, in recent times, it has been largely replaced by PDV due to the simplicity of alignment and versatility \cite{dolan2010accuracy}.

  In both VISAR and PDV, a Michelson interferometer is used with the target rear surface as one of the end mirrors. A wide-angle Michelson laser interferometer constructed out of a single mode laser with good temporal coherence is the main component of a VISAR. The reflected beam from the target rear is equally split into two in a free-space Michelson interferometer, and one of the arms is optically delayed using one or more etalon(s) to make the reference beam. The Doppler-shifted light interferes with the unshifted reference beam to produce a beat wave, the frequency of which is proportional to the instantaneous velocity of the shocks. Temporal resolution is achieved with a streak camera or a gated intensified CCD. The system can be adapted to targets of a highly or diffusely reflective back surface.

Recently, PDV is used more widely than VISAR. In PDV, interference occurs between light coming from the target and a separate reference beam as opposed to VISAR, where interference patterns are due to the overlapping of light from the target rear and a delayed replica of the same beam. PDV uses narrow linewidth  fiber lasers as the source in conjunction with a fiber optic Michelson interferometer. The fiber laser output is split into two, where one of the beams is used for monitoring the target movement and the other is used as a reference. Optical interference generated by combining these two beams is used for measuring the target displacement, and velocity information is gathered through time-frequency analysis. 	 

\begin{figure}[t]
\includegraphics[width=\linewidth]{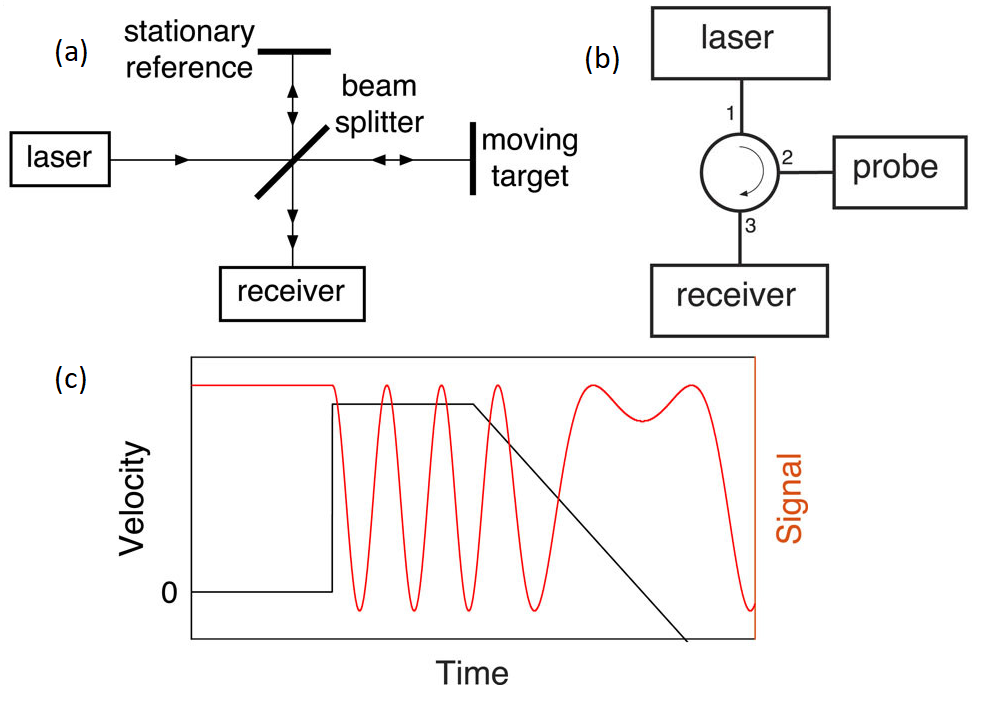}
\caption{\label{fig:12} (a) Michelson interferometer as a free-space equivalent to PDV, (b) scheme for conventional PDV system, and (c) typical PDV signal for time-dependent velocity. From \citealp{dolan2020extreme}.
}
\end{figure}

Schematics of the conventional PDV configuration, its free space equivalent, and a characteristic PDV signal for a dynamic velocity are given in Fig.~\ref{fig:12} \cite{dolan2020extreme}. The conventional PDV corresponds to a fiber optic Michelson interferometer with the rear of the target as one of the two mirrors (Fig.~\ref{fig:12}a). Free space equivalent of the fiber optic interferometer is shown in  Fig.~\ref{fig:12}b, where the laser light entering port 1 of a fiber circulator emerges from port 2 and travels to the rear of the target  and collects the Doppler-shifted reflection from the target. Light reflected from the target interferes with the retro-reflected light and outputs at port 3, and a receiver detects light amplitude. The receiver signal is constant when the target is at rest. For constant velocities of the target, the receiver detects a time-varying signal oscillating at the beat frequency given by $v_B = (2⁄\lambda)|v|$, where $v$ is target velocity and $\lambda$ is the wavelength of the unshifted probe. When the velocity is time dependent, the signal shape deviates from sinusoidal as shown in Fig.~\ref{fig:12}c at later times. The dynamic velocity of the target is extracted through time-frequency analysis. 

 The recent technological advances of fiber lasers transformed the capability of PDV to a useful shock-physics diagnostic. PDVs are capable of tracking of velocities from 0.01 mm/s to $\geq$ 10 km/s and is widely used in single-event measurements \cite{dolan2020extreme}. Most of PDV’s remaining challenges are in analysis/interpretation (multiple/overlapping frequencies, extreme accelerations, etc.)  and hence considerable development is required for truly robust analysis. A comprehensive description and analysis of all existing PDV variants is given in a recent review describing the merits, shortcomings, system requirements, and measured/expected signal shapes of five variants of conventional PDV \cite{dolan2020extreme}. Using simulation, \citet{chu2021time} proposed a time-lens PDV system for expanding the dynamic range of PDV and allowing the use of lower bandwidth electronics.  

\section {\label{sec:summary}Summary}
\begin{table*}[t]
\caption{\label{table:1} The measurable parameters using various optical diagnostic tools. Comments given highlight the pros and cons of each method.}
\begin{ruledtabular}
\begin{tabular}{|c|c|l|}
\hline
\textbf{Diagnostic}                                                            & \textbf{Measured LPP Parameters}                                                                                                                      & \multicolumn{1}{c|}{\textbf{Comments}}                                                                                                                                                                                                                                                                                                                                                                                                                                 \\ \hline
\begin{tabular}[c]{@{}c@{}}Optical emission \\ spectroscopy \end{tabular} & \begin{tabular}[c]{@{}c@{}}$T_e$, $T_{exc}$, $T_g(T_{vib}~ and ~T_{Rot})$,\\  $n_e$, $n_{atoms}$, kinetics of emitting \\ species\end{tabular} & \begin{tabular}[c]{@{}l@{}}Passive, direct/indirect, easy to perform, broadband \\ spectroscopy, complete or partial LTE is essential for plasma \\ characterization except for n$_e$  measurement  through Stark \\ broadening, plasma should be  optically thin, the spectral \\ resolution is constrained by instrumental resolution, Stark \\ studies require high spectral resolution, spatially and temporally \\ resolved studies are preferred. Line of sight averaging is an \\ issue, and Abel inversion is useful for obtaining accurate results.\end{tabular}                                                                                                          \\ \hline
Absorption spectroscopy                                                        & \begin{tabular}[c]{@{}c@{}}$T_{exe}$, $T_k$, Level populations, \\ $n_e$, $n_{atoms}$, $n_{ions}$, $n_{molecules}$\end{tabular}                       & \begin{tabular}[c]{@{}l@{}}Active method, direct/indirect, reduced pressure conditions\\ are preferred, monitors lower level population and is therefore \\ ideal for low-temperature plasma measurements, LTE assumption \\ required for plasma characterization, absolute number density \\ measurement possible, broadband  and narrow band sources \\ can be used, spectral resolution is constrained by detector for \\ broadband light sources, tunable laser sources provide high \\ spectral resolution although with limited bandwidth, \\ high-temporal resolution.\end{tabular} \\ \hline
\begin{tabular}[c]{@{}c@{}} LIF\end{tabular}           & $T_k$, $T_{exe}$, $n_e$, $n_{atoms}$, $n_{ions}$                                                                                                      & \begin{tabular}[c]{@{}l@{}}Active method, indirect, narrow band, high spectral \\ resolution, direct measurement of plasma parameters requires \\ calibration for absolute measurement, scattering issues,\\ optical depth and saturation considerations\end{tabular}                                                                                                                                                                                                                                                                                                                                                                                                          \\ \hline
Emission imaging                                                               & \begin{tabular}[c]{@{}c@{}}Morphology, specie \\ distribution, plume velocity, \\ instabilities\end{tabular}                                          & \begin{tabular}[c]{@{}l@{}}Passive method, direct, easy to perform, gated cameras with \\ high time resolution yield best results, narrowband filters and \\ ATOF provide various specie distributions.\end{tabular}                                                                                                                                                                                                                                                                                                                                                                                                             \\ \hline
\begin{tabular}[c]{@{}c@{}}LIF/absorption \\ imaging\end{tabular}              & \begin{tabular}[c]{@{}c@{}}Distribution of lower-state \\ population\end{tabular}                                                                     & \begin{tabular}[c]{@{}l@{}}Active method, direct, needs a probe laser whose wavelength \\ should be in resonance with a selected transition, time \\ resolution is provided by probe laser and/or detection scheme.\end{tabular}                                                                                                                                                                                                                                                                                                                                                                                                                                                                                                                     \\ \hline

Shadowgraphy                                                                   & \begin{tabular}[c]{@{}c@{}}Properties of the shocks \\ (velocity, temperature)\end{tabular}                                                           & \begin{tabular}[c]{@{}l@{}}Active and direct technique, different experimental \\ configurations, shorter pulse laser probing is preferred, \\ difficult to extract quantitative information. \end{tabular}                                                                                                                                                                                                                                                                                                                                                                                     \\ \hline
\begin{tabular}[c]{@{}c@{}}Schlieren \\ photography\end{tabular}               & \begin{tabular}[c]{@{}c@{}}Properties of the shocks, tracking \\ slow-moving particles\end{tabular}                                                   & \begin{tabular}[c]{@{}l@{}}Active and direct technique, broadband sources are preferred, \\ different experimental schemes, superior sensitivity compared \\ to shadowgraphy, provides only qualitative information.\end{tabular}                                                                                                                                                                                                                                                                                                                                                                                              \\ \hline
Interferometry                                                                 & $n_e$, $n_{atoms}$                                                                                                                                    & \begin{tabular}[c]{@{}l@{}}Active and direct method, many experimental configurations \\ available. Shorter pulsed and shorter wavelength laser \\ probing is suggested to overcome LPP gradients and critical \\ density issues, sensitivity depends on probe laser wavelength,\\ Abel's Inversion is necessary.\end{tabular}     

\\ \hline                                                                                    

\begin{tabular}[c]{@{}c@{}}Thomson \\ scattering\end{tabular}                 & \begin{tabular}[c]{@{}c@{}}  $T_e$, $T_{ion}$, $n_e$, ion and electron \\ distribution functions\end{tabular}                                           & \begin{tabular}[c]{@{}l@{}}Active method, highly accurate, spatially and temporally \\ resolved measurements, No LTE assumption required, \\complex experimental scheme,  high-power probe laser is \\ required due to low electron scattering cross-section.\end{tabular}                                                                                                                                                                                                                                                                                                                                                                                                                                                                                                                                                              \\ \hline
\begin{tabular}[c]{@{}c@{}}Moiré \\ deflectometry\end{tabular}                 & $n_e$                                                                                                                                                 & \begin{tabular}[c]{@{}l@{}}Active and direct method, suitable for measuring \\ plasmas with large density gradients, and low-density \\ coronal plasma.\end{tabular}                                                                                                                                                                                                                                                                                                                                                                                                                           \\ \hline
\begin{tabular}[c]{@{}c@{}}Velocimetry\\ (VISAR and PDV)\end{tabular}          & Shock compression                                                                                                                                     & \begin{tabular}[c]{@{}l@{}}Active and direct, interferomery-based technique, VISAR \\ setup and alignment is complicated, PDV is simpler to align \\ and easy to use.\end{tabular}                                                                                                                                                                                                                                                                                                                                                                                                                             \\ \hline
\end{tabular}
\end{ruledtabular}
\end{table*}The present review provides an overview of various optical plasma diagnostic tools that can be used to characterize laser produced plasmas, highlighting capabilities, limitations, and other experimental challenges of each method. The primary objective is to acquaint the reader with opportunities in the optical diagnostics of laser produced plasmas. The LPP is a very complex system whose properties are changing with space and time, and it is important to select the proper tool that is capable of delivering the physical property of the plasma one is aiming for. It is also very clear from the discussion that many tools can be used simultaneously for obtaining similar information; however, the accuracy of the measurement depends heavily on the associated assumptions and spatially and temporally weighted averaging. Table 1 summarizes the capabilities of each diagnostic tool, its measurement nature (direct or indirect/passive or active), associated assumptions, etc.

Among the many optical diagnostic tools, the emission spectroscopic tools are widely used because of their simplicity and cost-effective instrumentation. Although emission spectroscopy provides reasonable accuracy in the measurement of fundamental parameters, it is not useful for measuring the properties of the LPP at very early times and/or late times of its evolution. Optical probing methods such as Thomson scattering and interferometry give more accurate results at early times of plasma evolution, while LAS is better suited for late time characterization. Absorption spectroscopy employing tunable IR lasers (e.g., quantum cascade lasers) may be useful for measuring properties such as molecular density at very low temperatures.  

As Table~\ref{table:1} shows, each diagnostic tool has its pros and cons and should be considered as complementary. Each technique is useful for measuring certain parameters, but its use is limited to a certain time window during the LPP evolution due to the sensitivity issues of the selected measuring tool. Hence, multiple diagnostic tools are essential for a comprehensive insight into the entire plasma behavior. Many expanded capabilities of optical diagnostic tools in recent times are related to the improvements in laser technology and detector systems. For example, the availability of short-pulse lasers, as well as high-speed and ultrashort gating times for array detectors, provide measurements with higher temporal resolution which is extremely valuable for transient laser-plasma systems. Stable, user-friendly, and narrow linewidth tunable lasers are important for extending the use of active spectroscopic methods such as LAS and LIF, as is development of new spectroscopic approaches such as dual-frequency comb spectroscopy. The development of compact tabletop shorter wavelength light sources (EUV, soft X-ray etc.) could be very impactful for overcoming the critical density limitations seen in optical probing methods.  

This article highlights the basic principles of most common LPP optical characterization tools. There are other useful  techniques that use photons for plasma characterization which are not discussed here -  e.g., optical polarimetry \cite{Davies2014}, angular refractive refractometer \cite{follett2016plasma}, THz spectroscopy \cite{Herzer2018, Jamison2003}, Zeeman splitting \cite{Zeeman1984} etc. Besides, there exist an array of plasma diagnostic tools outside the optical regime (e.g., electrical and magnetic tools), which provide additional information about the kinetics of laser produced plasmas.   

\begin{acknowledgments}
This work was partially supported by the U.S. Department of Energy (DOE) National Nuclear Security Administration (NNSA) Office of Defense Nuclear Nonproliferation Research and Development (DNN R $\&$ D), and the Department of the Defense, Defense Threat Reduction Agency (DTRA) under award number HDTRA1-20-2-0001. The work of D.H.F was supported by DOE/NNSA under award Number DE-NA0003856, the University of Rochester, and the New York State Energy Research and Development Authority. Pacific Northwest National Laboratory is a multi-program national laboratory operated by Battelle for DOE under Contract DE-AC05-76RL01830. The content of the information given in this article does not necessarily reflect the position or the policy of the United States federal government, and no official endorsement should be inferred. K. K. A.   acknowledges the SERB (Grant No. TAR/2018/000340), and the UGC for financial assistance.  R.C.I acknowledges DST-FIST, CUSAT  through SMNRI-'21, and Chancellor's award for funding support. The authors are thankful to Dr. Pia Valdivia, Dr. Sean Stave and  Dr. Liz Kautz for their interest, help and support and  B. N. Balakrishna Prabhu and Catie Himes for editorial assistance.  
\end{acknowledgments}
\break



\bibliography{Main_Reference.bib}

\end{document}